\documentclass[11pt]{article}
\usepackage[utf8]{inputenc}
\usepackage[T1]{fontenc}
\usepackage{graphicx}
\usepackage{grffile}
\usepackage{longtable}
\usepackage{wrapfig}
\usepackage{rotating}
\usepackage[normalem]{ulem}
\usepackage{amsmath}
\usepackage{textcomp}
\usepackage{amssymb}
\usepackage{capt-of}
\usepackage{jheppub}
\usepackage{dsfont}
\hypersetup{colorlinks=true,linkcolor=blue}
\def\verytiny{\kern.08em}
\newlength\bshft
\def\fakebold#1{\setbox0=\hbox{$#1$}#1\kern-\wd0\kern\bshft#1\kern-\wd0\kern\bshft#1}
\abstract{Rational coefficients of special functions in scattering amplitudes are known to simplify on singular surfaces, often diverging less strongly than the naïve expectation. To systematically study these surfaces and rational functions on them, we employ tools from algebraic geometry. We show how the divergences of a rational function constrain its numerator to belong to symbolic powers of ideals associated to the singular surfaces. To study the divergences of the coefficients, we make use of $p\verytiny\text{-adic}$ numbers, closely related to finite fields. These allow us to perform numerical evaluations close to the singular surfaces in a stable manner and thereby characterize the divergences of the coefficients.  We then use this information to construct low-dimensional Ans\"atze for the rational coefficients. As a proof-of-concept application of our algorithm, we reconstruct the two-loop $0 \rightarrow q\bar q\gamma\gamma\gamma$ pentagon-function coefficients with fewer than 1000 numerical evaluations.}
\author[a]{Giuseppe De Laurentis,}
\emailAdd{giuseppe.de.laurentis@physik.uni-freiburg.de}
\affiliation[a]{Physikalisches Institut, Albert-Ludwigs-Universität at Freiburg, Hermann-Herder.Str.~3, D-79104 Freiburg, Germany}
\author[b]{Ben Page}
\emailAdd{ben.page@cern.ch}
\affiliation[b]{Theoretical Physics Department, CERN, Geneva, Switzerland}
\preprint{\begin{minipage}[t]{8cm}\begin{flushright} FR-PHENO-2022-03, \\ CERN-TH-2022-014\end{flushright}\end{minipage}}
\usepackage{color}
\usepackage{bbm}
\usepackage{enumitem}
\date{\today}
\title{\boldmath Ansätze for Scattering Amplitudes from  \(p\kern.08em\text{-adic}\) Numbers and Algebraic Geometry}
\hypersetup{
 pdfauthor={},
 pdftitle={\boldmath Ansätze for Scattering Amplitudes from  \(p\kern.08em\text{-adic}\) Numbers and Algebraic Geometry},
 pdfkeywords={},
 pdfsubject={},
 pdfcreator={Emacs 26.3 (Org mode 9.4.6)}, 
 pdflang={English}}
\begin{document}

\maketitle

\section{Introduction}
\label{sec:org6fa0859}

Precise theoretical predictions for collider experiments rely on increasingly
higher order and higher multiplicity calculations of scattering amplitudes. The
standard method of computation is to express the amplitudes in a basis of dimensionally-regulated master
integrals, reducing the calculation to determining the associated prefactors. These are rational functions of the external kinematics and of the dimensional regularization parameter. However, due to the algebraic
complexity of both intermediate stages and final results of analytic
calculations, this poses a considerable challenge. To combat the
difficulty of rational prefactor computations, in recent years it has become
commonplace to compute loop amplitudes numerically over so-called
``finite fields'' \cite{vonManteuffel:2014ixa,Peraro:2016wsq}, and subsequently obtain the
analytic form of the result by making use of an appropriate Ansatz.
By now there exist a number of advanced approaches for fitting specialized
Ansätze from numerical evaluations.
In cases where the target function is a rational function of an independent set of variables, then there exist ``functional reconstruction'' algorithms for an arbitrary number of variables
\cite{vonManteuffel:2014ixa,Peraro:2016wsq,Klappert:2019emp,Klappert:2020aqs}. 
In the multivariate case, these approaches reduce rational function
interpolation to the simpler polynomial case where either Newton
\cite{Peraro:2016wsq} or Vandermonde \cite{Klappert:2020aqs,Abreu:2021asb}
approaches are used. 
By now these are sufficiently well understood that there exist public implementations
\cite{Peraro:2019svx,Klappert:2019emp,Klappert:2020aqs}.
An important recent success of the Ansatz approach
is the calculation of a plethora of two-loop, five-point scattering
amplitudes -- both for fully massless configurations
\cite{Abreu:2018aqd,Chicherin:2018yne,Abreu:2019rpt,Chicherin:2019xeg,Badger:2018enw,Abreu:2018zmy,Abreu:2019odu,Abreu:2021oya,Badger:2019djh,Chawdhry:2019bji,Abreu:2020cwb,Chawdhry:2020for,Agarwal:2021grm,Agarwal:2021vdh,Chawdhry:2021mkw,Badger:2021imn}
and for configurations involving one massive particle
\cite{Badger:2021nhg,Abreu:2021asb,Badger:2021ega,Badger:2022ncb}. Furthermore, we have also seen
ground-breaking computations of three-loop four-point amplitudes
made possible by these tools \cite{Caola:2020dfu,Caola:2021rqz,Bargiela:2021wuy}.

In this work, we focus on an important problem found when applying the Ansatz
formalism to processes with a large number of scales.
Specifically, as one considers scattering amplitudes that depend on an increasing number of scales, the complexity of functional reconstruction approaches grows
exponentially. For example, the Ansätze used in the computation of the two-loop finite remainders for five-parton scattering
required \(\mathcal{O}(10^5)\) evaluations \cite{Abreu:2018zmy}, while those for four partons and a \(W\text{-boson}\)
required \(\mathcal{O}(10^6)\) evaluations \cite{Abreu:2021asb}. 
Despite these examples, there is growing evidence that more compact Ansätze for rational prefactors should exist.
Firstly, if we look towards highly
supersymmetric theories, we see that an Ansatz consisting of leading singularities made it possible to construct the full-color, two-loop, five-point amplitudes in
\(\mathcal{N}=4\) super-Yang-Mills theory with only 6 numerical evaluations
\cite{Abreu:2018aqd,Chicherin:2018yne}.
Secondly, in Ref.~\cite{Abreu:2018zmy}, it was observed that, when amplitudes are
expressed in a basis of ``pentagon functions'' \cite{Gehrmann:2018yef}, the
denominators of the rational prefactors can be derived from the symbol alphabet
of the associated integrals. This observation has led to efficient algorithms for determination
of the denominator factors \cite{Abreu:2018zmy,DeLaurentis:2019phz,Heller:2021qkz}.
Finally, it has been observed that two-loop, five-point amplitudes
in quantum chromodynamics (QCD) simplify when the rational prefactors are cast
in different types of partial-fraction decompositions, see
Refs.~\cite{Abreu:2019odu,DeLaurentis:2020qle,Badger:2021nhg}. The Le\u{\i}nartas
representation
\cite{leinartas1978factorization,raichev2012leinartas,Pak:2011xt,Meyer:2016slj}
has received particular attention with a number of algorithms for its
computation \cite{Abreu:2019odu,Boehm:2020ijp,Heller:2021qkz}.

Given this large body of evidence for the existence of compact Ansätze for
rational prefactors, our aim is to develop an approach to algorithmically construct such Ansätze
and thereby enable the analytic computation of two-loop, multi-scale amplitudes
with dramatically fewer numerical evaluations over finite fields.
To this end, we wish to exploit the well-known fact that gauge-theory amplitudes
admit more compact representations when expressed in terms of spinor-helicity
variables. This leads us to develop a framework based on the 
approach of Ref.~\cite{DeLaurentis:2019phz}, which has been applied to a
number of other amplitudes
\cite{DeLaurentis:2019vkf,Budge:2020oyl,Campbell:2021mlr}.
In this approach, one takes a perspective on the organization of rational
prefactors based on their behavior on singular surfaces.
Specifically, rational prefactors are studied numerically near surfaces where
one or more of the denominator factors vanish and this information is
incorporated into an Ansatz.
In order to set up this approach algorithmically, we formalize a number of its
ingredients using methods from computational algebraic geometry. These methods
have already found ample application in the scattering-amplitude literature
(see, for example, applications to integrand reduction
\cite{Badger:2012dp,Zhang:2012ce,Mastrolia:2012an} and integration-by-parts
relations \cite{Gluza:2010ws}).
Firstly, we interpret the spinor-helicity formalism in the language of algebraic
geometry, allowing us to use the tools of Gröbner bases to understand and solve
the problem of constructing linearly-independent polynomials of spinor brackets.
Secondly, we discuss how singular surfaces often have multiple
branches and we show how to systematically identify these branches by
constructing the primary decomposition of an ideal associated to the singular surface.
Thirdly, we show how the behavior of a rational function when approaching
surfaces where multiple denominators are singular is encoded in its analytic
structure.
Specifically, we show that the numerator of the rational function must belong to a certain 
ideal, controlled by the geometry of the singular surface. The relevant tool is
provided to us by the Zariski--Nagata theorem
\cite{Zariski1949,Nagata1962,EISENBUD1979157}, which tells us to consider the
so-called ``symbolic power'' of an associated ideal.

In order to determine the singular behavior of the rational prefactors, we
introduce a new numerical tool. Modern methods for two-loop amplitude
calculation rely on the absence of precision loss when working over finite
fields. However, finite fields
lack a concept of scale separation that is required to probe singular configurations.
To this end, we work in a middle ground provided to us by number theoretical
techniques: the \(p\verytiny\text{-adic}\) numbers (see
Ref.~\cite{Gouvea1997} for an introduction). While these objects are a rich
source of number theory, with their own notion of calculus, here we will only
scratch the surface and use their properties as a field.
One can regard them as bridging the gap between finite fields and floating-point
numbers: \(p\verytiny\text{-adic}\) numbers have natural expansions in powers of a prime
\(p\), and the first digit in such an expansion behaves like a finite field. This
set of numbers comes associated with a concept of size, which allows us to
perform numerical studies in singular configurations. 
At the same time, by working with large primes \(p\), there is a low probability
of numerical \(p\verytiny\text{-adic}\) calculations involving a spurious scale hierarchy
in intermediate stages. This makes it possible to control accidental precision
loss in numerical computation.
Combining this numerical tool with the algebro-geometric understanding, we then
present an algorithm for the construction of Ansätze for rational prefactors,
that can in principle be automated and applied to the computation of novel
two-loop scattering amplitudes.

This article is organized as follows. In Section \ref{sec:org2bd70b8}, we present algebro-geometric tools that allow us to understand functions
on spinor space and perform a systematic study of the singular varieties of
rational prefactors. Next, in Section \ref{sec:orgc26bb05},
we give an introduction to \(p\verytiny\text{-adic}\) numbers and explain how to generate
numerical phase-space points on or near singular varieties. Thereafter, in
Section \ref{sec:orgcdcb75d} we collect the theoretical work of the
previous section into an algorithm to generate compact Ansätze which leverage the
singularity information. In Section \ref{sec:application_to_photons} we make an example
application of the algorithm to the \(0 \rightarrow q\bar q\gamma\gamma\gamma\)
finite remainder coefficients at two loops. Finally, we summarize and conclude
in Section \ref{sec:summary}.

\section{Algebraic Geometry and Spinor Space}
\label{sec:org2bd70b8}
\label{sec:SpinorHelicityDefinitions}

Scattering amplitudes are transcendental functions
of the external kinematics, which are
typically evaluated on the set of
four-momenta associated to the external states. For an \(n\text{-point}\) process,
these are a collection of momenta \(\{k_1, \ldots, k_n\}\) which satisfy on-shell
and momentum-conservation relations, that is
\begin{equation}
k_i^2 = m_i^2 \quad \; \mathrm{and} \quad \; 0 = \sum_{i=1}^n k_i^\mu \, ,
\end{equation}
where we use the all-outgoing convention.
In the case of massless scattering, where \(m_i=0\),
it is natural to employ spinor variables
instead of Mandelstam variables to describe scattering
amplitudes. The connection between a massless four momentum \(k_i\) and a
pair of Weyl spinors \((\lambda_i, \tilde{\lambda}_i)\), is
made through the relation
\begin{equation}\label{eq:four-mom-spinor-relation}
k_{i\mu} \sigma^{\mu\dot{\alpha} \alpha} = \tilde{\lambda}^{\dot{\alpha}}_i \lambda_i^\alpha \, ,
\end{equation}
where \(\sigma^{\mu\dot{\alpha} \alpha} = (\mathds{1}, \vec\sigma)\) denotes an Infeld--Van
der Waerden symbol, and \(\vec\sigma\) are the three Pauli matrices. We take the metric on spinor space
to be the \(2\times 2\) Levi-Civita symbol \(\smash[b]{\epsilon^{\alpha\beta}=\epsilon^{\dot\alpha\dot\beta}=i\sigma_2}\).
The metric with lowered indices is then \(\smash[b]{\epsilon_{\alpha\beta}=\epsilon_{\dot\alpha\dot\beta}=(i\sigma_2)^T}\), such that \(\smash[b]{\epsilon_{\alpha\beta}\epsilon^{\beta\gamma}=\delta_\alpha^\gamma}\), where \(\delta\) is the Kronecker delta.
Raising and lowering of the indices is achieved by contraction with the metric,
that is~\(\smash[b]{\lambda_{i\alpha} = \epsilon_{\alpha\beta}\lambda_i^\beta}\)
and~\(\smash[b]{\tilde{\lambda}_{i\dot{\alpha}} = \epsilon_{\dot{\alpha}\dot{\beta}}\tilde{\lambda}_i^{\dot{\beta}}}\).
Spinor-helicity variables trivialize
on-shell relations, while momentum conservation becomes a quadratic relation
\begin{equation}
0 = \sum_{i=1}^n \tilde{\lambda}_{i}^{\dot{\alpha}} \lambda_{i}^{\alpha}  \, .
\label{eq:spinorMomentumConservation}
\end{equation}
Invariant quantities can be built by contracting the spinors
in so-called spinor \textit{brackets}. 
We define them through the following contractions 
\begin{align}
\langle i j \rangle = \lambda^\alpha_i \lambda_{j\alpha} \quad \; \text{and} \quad \; [i j] = \tilde\lambda_{i\dot\alpha}\tilde\lambda^{\dot\alpha}_j \, ,
\label{eq:spinorBrackets}
\end{align}
where the Einstein summation convention is implied over the spinor indices. Furthermore, we will make use of simple spinor chains, specifically we define
\begin{align}
\begin{split}
    \langle i | j+k | l] &= \langle ij \rangle [jl] + \langle i k \rangle [kl] \, , \\
    [ i | j+k | l \rangle &= [ij]\langle jl \rangle + [ik] \langle k l \rangle \, , 
\end{split}
\end{align}
as well as 
\begin{align}
\begin{split}
    \langle i | j+k | l+m | n\rangle &= \langle i | j+k | l]\langle ln \rangle + \langle i | j+k | m]\langle mn \rangle \, , \\
    [ i | j+k | l+m | n ] &= [ i | j+k | l \rangle [ ln ] + [ i | j+k | m \rangle [ mn ] \, .
\end{split}
\end{align}
For brevity, we will often denote an \(n\text{-point}\) phase-space
point as \((\lambda, \tilde{\lambda})\).

Physical functions of spinor variables satisfy a number of
properties. Specifically, they have well-defined mass dimension and
little-group weights. 
Working with some function \(\mathcal{E}(\lambda,
\tilde{\lambda})\) with well-defined mass dimension means that if we
uniformly scale all of the spinors by \(z\) then we find that
\begin{equation}
    \mathcal{E}(z \lambda_1, z \lambda_2, \ldots, z \tilde{\lambda}_1, z\tilde{\lambda}_2, \ldots) = z^{2 [\mathcal{E}]}\mathcal{E}(\lambda_1, \lambda_2, \ldots, \tilde{\lambda}_1, \tilde{\lambda}_2 \ldots) \, ,
\label{eq:MassDimensionDefinition}
\end{equation}
where \([\mathcal{E}]\) is the mass dimension of \(\mathcal{E}\). If \(\mathcal{E}\)
has well-defined little-group weight \(k\) then this means that if we scale
\(\lambda_k\) by \(z\) and \(\tilde{\lambda}_k\) by \(1/z\) we find
\begin{equation}
    \mathcal{E}(\ldots, z \lambda_k, \ldots, \tilde{\lambda}_k / z, \ldots) = z^{\{\mathcal{E}\}_k}\mathcal{E}(\ldots, \lambda_k, \ldots, \tilde{\lambda}_k, \ldots) \, ,
\label{eq:LittleGroupScalingDefinition}
\end{equation}
where \(\{\mathcal{E}\}_k\) is the \(k^\text{th}\) little-group weight of
\(\mathcal{E}\). Importantly, this rescaling does not affect the validity of Eq.~\eqref{eq:four-mom-spinor-relation}. 
We will always work with functions \(\mathcal{E}\) with well-defined mass
dimension and little-group weights.
Finally, note that for such functions, it can be useful to compute the mass-dimension
and little-group weights numerically via
Eqs.~\eqref{eq:MassDimensionDefinition} and
\eqref{eq:LittleGroupScalingDefinition}.

Scattering amplitudes have well-defined mass dimension
and little-group weights. The mass dimension depends only on the 
multiplicity of the process. For an
\(n\text{-point}\) amplitude \(\mathcal{A}_n\) it is well known that the
mass dimension is given by
\begin{equation}
    [\mathcal{A}_n] = 4 - n \, .
\end{equation}
The little-group weights depend on the helicity states\footnote{As
standard, massless \(s\text{-spin}\) states will have helicity \(\pm s\),
i.e.~\(h=0\), \(\pm 1/2\) and \(\pm 1\) for scalars, spin-1/2 fermions
and vectors respectively.} of the scattered particles, more precisely
the \(k^\text{th}\) little-group weight depends on the helicity state of
the \(k^\text{th}\) particle, denoted as \(h_k\). In the all-outgoing
convention, the \(k^\text{th}\) little-group weight is given by
\begin{equation}
    \{\mathcal{A}_n\}_k = -2\,h_k \, .
\end{equation}
Beyond tree level, when working in dimensional regularization, a scattering
amplitude can be decomposed as a linear combination of so-called ``master
integrals''. Such a decomposition can be written as
\begin{equation}
    \mathcal{A}^{(l)}_n = \sum_{i} \mathcal{B}_i(\lambda, \tilde{\lambda}, \epsilon) \mathcal{I}_i(\lambda, \tilde{\lambda}, \epsilon)\, ,
\label{eq:amplitudeMasterDecomposition}
\end{equation}
where the \(\mathcal{B}_i\) are rational functions of the spinors
 \((\lambda, \tilde{\lambda})\) and of the dimensional
regulator \(\epsilon\), and the \(\mathcal{I}_i\) are transcendental
functions thereof. 
It is well understood that amplitudes in gauge theory diverge in a universal way
 \cite{Catani:1998bh, Becher:2009cu, Gardi:2009qi,Becher:2009qa}, (see Ref.~\cite{Agarwal:2021ais} for a recent review). Specifically, after renormalization,
 these divergences can be written in terms of lower loop amplitudes and
 universal operators. That is, one can write
\begin{equation}
   \mathcal{A}^{(l)}_{n,R} = \sum_{l' = 0}^{l-1} {\bf I}^{(l-l')} \mathcal{A}^{(l')}_{n,R} + \mathcal{H}_n^{(l)} + \mathcal{O}(\epsilon) \, ,
  \label{eq:AmplitudeDivergence}
\end{equation}
where \(\mathcal{A}^{(l)}_{n,R}\) is the renormalized \(l-\text{loop}\) amplitude and
we have introduced the so-called ``finite remainder'' \(\mathcal{H}_n^{(l)}\),
which captures the new information at each perturbative order. In practice, one
computes the finite remainder by inserting the \(\epsilon\) expansion of the
master integrals into Eq.~\eqref{eq:amplitudeMasterDecomposition} and
subtracting the lower loop contributions in
Eq.~\eqref{eq:AmplitudeDivergence}.
The resulting expression for the finite remainder can be expressed in a basis of
special functions. That is, in general we can write
\begin{equation}
    \mathcal{H}^{(l)}_n = \sum_{i} \mathcal{C}_i(\lambda, \tilde{\lambda}) \mathcal{F}_i(\lambda, \tilde{\lambda}) \, ,
\label{eq:remainderFunctionDecomposition}
\end{equation}
where the \(\mathcal{C}_i\) are rational functions of the spinors and
the \(\mathcal{F}_i\) are special functions of the spinors.
In this work, we consider the \(\mathcal{C}_i\) in common denominator form. Specifically, we write 
\begin{equation}
  \mathcal{C}_i(\lambda, \tilde{\lambda}) 
    = \frac{\mathcal{N}_i(\lambda, \tilde{\lambda})}
      {\prod_{j=1}^{n_i} \mathcal{D}_{j}(\lambda, \tilde{\lambda})^{q_{ij}}} \, ,
  \label{eq:CommonDenominatorForm}
\end{equation}
where \(n_i\) is the number of denominator factors and the \(\mathcal{N}_i\) and \(\mathcal{D}_j\) are polynomials of spinors.
As is well known, the amplitude only picks up a little-group rescaling under Lorentz transformations
and so it can only depend on the spinors indirectly through the spinor
brackets
of Eq.~\eqref{eq:spinorBrackets}. In practice, it is trivial to choose the
basis of transcendental functions \(\mathcal{F}_i\) to also have this
property, and so the coefficient functions inherit it as
well. Importantly, the \(\mathcal{N}_i\) and \(\mathcal{D}_j\) all have
well-defined mass dimension and little-group weights.

For the rest of this work, we shall work in a framework where we are able to
numerically evaluate the \(\mathcal{C}_i\) over an arbitrary field. In practice, this may be
when one has an explicit analytic form available, or an
appropriate algorithm to compute the \(\mathcal{C}_i\). Our aim
is then to use this numerical information in an efficient way to determine the
analytic form of the functions \(\mathcal{C}_i\).

\subsection{Rudiments of Algebraic Geometry}
\label{sec:org4a8329e}
\label{AlgebraicGeometryBasics}

The coefficients \(\mathcal{C}_i\) in a scattering amplitude have been
introduced in the previous section as ratios of polynomials,
which are to be evaluated on inputs that satisfy momentum-conservation
relations.
In this section, we introduce
basic technologies of algebraic geometry which will allow us to
understand polynomials in this context in detail. We
intend our presentation to be self-contained and we refer the
reader to Refs.~\cite{cox1994ideals,Zhang:2016kfo,becker2012groebner} for an
introductory account of the requisite algebraic geometry.

\subsubsection{Polynomials, Ideals and Varieties}
\label{sec:orgd3afc14}
The central object of study will be polynomials in spinor variables.
Therefore, we consider the \textbf{polynomial ring} of spinor variables for
\(n\) massless particles,
\begin{equation}\label{eq:covariant-scalar-ring}
   S_n = \mathbb{F}\bigl[
\lambda_{10}, \lambda_{11}, \ldots, \lambda_{n0}, \lambda_{n1}, \tilde{\lambda}_{1 {\dot 0}}, \tilde{\lambda}_{1 {\dot 1}}, \ldots, \tilde{\lambda}_{n {\dot 0}}, \tilde{\lambda}_{n {\dot 1}}
\bigr] \, ,
\end{equation}
where \(\mathbb{F}\) is the coefficient field and the variables are the various \(\lambda_{i \alpha}\)
and  \(\tilde{\lambda}_{j \dot{\alpha}}\).
All polynomials in spinor variables are elements of \(S_n\). For example, the spinor
brackets defined in Eq.~\eqref{eq:spinorBrackets} can be identified as elements of \(S_n\). Furthermore, constraints on the spinors, such as momentum conservation
\eqref{eq:spinorMomentumConservation} or being on a particular surface,
are expressed using elements of \(S_n\).  
Here, and throughout this
work, we shall abstract over the field \(\mathbb{F}\).  In practice,
we can consider \(\mathbb{F}\) to be the rational numbers \(\mathbb{Q}\), the real
numbers \(\mathbb{R}\), the complex numbers \(\mathbb{C}\), a finite field
\(\mathbb{F}_p\) or the \(p\verytiny\text{-adic}\) numbers \(\mathbb{Q}_p\), which we will
discuss in Section \ref{sec:padicIntro}. For
theoretical considerations, such as considering the geometry, we will always work in an algebraically
closed field such as the complex numbers. For practical calculations,
we will be working over finite fields or \(p\verytiny\text{-adic}\)
numbers.  Throughout this work, we will take this polynomial
perspective as our foundation. This perspective explicitly breaks
Lorentz covariance in intermediate stages of our calculation, as we work in a
given frame. Furthermore, the ring \(S_n\) contains unphysical polynomials, such
as ones without well-defined mass dimension and little-group weight. We will
return to the question of imposing these constraints in Section
\ref{sec:IndependentFunctions}.

\paragraph{Ideals.}
The key algebraic object that we use is a so-called \textbf{ideal}. We will work with
rings, such as polynomial rings, in which ideals are finitely generated. 
Specifically, we consider a set of elements \(\{p_1, \ldots, p_k\} \in A\) called
\textbf{generators} and define an ideal of \(A\) as
\begin{equation}
    \bigl \langle p_1, \ldots, p_k \bigr \rangle_{A} 
    \, = \,
    \left\{ \sum_{i=1}^k a_i p_i, \, \, a_i \in A \right\} .
\label{eq:idealDefinition}
\end{equation}
Here, as we will work in a number of rings, we extend the
\(\langle \ldots \rangle\) notation of Ref.~\cite{cox1994ideals} with a
subscript to denote the ring under consideration. When discussing an ideal we typically label it as \(J\) or \(K\). 
From the definition in Eq.~\eqref{eq:idealDefinition}, it is clear that an ideal always
forms a subset of the ring \(A\). In the case where the subset is proper,
that is when we have an ideal \(J\) such that \(J \subsetneq A\), we say that \(J\) is a
\textbf{proper ideal}.
We refer to the set
\(\{p_1, \ldots, p_k\}\) as a generating set of the ideal. In
practice we will always consider physical generating sets,
i.e.~sets in which each \(p_i\) has well-defined mass dimension and little-group weight. Furthermore, we will often consider ideals that are generated
by multiple elements which can be grouped into an object with some open spinor
index. In this case we will use a natural shorthand where we do not write the
individual generators, but only the object with an open index. As a simple
example consider
\begin{equation}
\langle \lambda_{j \alpha} \rangle_{S_n} \overset{!}{=} \langle \lambda_{j 0}, \lambda_{j 1} \rangle_{S_n} \, .
\end{equation}
Multiple generating sets may correspond to the same
ideal, i.e.~they are not unique, and different generating sets
of the same ideal may have a different number of elements. For any
ideal \(J\), there exist generating sets with a minimal number of
elements and any such generating set is called a \textbf{minimal generating set}. 
The size of a minimal generating set is denoted by
\begin{equation}\label{eq:minimal_generating_set}
\mu(J) = \mathrm{min}\left(\big\{ |G| \,\, : \,\, G \text{ is a generating set of } J \big\}\right) ,
\end{equation}
where \(|G|\) denotes the number
of elements of the generating set \(G\).  While still not unique, we
will always present ideals through minimal generating sets. For ideals generated by
homogeneous polynomials, such as those which we consider, minimal
generating sets can be determined algorithmically\footnote{For example,
one can find such an algorithm implemented in the computer algebra
system \texttt{Singular} \cite{DGPS} under the \texttt{minbase} command.}.  In
practice, we will always have a generating set of the ideal at
hand. A trivial example of an ideal is the ideal generated by the zero element
of the ring. This is the set containing only the zero element, that is
\begin{equation}
    \langle 0 \rangle_A = \{0\} \, .
\end{equation}
As algebraic objects, ideals have natural algebraic operations associated to
them. For example, we will make use of the \textbf{ideal sum}, which we define through
\begin{equation}
    \langle p_1, \ldots, p_k \rangle_A + \langle q_1, \ldots, q_l \rangle_A = \langle p_1, \ldots, p_k, q_1, \ldots, q_l \rangle_A \, .
\end{equation}
Furthermore, one can take the \textbf{ideal product}.
Given two ideals \(J = \langle p_{1}, \ldots, p_a \rangle_A\) and \(K = \langle
q_{1}, \ldots, q_{b} \rangle_A\), we define the ideal product \(JK\) as
\begin{equation}
JK = \langle  p_i q_j \,\, : \,\, 1 < i \le a \, , \; 1 < j \le b \rangle_A \, ,
\label{eq:IdealProductDefinition}
\end{equation}
that is, the generators of \(JK\) are the products of the generators of \(J\)
and \(K\). It is clear that the ideal product is commutative, i.e.~\(JK = KJ\).
It will also be useful to consider 
the \textbf{ideal power} \(J^k\), which we define recursively through
\begin{equation}
   J^0 = \langle 1 \rangle_A \quad \; \text{and} \quad \; J^k = J J^{k-1} \, .
\label{eq:IdealPowerDefinition}
\end{equation}

When working with an ideal \(J\) in \(S_n\), we will often be interested in other
ideals which can be constructed from \(J\) by parity or permutations of the
associated spinors.
We define a permuted ideal through
\begin{equation}
\label{eq:IdealPermutation}
 J(\sigma(1)\dots\sigma(n)) 
 = J|_{\lambda_i \rightarrow \lambda_{\sigma(i)}, \, \tilde \lambda_i \rightarrow \tilde \lambda_{\sigma(i)}} \, ,
\end{equation}
where \(\sigma\) is a permutation of \(\{1, \ldots, n\}\).
We will also consider the parity conjugate ideal \(\overline{J}\) defined by a
swap of the \(\lambda\) and \(\tilde{\lambda}\) spinors, that is
\begin{equation}
\label{eq:IdealParity}
 \overline{J} = J|_{\lambda_\alpha \leftrightarrow \tilde\lambda_{\dot\alpha}} \, .
\end{equation}
In practice, one computes generating sets of the these ideals by applying the
permutation/parity conjugation to the generators of \(J\).

\paragraph{Algebraic Varieties.}
Now, note that for \(n\text{-point}\) spinor space, we can regard the tuple of
spinor variables \(\{\lambda_{10}, \lambda_{11}, \ldots, \lambda_{n0},
\lambda_{n1}, \tilde{\lambda}_{1 {\dot 0}}, \tilde{\lambda}_{1 {\dot 1}},
\ldots, \tilde{\lambda}_{n {\dot 0}}, \tilde{\lambda}_{n {\dot 1}}\}\) as taking
values in the \(4n\text{-dimensional}\) space \(\mathbb{F}^{4n}\).
Physical spinors are constrained to satisfy momentum
conservation according to Eq.~\eqref{eq:spinorMomentumConservation}. It is therefore natural to
consider the set of solutions of momentum conservation in \(\mathbb{F}^{4n}\),
which defines a so-called \textbf{algebraic variety}.
In general, we can associate a variety to any ideal in \(S_n\). That is, given an
ideal \(J =\langle p_1, \ldots, p_k \rangle_{S_n}\), the associated algebraic
variety is defined as
\begin{equation}
   V(\langle p_1, \ldots, p_k\rangle_{S_n}) = \Big\{ (\lambda, \tilde{\lambda}) \in \mathbb{F}^{4n} \, : \, p_i(\lambda, \tilde{\lambda}) = 0 \,\, \mathrm{for} \,\, 1 \le i \le k \Big\} \, .
\label{eq:VarietyDefinition}
\end{equation}
From this definition it is clear that \(V(J) \subseteq \mathbb{F}^{4n}\) for any ideal \(J\). 
While we have defined an algebraic variety over an arbitrary field \(\mathbb{F}\),
many powerful theorems of algebraic geometry can be applied only when \(\mathbb{F}\) is an
algebraically closed field, such as the complex numbers. In
this paper, we will always work over these fields when considering geometry.
We remark that the definition of a variety in Eq.~\eqref{eq:VarietyDefinition} allows varieties to be
``reducible'', an important fact we shall return to in detail in Section
\ref{sec:IrreducibleSingularVarieties}.
We note two trivial cases: \(V(\langle 1 \rangle_{S_n})\) corresponds to the empty
variety and \(V(\langle 0 \rangle_{S_n})\) corresponds to all of \(\mathbb{F}^{4n}\).

In the same way that we have just associated a variety to an ideal, we can
naturally associate an ideal to a variety. Specifically, for a variety \(U\) in
\(\mathbb{F}^{4n}\) it turns out that the set of polynomials that vanish on \(U\)
forms an ideal, which is defined as
\begin{equation}
   I(U) = \Big\{p \in S_n \,:\,p(\lambda, \tilde{\lambda}) = 0 \,\, \text{for all}\,\, (\lambda, \tilde{\lambda}) \in U \Big\} \, .
\end{equation}

\paragraph{Application to Momentum Conservation.}
To understand these ideas in a physical context, consider the
ideal of \(S_n\) generated by the four momentum-conservation polynomials
of Eq.~\eqref{eq:spinorMomentumConservation}, which we denote as
\begin{equation}\label{eq:momentum_conservation_ideal}
    J_{\Lambda_n}  = \left\langle \sum_{i=1}^n \lambda_{i \alpha} \tilde{\lambda}_{i \dot{\alpha}} \right\rangle_{S_n}.
\end{equation}
We will refer to \(J_{\Lambda_n}\) as ``the momentum-conservation ideal''. Physically,
\(J_{\Lambda_n}\) is the set of all polynomials in spinor variables which are
rewritings of zero.
The associated variety is the set of points in spinor space that satisfy momentum
conservation. We dub this the ``momentum-conservation variety'' and it is
denoted as \(V(J_{\Lambda_n})\).
All varieties of interest in this work will be sub-varieties of
\(V(J_{\Lambda_n})\), as physical configurations of spinors must satisfy momentum conservation.
We note that 
\begin{equation}\label{eq:radical}
   J_{\Lambda_n} = I(V(J_{\Lambda_n})) \, .
\end{equation}
That is, the momentum-conservation ideal contains all polynomials which vanish
on the momentum-conservation variety.
In an algebraically closed field, the operation in Eq.~\eqref{eq:radical} of
taking the ideal associated to the variety associated to an ideal corresponds to
taking the \textbf{radical} of an ideal \cite[Chapter 4]{cox1994ideals}, which we will
denote as
\begin{equation}
   \sqrt{J} = I(V(J)) \, .
  \label{eq:radicalNotation}
\end{equation}
If it is the case that \(\sqrt{J} = J\), then the ideal \(J\) is said to be
``radical'' (see Appendix \ref{app:AlgGeoGlossary} for the
algebraic definition). Therefore, we see that
Eq.~\eqref{eq:radical} says that \(J_{\Lambda_n}\) is radical.

\subsubsection{Independent Sets and Dimension}
\label{sec:orgee14215}
\label{sec:IndependentSets}    
A natural question to ask of any geometric structure is its dimension.
Varieties, as surfaces defined by algebraic equations, indeed have a concept of
dimension that we can associate to them. Furthermore, one can also associate a
concept of dimension to an ideal. In this section, we introduce the 
concepts relevant for our work and refer the reader to Ref.~\cite[Section 6.3]{becker2012groebner} 
for a deeper treatment.

In order to ease the discussion, we will work over the polynomial ring
\(\mathbb{F}[X_1, \ldots, X_n]\). We will denote the collection of variables as
\(\underline X = \{X_1, \dots, X_n\}\). We will further denote a subset of the
variables as \(\underline{Y} \subseteq \underline{X}\).
To begin phrasing the question of dimension we ask if the
variables \(\underline{Y}\) can be chosen independently on the variety \(V(J)\).
The important observation is that the answer will be `no' if there is some
polynomial in the ideal \(J\) which depends only on the variables \(\underline{Y}\).
If there is no such polynomial, then the variables are not constrained in terms
of each other, and so the variables \(\underline{Y}\) can be chosen independently.
We can formally state the question of the existence of such a polynomial by
considering the associated \textbf{elimination ideal} defined as\footnote{This can be computed via
Gröbner basis methods, see e.g. Ref.~\cite[Chapter 3]{cox1994ideals}. Nevertheless,
computation of the elimination ideal can be avoided when computing the dimension.}
\begin{equation}
    J_{\underline{Y}} = J \cap F[\underline{Y}] \, ,
    \label{eq:EliminationIdealDefinition}
\end{equation}
that is, the intersection of the ideal \(J\) with the set of all
polynomials which depend only on the variables \(\underline{Y}\). 
An \textbf{independent set} \(\underline{Y}\) of a proper ideal
\(J\) is defined by requiring that the associated elimination ideal \(J_{\underline{Y}}\) contains
only the zero element, that is
\begin{equation}\label{eq:definition-independent-set}
 J_{\underline{Y}} = \{0\} \quad \Rightarrow \quad \underline{Y} \text{ is an independent set of } J \, .
\end{equation}
Furthermore, an independent set is said to be \textbf{maximally}
independent if there exists no other independent set which contains it. That is, 
\begin{equation}\label{eq:maximally_independent_set}
\underline Y \text{ is maximally independent if } \nexists \;\, \underline{Y'} \supset \underline{Y}\,, \, \text{ with }  \underline{Y'} \text{ and }  \underline{Y} \text{ independent sets} \, .
\end{equation}
Note that, in the general case, not all maximally independent sets
need be of the same length. 

With the definition of independent sets in hand, we can now discuss dimension.
Specifically, for a proper ideal \(J\) of a polynomial ring
\(\mathbb{F}[\underline{X}]\), the dimension of \(J\) is defined
as
\begin{equation}\label{eq:dimension}
   \dim(J) = \text{max}\left(\{|\underline{Y}| \,\, : \,\, \underline{Y} \, \text{ is an independent set of } \, J \}\right) \, .
\end{equation}
That is, the dimension of the ideal is the length of the largest
independent set. It is clear that this length is unique, as
there will always exist at least one independent set (the empty set) and we take
the length of the largest independent set. The \textbf{dimension} of the variety
associated to \(J\) is defined as
\begin{equation}
    \dim\big(V(J)\big) = \dim(J) \, .
\end{equation}
To build intuition, consider the trivial ideal \(\langle 0
\rangle_A\). In the case where \(A\) is a polynomial ring, as there are no constraints,
\(\dim\big(\langle 0\rangle_A\big)\) naturally coincides with the number of variables.
As a second example, consider a case where \(V(J)\) corresponds to a finite set of
points, then all variables are fixed on each point and so there is no non-empty
independent set. One thus finds that \(\dim(J)=0\). Naturally, \(J\) is called a
\textbf{zero-dimensional ideal}.
Importantly, efficient algorithms exist to compute both the maximally independent sets of an ideal \(J\) and \(\mathrm{dim}(J)\)
given a Gröbner basis of \(J\), see e.g.~Ref.~\cite[Proposition~9.29]{becker2012groebner}\footnote{Implementations of algorithms to compute maximally independent sets and
dimensions of ideals can be found in computer algebra systems such as
\texttt{Singular}.}.

Finally, we introduce the notion of \textbf{codimension} of an ideal. Specifically,
for a proper ideal \(J\) in a ring \(A\), we define
\begin{equation}
    \mathrm{codim}(J) = \dim(\langle 0 \rangle_A) - \dim(J) \, .
  \label{eq:CodimensionDefinition}
\end{equation}
With this language then we see that
\begin{equation}
    \mathrm{codim}(J_{\Lambda_n}) = 4 \, .
\label{eq:MomentumConservationCodimension}
\end{equation}
Intuitively, this can be regarded as the number of constraints imposed by the
generators of an ideal. We remark that the number of constraints may be less
than the number of generators. If the codimension of an ideal is equal to the
length of its minimal generating set, we say that this ideal has \textbf{maximal
codimension}, that is
\begin{equation}\label{eq:maximal_codimension}
\mu(J) = \text{codim}(J) \,\,\, \Rightarrow \,\,\, J \text{ is of maximal codimension.}
\end{equation}
For example, the momentum-conservation ideal \(J_{\Lambda_n}\) is of
maximal codimension.

\subsubsection{Gröbner Bases}
\label{sec:orgbbd96b4}
\label{GroebnerBases}    

In order to make practical use of the concepts that we present in this
paper, two major tools are polynomial reduction and Gröbner bases. Here we
review these objects in order to set up notation but, as these are common tools
in the particle physics literature, we refer the reader to Ref.~\cite{cox1994ideals}
for a pedagogical introduction.
As we work in a number of polynomial rings, in this section we shall maintain the generic
notation introduced in the previous section. 
A polynomial ring \(\mathbb{F}[\underline X]\), with \(\underline{X} = \{X_1, \ldots, X_n\}\), can be viewed as a (countably) infinite-dimensional vector
space -- the direct sum of one-dimensional spaces corresponding to the
monomials, which we denote as
\begin{equation}
    \underline{X}^{\underline{\alpha}} = \prod_{i=1}^n X_i^{\alpha_i} \, ,
\end{equation}
where \(\underline{\alpha} \in \mathbb{Z}_{\ge 0}^n\). That is, each \(\alpha_i\) is a non-negative integer. A polynomial \(p\) in \(\mathbb{F}(\underline{X})\) takes the form
\begin{equation}
    p = \sum_{\underline{\alpha} \in \mathbb{Z}_{\ge 0}^n} c_{\underline{\alpha}} \underline{X}^{\underline{\alpha}}, \qquad c_{\underline{\alpha}} \in \mathbb{F} \, ,
\end{equation}
where only a finite number of \(c_{\underline{\alpha}}\) are non-zero. A useful
structure to put on the space is a so-called
\textbf{monomial ordering}, denoted by \(\succeq\). This is a (total) ordering of the
exponents \(\underline{\alpha}\) of the monomials. Common orderings are
``lexicographic'' and ``degree reverse lexicographic''. Both of these orderings depend on an
underlying ordering of the variables \(\underline{X}\). Unless otherwise stated,
 throughout this work we use the degree reverse lexicographic ordering.
Given an ordering \(\succeq\), one can organize the terms of any polynomial and
thereby define a \textbf{lead monomial}, given by
\begin{equation}\label{eq:lead_monomial}
    \mathrm{LM}(p) = \underline{X}^{\underline{\beta}} \, , \;\, \text{where} \;\; \underline{\beta} = \mathrm{max}_{\succeq}\left(\{\underline{\alpha} : c_{\underline \alpha} \ne 0\}\right) \, ,
\end{equation}
where the maximum is taken over the set with respect to the ordering \(\succeq\).
An important application of the lead monomial of a polynomial is to define the
concept of reducibility of one polynomial by another. Specifically, one says that
\(p\) is \textbf{reducible} by \(h\) if the lead monomial of \(h\) is a factor of the lead monomial of \(p\), that is
\begin{equation}
    \mathrm{LM}(p) \,\, | \,\, \mathrm{LM}(h) \quad \Rightarrow \quad p \text{ is reducible by } h \, ,
\label{eq:ReducibilityDefinition}
\end{equation}
where we use \(x\, | \, y\) to denote that \(y\) is a factor of \(x\). If \(y\) does not factor \(x\) we write \(x \nmid y\).
If the lead monomial of \(h\) is not a factor of the lead monomial of \(p\) then we say that \(p\) is \textbf{irreducible} by \(h\).
Given a set of generators \(H = \{h_1, \ldots, h_k\}\) of an ideal \(J\), one can
then discuss \textbf{polynomial reduction}. Specifically, it turns out that one
can always write a polynomial \(p\) as
\begin{equation}\label{eq:poly_reduction}
    p = q_1 h_1 + \ldots + q_k h_k + \Delta_{H}(p) \, ,
\end{equation}
where \(\Delta_{H}(p)\) is irreducible by any of the \(h_i\). The object
\(\Delta_{H}(p)\) is of fundamental importance and is known as the \textbf{remainder
modulo \(H\)}. An important feature of remainders is that they are linear
combinations of monomials that are irreducible by the given generating set,
that is
\begin{equation}
    \Delta_{H}(p) = \sum_{{\underline{\beta}} \in \mathrm{irreds}(H) } d_{\underline{\beta}} \underline{X}^{\underline{\beta}}\, , \;\, \text{where} \;\, \mathrm{irreds}(H) = \{\underline{\beta} \; : \; \underline{X}^{\underline{\beta}} \,\, \nmid \mathrm{LM}(h) \; \forall \; h \, \in \, H\} \, .
\label{eq:RemainderRepresentation}
\end{equation}
Here \(\mathrm{irreds}(H)\) is the set of exponents whose associated monomial
is not a (polynomial) multiple of the lead monomial of any element of the set \(H\).
The aim of introducing the remainder \(\Delta_H(p)\) is to define a canonical form of
\(p\) when working modulo elements of the ideal \(J\).
However, it turns out that the remainder modulo \(H\) is
not uniquely determined by the ordering \(\succeq\) and the ideal \(J\) that it generates. It also
depends on the details of the set \(H\). Specifically, if a polynomial is reducible by an element of the ideal
\(J\) it may not be reducible by an element of the generating set \(H\).
However, given an ordering \(\succeq\), there exist special generating sets of \(J\)
that do uniquely determine the remainder. These are known as \textbf{Gröbner bases}. We
denote a Gröbner basis of an ideal \(J\) as \(\mathcal{G}(J)\). General
algorithms exist to compute Gröbner bases, which are implemented in many
computer algebra systems. Remainders modulo a Gröbner basis have the
important property that if \(p\) is in the ideal, then the remainder is zero. That is, 
\begin{equation}
    p \in J \quad \Leftrightarrow \quad \Delta_{\mathcal{G}(J)}(p) = 0 \, .
\label{eq:RemainderMembershipCondition}
\end{equation}

\paragraph{Organizing Vector Spaces by Ideals.}
\label{sec:orgd1063c6}

A useful application of Gröbner bases is to split a subspace of a polynomial
ring into a subspace that belongs to an ideal, and a remaining subspace.
Specifically, consider the polynomial ring \(\mathbb{F}[\underline{X}]\), an
ideal \(J\) of \(\mathbb{F}[\underline{X}]\) and a finite-dimensional vector
space \(W\) that is a subspace of \(\mathbb{F}[\underline{X}]\). Using \(J\), one
can split the space \(W\) into a direct sum as
\begin{equation}
    W \cong (W \cap J) \oplus W / (W \cap J) \, .
   \label{eq:SpaceDecompositionIsomorphism}
\end{equation}
The left summand \((W \cap J)\) is the subspace of \(W\) formed by all elements
that are also elements of \(J\). The right summand \(W / (W \cap J)\) is the quotient of \(W\) by
this subspace. \(W / (W \cap J)\) can be considered as the space of elements of \(W\)
modulo the elements of \(J\). To make
practical use of the decomposition in
Eq.~\eqref{eq:SpaceDecompositionIsomorphism}, we now discuss how to find
a basis of the two summand spaces.

Let us first consider how to find a basis of \(W \cap J\) given a basis \(\Omega
   = \{ \Omega_1, \ldots, \Omega_{\dim(W)} \}\) of \(W\).
We recall from Eq.~\eqref{eq:RemainderMembershipCondition}
that all elements of \(J\) have zero remainder modulo \(\mathcal{G}(J)\). 
As polynomial division acts linearly on \(W\), we consider the remainders of
the basis elements, \(\Delta_{\mathcal{G}(J)}(\Omega_j)\). Recall that these
remainders can be expressed in terms of monomials irreducible by \(\mathcal{G}(J)\), that is
\begin{equation}
    \Delta_{\mathcal{G}(J)}(\Omega_j) =  \sum_{\underline{\beta_i} \in \mathrm{irreds}(\mathcal{G}[J])}   \Delta_{ij} \! \left({\mathcal{G}[J]}, \Omega \right) \underline{X}^{\underline{\beta_i}} \, ,
    \label{eq:RemaindersInMonomialBasis}
\end{equation}
where \(\Delta_{ij} \! \left({\mathcal{G}[J]}, \Omega \right)\) is the
\(\mathbb{F}\text{-valued}\) matrix of coefficients of the remainder of
\(\Omega_j\) when expressed in terms of the monomials \(\underline{X}^{\underline{\beta_i}}\).
Note that for ideals that are not zero
dimensional, \(\mathrm{irreds}(\mathcal{G}[J])\) is an infinite set. However,
as in practice the degree of \(\Omega_j\) is bounded, the sum is always finite.
Clearly, one can linearly express the remainder of any element of \(W\) in
terms of the \(\Delta_{\mathcal{G}(J)}(\Omega_j)\).
Therefore, we see that any element \(w\) of \(W \cap J\) takes the form
\begin{equation}
   w = \sum_{j=1}^{\dim(W)} c_j \Omega_j, \quad \text{such that} \quad
   \sum_{j = 1}^{\dim(W)} \Delta_{ij}\!\left({\mathcal{G}[J]}, \Omega \right) c_j = 0 \, ,
   \label{eq:LinearizedIntersection}
\end{equation}
where \(c_j \in \mathbb{F}\). Eq.~\eqref{eq:LinearizedIntersection} states
that the \(c_j\) live in the nullspace of the \(\mathbb{F}\text{-valued}\) matrix
\(\Delta_{ij}\!\left({\mathcal{G}[J]}, \Omega \right)\). A basis of this
nullspace can be computed with standard linear algebra techniques.
Through Eq.~\eqref{eq:LinearizedIntersection}, we then arrive at a basis
of \(W \cap J\).

Next, we consider how to construct a set of elements of \(W\) that form a
basis of \(W/(W \cap J)\) when considered modulo elements of \(W \cap J\). 
Specifically, we show that one can choose a subset of the basis elements of
\(W\) using standard linear-algebra techniques.
To see this, note that \(W/(W \cap J)\) is isomorphic to the space spanned by
the remainders modulo \(\mathcal{G}(J)\) of the elements of \(W\).
Furthermore, considering \(i\) as a row index and \(j\) as a column index, this
space is isomorphic to the column space of the matrix
\(\Delta_{ij}\!\left({\mathcal{G}[J]}, \Omega \right)\).
We then see that
\begin{equation}
   W/(W \cap J) \,\, \cong \,\, \mathrm{span}_{\mathbb{F}} \left(\Big\{ \Omega_j   \quad \text{such that} \quad {j \in  \mathrm{pivots}\left[\Delta_{ij}\!\left({\mathcal{G}[J]}, \Omega \right)\right]} \Big\}\right) \, .
  \label{eq:RemainderQuotientSpaceDefinition}
\end{equation}
That is, a basis of \(W/(W \cap J)\) can be chosen as the subset of \(\Omega\)
corresponding to the pivot columns of the matrix
\(\Delta_{ij}\!\left({\mathcal{G}[J]}, \Omega \right)\). 
Note that constructing \(W/(W \cap J)\) in this way gives a true subspace of
\(W\), rather than one up to isomorphism. Therefore, with the construction in
Eq.~\eqref{eq:RemainderQuotientSpaceDefinition},
Eq.~\eqref{eq:SpaceDecompositionIsomorphism} is an equality.
To compute the set of pivot indices, one can use the standard technique
where the pivot indices are read from the row-reduced echelon form of
\(\Delta_{ij}\!\left({\mathcal{G}[J]},
   \Omega \right)\). We note that the subset of \(\Omega\) that is chosen as a basis
by this algorithm depends on the ordering of the elements of \(\Omega\).
Specifically, elements of \(\Omega\) that occur earlier in the set are
prioritized.

\subsubsection{Quotient Rings}
\label{sec:org3290d0f}
\label{sec:quotientRings}    

When considering physical polynomials in spinor variables, i.e.~polynomials subject to 
momentum conservation, it is easy to see that the polynomial ring \(S_n\) is redundant. Specifically, we wish to consider a number
of elements of \(S_n\) as equivalent: those which can be converted
into each other by application of the momentum-conservation identity. 
In numerical applications, where one only has access to evaluations of functions
on points \((\lambda, \tilde{\lambda}) \in V(J_{\Lambda_n})\), this is essential:
any two polynomials \(p, q \in S_n\) that are equivalent under momentum
conservation will evaluate to the same value on such a point.
Hence, the momentum-conservation ideal induces an \textbf{equivalence class} of polynomials: we
wish to consider two polynomials in \(S_n\) which can differ by some element of
\(J_{\Lambda_n}\) as equivalent. That is, for \(p, q \in S_n\) 
\begin{equation}
    p \sim q \iff p-q \in J_{\Lambda_n} \, .
\label{eq:QuotientRingEquivalence}
\end{equation}
Working in a polynomial ring up to equivalence by an
ideal means that we work in a \textbf{quotient ring}. Specifically, all independent
spinor polynomials given momentum conservation belong to the quotient ring 
\begin{equation}\label{eq:spinor_quotient_ring_definition}
 R_n = S_n / J_{\Lambda_n} \, .
\end{equation}
Returning to Eq.~\eqref{eq:QuotientRingEquivalence}, both \(p\) and \(q\) belong
to the same equivalence class in \(R_n\), and we say that \(p\) and \(q\) are
\textbf{representatives} of this equivalence class. 
In order to represent elements of quotient rings, one can make use of Gröbner
bases. Specifically, if two elements \(p\) and \(q\) are
equivalent then their difference belongs to the ideal \(J_{\Lambda_n}\) and so
\begin{equation}
   \Delta_{\mathcal{G}(J_{\Lambda_n})}(p - q) = 0 \, .
\end{equation}
Rearranged, this means that the remainders of \(p\) and \(q\) are equal. Therefore,
elements of a quotient ring are uniquely (canonically) represented by their
remainders modulo a Gröbner basis.
For a recent application of polynomial quotient rings in other areas of
particle physics see Ref.~\cite{Henning:2017fpj}. In this work, we will refer to \(R_n\)
as the set of ``physically inequivalent'' polynomials.

\paragraph{Ideals and Varieties in Quotient Rings.}
\label{sec:orga5642e4}
It is clear from the definition in Eq.~\eqref{eq:idealDefinition} that we can consider ideals in
polynomial quotient rings. 
Ideals in polynomial quotient rings are also finitely generated and one
can represent generators by a representative of the equivalence class.
Let us denote a polynomial ring by \(A\), an ideal of \(A\) by \(J\) and consider the polynomial quotient ring \(A/J\). 
It turns out that, for computations involving an ideal \(K\) of \(A/J\), we can
perform the computation using an ideal in the polynomial ring \(A\) which corresponds to \(K\).
Therefore, we are able to continue to use Gröbner basis technology when working
with polynomial quotient rings.
Let us concretize the correspondence as follows.
Let us denote the generators of \(J\) as \(\{ q_1, \ldots, q_l \}\). We introduce a map \(\pi_{A, A/J}\) which takes an ideal of \(A\) to an ideal of
the quotient ring \(A/J\). Explicitly we have 
\begin{equation}
    \pi_{A, A/J} \left(\, \left\langle p_1, \ldots, p_k \right\rangle_{A}\, \right) = \langle p_1, \ldots, p_k \rangle_{A/J} \, ,
\end{equation}
where the \(p_i\) on the right hand side are understood as representatives of the equivalence class.
The map \(\pi_{A, A/J}\) is many-to-one, but if one restricts the domain to ideals \(K\) of \(A\)
which contain \(J\), then the map is one-to-one~\cite[Lemma 1.63]{becker2012groebner}. 
Given this restriction, the \(\pi_{A, A/J}\) has a unique inverse given by
\begin{equation}
    \pi^{-1}_{A, A/J} \left(\, \langle p_1, \ldots, p_k \rangle_{A/J}\, \right) = \left\langle p_1, \ldots, p_k, q_1, \ldots, q_l \right\rangle_{A} \, . 
\end{equation}
Consider \(R_n\), this means that, given an ideal \(J\) of \(R_n\), one appends the
generators of \(J_{\Lambda_n}\) to find the corresponding ideal in \(S_n\).
The map \(\pi_{A, A/J}\) has a number of applications. For instance, given a
representative \(q\) of an element of \(A/J\) and an ideal \(K\) of \(A/J\), we define
the remainder of \(q\) modulo a Gröbner basis of \(K\) through
\begin{equation}
    \Delta_{\mathcal{G}(K)}(q) = \Delta_{\mathcal{G}( \pi^{-1}_{A, A/J}[K])}(q) \, ,
\end{equation}
where the right hand side is again a representative of an element of \(A/J\). This
allows us to apply the vector space organization technology of Section
\ref{GroebnerBases} also in quotient rings.
Moreover, \(\pi_{A, A/J}\) induces a definition of dimension of ideals in the quotient ring. Specifically, we define
\begin{equation}
    \dim\left(\langle p_1, \ldots, p_k \rangle_{A/\langle q_1, \ldots, q_l \rangle_A}\right) = \dim\left(\left\langle p_1, \ldots, p_k, q_1, \ldots, q_l \right\rangle_{A} \right) \, .
\end{equation}
Recalling the definition of codimension in Eq.~ \eqref{eq:CodimensionDefinition} and the
codimension of momentum conservation in Eq.~ \eqref{eq:MomentumConservationCodimension},
we see that the dimension of \(\langle 0 \rangle_{R_n}\) is given by
\begin{equation}
    \mathrm{dim}(\langle 0 \rangle_{R_n}) = 4n - 4 \, .
\end{equation}
Furthermore, \(\pi_{A, A/J}\) gives a natural
way to understand the geometry of ideals in a quotient ring. Specifically, we
can apply the correspondence and consider the variety associated to the associated ideal in \(S_n\), i.e.~we define
\begin{equation}
    V\left(\langle p_1, \ldots, p_k \rangle_{R_n}\right) = 
    V\Big(\Big\langle p_1, \ldots, p_k, \sum_{i=1}^n \lambda_{i \alpha} \tilde{\lambda}_{i \dot{\alpha}} \, \Big\rangle_{S_n} \, \Big) \, .
\end{equation}
Therefore, all varieties associated to ideals in \(R_n\) are sub-varieties of the of
the momentum-conservation variety \(V(J_{\Lambda_n})\). 

Finally, we point out that we will also make use of ideals of \(R_n\) generated by the
application of a permutation or parity operation as denoted in
Eq.~\eqref{eq:IdealPermutation}. Similar to the \(S_n\) case, one can compute generating sets
of these ideals by applying the permutation and/or parity operation to the
generators in \(R_n\). This follows as the momentum-conservation ideal
\(J_{\Lambda_n}\) is invariant under permutations and parity.

\subsection{Linearly Independent Polynomials in Spinor Brackets}
\label{sec:orgff4fa04}
\label{sec:IndependentFunctions}

In the previous subsection, we set up an algebro-geometric framework to
understand spinor space. 
We have seen that all polynomial functions on \(V(J_{\Lambda_n})\) are contained
in \(R_n\).
In practice, when discussing scattering amplitudes, we are only interested in a
subset of these functions: those which are Lorentz invariant up to a little-group rescaling.
In this sense, the ring \(R_n\) is a superset of
the polynomial functions relevant for scattering amplitudes. 
Furthermore, \(R_n\) is the set of
spinor polynomials up to equivalences induced by momentum
conservation. For the purposes of making an Ansatz, it is necessary that there are
no linear dependencies between the Ansatz elements. In this section, we discuss
how we resolve these two issues.

\paragraph{The Bracket Subring.}
\label{sec:org77e46d4}
Our aim is to understand the set of physically inequivalent polynomial functions which are Lorentz invariant, up to a little-group rescaling. 
These form a subset of \(R_n\), which we denote by
\begin{equation}
    \mathcal{R}_n = \{\, a \in R_n \,\, : \,\, \Lambda(a) =  Z_{(\Lambda, a)} \, a \,\} \, ,
\label{eq:LorentzInvariantSubring}
\end{equation}
 where \(\Lambda\) is a Lorentz transformation which is continuously connected to
 the identity and \(Z_{(\Lambda, a)}\) is an element of \(\mathrm{GL}(1)\)
 corresponding to a little-group rescaling of \(a\) when acted on by \(\Lambda\).
 One can show that \(\mathcal{R}_n\) is a ring.
 Therefore, as a subset of \(R_n\) which is also a ring, \(\mathcal{R}_n\) is a \textbf{subring} of \(R_n\).
 From a physical perspective, it is clear that \(\mathcal{R}_n\) is composed of
 polynomials which can be described in terms of spinor brackets. Therefore, we
 refer to \(\mathcal{R}_n\) as the bracket subring.
In order to work with \(\mathcal{R}_n\) in practice, it is convenient to
 reformulate it in a way that manifests the Lorentz transformation properties of
 its elements. To this end, we will observe that we can describe \(\mathcal{R}_n\) as a
 polynomial quotient ring. In this formulation, one can then use Gröbner basis
 technology for standard operations, such as finding a canonical form of
 elements of \(\mathcal{R}_n\), checking equivalence of elements of \(\mathcal{R}_n\) and intersecting a subspace of \(\mathcal{R}_n\) with an ideal of \(\mathcal{R}_n\).

Let us consider a polynomial ring where we label the variables by the independent
spinor brackets for \(n\) particles,
\begin{equation}
   \mathcal{S}_n = \mathbb{F}\bigl[ \, \langle 1  2 \rangle, \langle 1  3 \rangle, \ldots, \langle (n-1) n \rangle, [12], [13], \ldots, [(n-1) n] \bigr] \, .
\end{equation}
 We note that \(\mathcal{S}_n\) is a polynomial ring in \(2 \binom{n}{2}\) variables,
 that is we choose our variables to be \(\langle ij \rangle\) and \([ij]\) for \(i <
 j\). We stress that, in the context of \(\mathcal{S}_n\), the spinor brackets are to be
 considered as variables and not as polynomials in \(R_n\).
It can be shown (see Appendix
 \ref{LorentzInvarianceProof}) that \(\mathcal{R}_n\) is isomorphic to a polynomial
 quotient ring as
 \begin{equation}
\label{eq:invariant_qring}
     \mathcal{R}_n \cong 
      \mathcal{R}_n^{(q)} \, , \quad \text{where} \quad \mathcal{R}_n^{(q)} = \mathcal{S}_n/(\mathcal{J}_{\Lambda_n} + \mathcal{K}_{\Lambda_n} + \overline{\mathcal{K}}_{\Lambda_n})\, \\
 \end{equation}
and
\begin{align}
  \mathcal{J}_{\Lambda_n} &= \Bigg\langle {\sum_{\substack{j=1 \\ j \ne i,k}}^n} \langle ij \rangle [jk] \,\, : \,\, 1 \le i \le n, \, 1 \le k \le n \Bigg\rangle_{\mathcal{S}_n} \, , \label{eq:invariant_momentum_conservation_ideal}\\
  \mathcal{K}_{\Lambda_n} &= \big\langle \langle ij\rangle\langle kl \rangle + \langle ik\rangle\langle lj \rangle + \langle il\rangle\langle jk \rangle  \,\, : \,\, 1 \le i < j < k < l \le n \big\rangle_{\mathcal{S}_n} \, ,\\
  \overline{\mathcal{K}}_{\Lambda_n} &= \big\langle \hspace{0.9mm} [ij][kl] \hspace{0.9mm} + \hspace{0.9mm} [ik][lj] \hspace{0.9mm} + \hspace{0.9mm} [il][jk] \hspace{0.9mm} \,\, : \,\, 1 \le i < j < k < l \le n \big\rangle_{\mathcal{S}_n} \, .
\label{eq:schouten_ideal}
\end{align}
Here, for ease of notation, for brackets with \(j \ge i\) we make use
of the identities
\begin{equation}
    \langle ji \rangle = - \langle ij \rangle \quad \; \text{and} \quad \; [ji] = -[ij] \, .
\end{equation}
Physically, \(\mathcal{J}_{\Lambda_n}\) is the set of relations between spinor
brackets generated by the momentum-conservation identities, and \(\mathcal{K}_{\Lambda_n}\) and \(\overline{\mathcal{K}}_{\Lambda_n}\) are the
set of relations generated by the Schouten identities.
Eq.~\eqref{eq:invariant_qring} says that \(\mathcal{R}_n^{(q)}\) is the set of inequivalent
spinor bracket polynomials under this set of identities. 

It is natural to ask what happens if we consider of the set of elements of
an ideal \(J\) of \(R_n\) that can be expressed in terms of spinor brackets.
Mathematically, we are inquiring about the object \(J \cap \mathcal{R}_n\).
Importantly, it can be shown that \(J \cap \mathcal{R}_n\) is an ideal of \(\mathcal{R}_n\).
To be able to perform practical computations with \(J \cap \mathcal{R}_n\), we
wish to find the ideal in \(\mathcal{R}_n^{(q)}\) to which \(J \cap \mathcal{R}_n\)
maps.
Specifically, we need to be able to construct a generating set of this ideal.
To this end, we will make use of the correspondence between ideals of
polynomial rings and polynomial quotient rings discussed in Section \ref{sec:quotientRings}
and begin by working with the polynomial rings \(\mathcal{S}_n\) and \(S_n\).
Consider the augmented polynomial ring
\begin{equation}
    \Sigma_n = \mathbb{F}[\langle 12 \rangle, \ldots, \langle (n-1) n\rangle, [12], \ldots, [(n-1)n], \lambda_{1 0}, \lambda_{1 1}, \ldots, \tilde{\lambda}_{1 \dot{0}}, \tilde{\lambda}_{1 \dot{1}}, \ldots] \, ,
\end{equation}
that is, a polynomial ring whose variables are both the spinor brackets and the spinor variables. 
It is clear that \(\Sigma_n\) contains both \(\mathcal{S}_n\) and \(S_n\) as
subrings. Given an ideal \(J = \langle p_1, \ldots p_k \rangle_{S_n}\) we construct
the ideal
\begin{equation}
    \kappa[J] = \left\langle p_1, \ldots, p_k, \langle 12 \rangle - \big(\lambda_{1 0}\lambda_{2 1} - \lambda_{2 0} \lambda_{1 1}\big), \ldots, [12] - \big(\tilde{\lambda}_{1 \dot{0}}\tilde{\lambda}_{2 \dot{1}} - \tilde{\lambda}_{2 \dot{0}} \tilde{\lambda}_{1 \dot{1}}\big), \ldots \right\rangle_{\Sigma_n} \, .
\end{equation}
Here, \(\kappa[J]\) is generated by the generators of the ideal \(J\), as well as by the relations between the spinor brackets and spinor
variables\footnote{A very similar setup
can be found in constructing ``algebraic dependence relations'' in the Le\u{\i}nartas
algorithm \cite{leinartas1978factorization, raichev2012leinartas, Meyer:2016slj}.}. It can be shown (see Appendix \ref{LorentzInvarianceProof}) that
the set of elements of \(J\) that can be written in terms of spinor brackets correspond to the ideal 
of \(\mathcal{S}_n\) given by
\begin{equation}
    \kappa[J] \cap \mathcal{S}_n \, .
\end{equation}
This intersection is an example of elimination of variables and can be computed in practice via Gröbner basis techniques, see e.g.~Section 2.4.3 of Ref.~\cite{Zhang:2016kfo}. 
We can then use the correspondence map to understand the ideals of \(\mathcal{R}_n^{(q)}\) associated to ideals of \(R_n\).
Combining this with the isomorphism in Eq.~\eqref{eq:invariant_qring},
for an ideal \(J\) of \(R_n\) we have that
\begin{equation}\label{eq:conversion_covariant_invariant_elimination_ideal}
     J \cap \mathcal{R}_n \cong \pi_{\mathcal{S}_n, \mathcal{R}_n^{(q)}} \left(\kappa[ \pi_{S_n, R_n}^{-1}(J)] \cap \mathcal{S}_n\right) \, ,
 \end{equation}
 where the right hand side is the ideal in the polynomial quotient ring formulation of \(\mathcal{R}_n\).
As a first example of this technology, one can consider the situation where we
wish to find the ideal in \(\mathcal{S}_n\) corresponding to \(\langle 0
\rangle_{S_n}\).
 A Gröbner basis calculation shows that 
\begin{equation}
   \kappa[\langle 0 \rangle_{S_n}] \cap \mathcal{S}_n = \mathcal{K}_{\Lambda_n} + \overline{\mathcal{K}}_{\Lambda_n} \, ,
\end{equation}
that is, we have only generated the Schouten identities. A less trivial example is
to find the ideal in \(\mathcal{S}_n\) which corresponds to \(J_{\Lambda_n}\). One
finds that 
\begin{equation} 
   \kappa[ J_{\Lambda_n} ] \cap \mathcal{S}_n = \mathcal{J}_{\Lambda_n} + \mathcal{K}_{\Lambda_n} + \overline{\mathcal{K}}_{\Lambda_n} \, .
\end{equation} 
Here, we now pick up both momentum-conservation and Schouten identities.

\paragraph{Physical Polynomial Space.}
\label{sec:org5c679a7}
In this work, our aim is to construct compact Ansätze for the rational prefactors. 
So far, we have discussed the polynomials relevant for the numerators of rational
prefactors in scattering amplitudes as living in the spinor bracket ring 
\(\mathcal{R}_n\). However, this is an infinite dimensional vector space, and so
this information is insufficient for the construction of a finite Ansatz.
Nevertheless, physical polynomials, such as numerators of rational prefactors,
have well-defined mass dimension and little-group weight. This leads us to
define the space of independent bracket polynomials with a well-defined mass
dimension \(d\) and little-group weights \(\phi_k\),
\begin{equation}
    \mathcal{M}_{d, \vec{\phi}} = \Big\{ a \in \mathcal{R}_n \; : \; [a] = d , \, \mathrm{and} \; \{a\}_k = \phi_k \Big\} \, .
    \label{eq:PhysicalBracketSpace}
\end{equation}
Note that as the mass dimension \(d\) is fixed, \(\mathcal{M}_{d, \vec{\phi}}\) is a
finite-dimensional vector space over~\(\mathbb{F}\). 
If we can find a basis of \(\mathcal{M}_{d, \vec{\phi}}\), we can use this basis
as an Ansatz for the numerator polynomial.
Furthermore, any Ansatz for the numerator polynomial must be expressible in terms of a basis of  \(\mathcal{M}_{d, \vec{\phi}}\). 
Therefore, this basis is a natural starting point for refined Ansätze with
special properties.
We will now describe an algorithm to construct a basis of
\(\mathcal{M}_{d, \vec{\phi}}\).
There are two problems we need to solve.
First, we must construct elements of \(\mathcal{R}_n\) which are linearly
independent. It is clear from the definition of \(\mathcal{R}_n^{(q)}\) that monomials
in \(\mathcal{M}_{d, \vec{\phi}}\) are related by momentum-conservation and
Schouten identities. 
Second, we must impose the constraints of fixed mass dimension and little-group
weight. We note that other methods have been put forward to build a basis of
\(\mathcal{M}_{d, \vec{\phi}}\), see
e.g.~Ref.~\cite{Huber:2021vnc,DeAngelis:2022qco}.
Our methods make use of general features of the algebra of
polynomials and we expect them to have wide applicability to many
problems.
The approach that we employ here was previously also used in
Ref.~\cite{Zhang:2012ce}, where the problem was finding a linearly independent set
of monomials in loop momentum on a given generalized unitarity cut.

As \(\mathcal{R}_n\) is isomorphic to a polynomial quotient ring in spinor brackets, all elements can be expressed
as linear combinations of monomials in the spinor brackets, which we denote as
\begin{equation}
  m_{(\alpha, \beta)} = \prod_{j = 1}^n \prod_{i = 1}^{j-1} \langle ij \rangle^{\alpha_{ij}} [ij]^{\beta_{ij}} \, ,
\label{eq:SpinorMonomialDefinition}
\end{equation}
where \(\alpha_{ij}\) and \(\beta_{ij}\) belong to \(\mathbb{Z}_{\ge 0}\). It will
turn out that we can pick a subset of the monomials in spinor brackets as basis
elements. Specifically, we will show that
\begin{equation}
    \mathcal{M}_{d, \vec{\phi}} = \mathrm{span}_{\mathbb{F}}\left( M_{d, \vec{\phi}} \right) \quad \text{where } \quad M_{d, \vec{\phi}} = \left\{ m_{(\alpha, \beta)} \quad \text{such that} \quad (\alpha, \beta) \in X_{d, \vec{\phi}} \right\}\, .
    \label{eq:PhysicalBracketSpaceSpan}
\end{equation}
That is, \(\mathcal{M}_{d, \vec{\phi}}\) is the set of all
\(\mathbb{F}\text{-linear}\) combinations of the elements of \(M_{d, \vec{\phi}}\),
the set of spinor bracket monomials whose exponents lie in the finite set \(X_{d,
\vec{\phi}}\).
Our task is to determine the set of exponents \(X_{d,
\vec{\phi}}\) such that the associated monomials have mass dimension \(d\),
little-group weights \(\vec{\phi}\) and are linearly independent elements of
\(\mathcal{R}_n\).

To begin, we discuss the structure of \(\mathcal{R}_n\) as an
\(\mathbb{F}\text{-vector}\) space. As it is a ring, it is an infinite-dimensional vector space.
Physically, a basis of \(\mathcal{R}_n\) as an \(\mathbb{F}\text{-vector}\) space
gives a set of linearly independent polynomials in spinor brackets when
one takes into account the momentum-conservation and Schouten identities.
To resolve these identities, consider a polynomial quotient ring \(A/J\), where \(A\) is a polynomial ring over the field \(\mathbb{F}\) and \(J\) is an ideal of \(A\).
We recall from Section \ref{GroebnerBases} that elements of \(A/J\) can be uniquely
expressed as an \(\mathbb{F}\text{-linear}\) combination of monomials that are
irreducible by the Gröbner basis \(\mathcal{G}(J)\). 
Therefore, we see that the monomials which are irreducible by \(\mathcal{G}(J)\) form a basis
of \(A/J\) as an \(\mathbb{F}\text{-vector}\) space~\cite[Chapter 4.3, Proposition 4]{cox1994ideals}. 
Recall from Eq.~\eqref{eq:invariant_qring} that \(\mathcal{R}_n\) is isomorphic to the polynomial quotient ring
\(\mathcal{R}_n^{(q)} = \mathcal{S}_n / (\mathcal{J}_{\Lambda_n} + \mathcal{K}_{\Lambda_n} + \overline{\mathcal{K}}_{\Lambda_n})\).
Therefore, viewing \(m_{(\alpha, \beta)}\) as an element of \(\mathcal{R}_n^{(q)}\), we require that
\begin{equation}
    m_{(\alpha, \beta)} \nmid \mathrm{LM}(g) \quad \text{for all} \quad g \in \mathcal{G}(\mathcal{J}_{\Lambda_n} + \mathcal{K}_{\Lambda_n} + \overline{\mathcal{K}}_{\Lambda_n}) \, ,
    \label{eq:irreducibilityConstraint}
\end{equation}
where \(\mathcal{G}(\mathcal{J}_{\Lambda_n} + \mathcal{K}_{\Lambda_n} + \overline{\mathcal{K}}_{\Lambda_n})\) is the
Gröbner basis associated to momentum-conservation and Schouten identities.
Note that the statement that a polynomial is reducible can be
stated as a set of simultaneous linear inequalities. 
That is, all of the elements of \((\alpha, \beta)\) must be greater than or equal
to the corresponding entry in the exponent of \(\mathrm{LM}(g)\).
The irreducibility constraint,
Eq.~\eqref{eq:irreducibilityConstraint}, is the complement of this.
In summary, monomials \(m_{(\alpha, \beta)}\) which satisfy the irreducibility
constraint form a basis of \(\mathcal{R}_n\) as an \(\mathbb{F}\text{-vector}\) space.

The constraints of fixed little-group weight and mass dimension translate
to linear constraints on the exponents \((\alpha, \beta)\). First, we consider
mass dimension: all spinor brackets have unit mass dimension, so we
can easily write the mass dimension of a monomial \(m_{(\alpha, \beta)}\) as
\begin{equation}
    [m_{(\alpha, \beta)}] = \sum_{j = 1}^n \sum_{i = 1}^{j-1} \left( \alpha_{ij} + \beta_{ij} \right) = d \, .
\label{eq:massDimensionConstraint}
\end{equation}
Next, we consider little-group weight of a monomial \(m_{(\alpha, \beta)}\). It is clear that
\begin{equation}
    \{ m_{(\alpha, \beta)} \}_{k} = \sum_{j = 1}^n \sum_{i = 1}^{j-1} \left( \alpha_{ij} \{ \langle ij \rangle \}_k + \beta_{ij} \{ [ ij ] \}_k \right) = \phi_k \, ,
\label{eq:littleGroupConstraint}
\end{equation}
where
\begin{equation}
     \{ \langle ij \rangle \}_k  = \delta_{ik} + \delta_{jk} \qquad \mathrm{and} \qquad \{ [ ij ] \}_k  = - \left( \delta_{ik} + \delta_{jk}\right)
\end{equation}
are the little-group weights of the angle and square brackets
respectively.
It is interesting to note that, as the \(\alpha_{ij}\) and
\(\beta_{kl}\) are non-negative, Eqs.~\eqref{eq:massDimensionConstraint} and
\eqref{eq:littleGroupConstraint} cut out a convex polytope.

Together with the irreducibility constraints
\eqref{eq:irreducibilityConstraint}, the constraints of mass dimension
\eqref{eq:massDimensionConstraint} and little-group weights
\eqref{eq:littleGroupConstraint} on \((\alpha, \beta)\) define the set of
exponents \(X_{d, \vec{\phi}}\).
To solve these equations, first note that the exponents \((\alpha, \beta)\) are
non-negative integers. The space cut out by our equations is bounded, and therefore
\(X_{d, \vec{\phi}}\) is finite. Solving these equations is then reduced
to enumerating their solutions.
Efficient algorithms to enumerate such non-negative integer solutions are commonly implemented in 
computer algebra systems\footnote{For example, in the computer algebra system
\texttt{Mathematica} one can simply apply the \texttt{Solve} function, requiring
the solution domain to be the \texttt{NonNegativeIntegers}.}.

\subsection{Geometry of Singular Varieties}
\label{sec:org32bcb1d}
\label{sec:IrreducibleSingularVarieties}

Let us consider the rational functions \(\mathcal{C}_i\) from
Eq.~\eqref{eq:CommonDenominatorForm} and in particular the set of
all their possible denominator factors \(\mathcal{D} = \{ \mathcal{D}_1, \ldots, \mathcal{D}_{n_i} \}\). In general, this set
will depend on the specifics of the considered external
kinematics, together with the definitions of the functions \(\mathcal{F}\).
As an example of typical elements of the set \(\mathcal{D}\) we can consider all spinor
brackets from Eq.~\eqref{eq:spinorBrackets}.
We wish to study the behavior of the rational coefficients 
when considered near varieties on which some subset of the denominators
\(\mathcal{D}\) vanishes.
We dub these varieties ``singular varieties''\footnote{The term ``singular'' here
refers to a property of rational functions on the variety, not to a topological
property of the variety itself.}. Conventional examples are configurations where
external particles become soft or collinear. We denote a singular variety as
\begin{equation}
   U_{\vec{\gamma}} = V \big(\langle \mathcal{D}_{\gamma_1}, \ldots, \mathcal{D}_{\gamma_m}\rangle_{R_n} \big) \, ,
\label{eq:SingularVarietyDefinition}
\end{equation}
for some subset of the denominators \(\{\mathcal{D}_{\gamma_1}, \ldots,
\mathcal{D}_{\gamma_m}\} \subseteq \mathcal{D}\). We note that, by
definition, the variety \(U_{\vec{\gamma}}\) corresponds to an ideal
\begin{equation}
   J_{\vec{\gamma}} = \langle \mathcal{D}_{\gamma_1}, \ldots, \mathcal{D}_{\gamma_m}\rangle_{R_n} \, ,
\label{eq:SingularIdealDefinition}
\end{equation}
which we dub a ``singular ideal''. In practice, we will only be considering ideals generated by one or two denominator factors, i.e.~the
cases \(m = 1\) or \(m = 2\).

A key feature of the algorithm we present in this paper will be to
generate numerical configurations of spinors which lie close to singular
varieties in order to determine how fast a given rational expression diverges
close to the singular variety. This procedure is complicated by the
fact that the singular varieties may branch and that the degree of
divergence may differ close to different branches of the same
variety. Therefore, we will find it necessary to be able to control
which branch we are approaching numerically.
The remainder of this section reviews in general terms the
geometric and algebraic concepts related to branching. We refer the reader to 
Section 4 of Ref.~\cite{cox1994ideals} for more details.

\paragraph{Irreducible Varieties.}
\label{sec:orgdf6736d}
The key geometric concept related to branching is that of
reducibility of a variety. The object that we wish to consider is that
of an \textbf{irreducible variety}. A variety \(U\) is defined to be irreducible if
\begin{equation}\label{eq:IrreducibleVariety}
U = U_1 \cup U_2 \;\; \Rightarrow \;\l U_1 = U \;\; \text{or} \;\; U_2 = U \, .
\end{equation}
In our case, the varieties \(U_{\vec{\gamma}}\) in Eq.~\eqref{eq:SingularVarietyDefinition} may
well be reducible. A reducible variety \(U\) can be written as a
proper union of sub-varieties and there exists a \textbf{minimal decomposition}
\begin{equation}
   U = \bigcup_{k=1}^{n_B(U)} U_k \, ,
   \label{eq:SingularVarietyReducibility}
\end{equation}
where \(n_B(U)\) is the number of varieties in the decomposition, each \(U_k\) is
irreducible and \(U_i \not\subseteq U_j\) for all \(i \neq j\).
We call each \(U_k\) a ``branch'' of \(U\) and this last condition is that no
branch is contained within another. A minimal decomposition is unique, up
to the order of the branches \(U_k\) \cite[Section 4.6, Theorem 4]{cox1994ideals}.

Importantly, the reducibility of a variety may not be manifest from
the set of equations used to define it. Our task is now to discuss how to understand the decomposition of 
Eq.~\eqref{eq:SingularVarietyReducibility} algebraically so that one can perform the decomposition systematically.

\paragraph{Ideals Associated to Irreducible Varieties.}
\label{sec:orga48cfbb}
To understand the decomposition of varieties in an algebraic fashion,
let us start by considering an ideal \(J\), its associated variety \(U =
V(J)\) and its minimal decomposition \(U = \bigcup_{k}^{n_B(U)} U_k\). It is natural
to consider the ideal associated to each branch \(U_k\)
\begin{equation}\label{eq:prime_ideal_from_variety_branch}
    P_k = I(U_k) \, .
\end{equation}
Here we have judiciously labeled the ideal as \(P\), which hints at the
fact that the ideal associated to an irreducible variety is \textbf{prime}
(see Appendix \ref{app:AlgGeoGlossary} for the algebraic definition). 

A reasonable expectation could be to express \(J\) in terms of the
\(P_k\)'s. However, recall that in general \(J \neq I(V(J))\), as this 
requires the ideal to be radical (see Eq.~\eqref{eq:radicalNotation}).  In fact, it
turns out that the algebraic analogue of a minimal decomposition of
\(U\), called a \textbf{minimal primary decomposition}, expresses \(J\) as 
\begin{equation}\label{eq:PrimaryDecomposition}
  J = \bigcap_{l=1}^{n_Q(J)} Q_l \, , 
\end{equation}
where each \(Q_l\) is \textbf{primary} (see Appendix \ref{app:AlgGeoGlossary} for the definition),
all \(\sqrt{Q_l}\) are distinct, and no \(Q_l\) can be removed from
the intersection without changing the result, i.e.\(\;Q_m \nsupseteq \bigcap_{l\neq m} Q_l\). We stress
that the intersection of ideals should be viewed considering the
ideals as infinite sets of polynomials.  We call each \(Q_l\) a primary
component of \(J\) and we denote the number of primary components as
\(n_Q(J)\).  The radical of each primary component \(Q_l\) is a prime
ideal
\begin{equation}\label{eq:PrimePrimaryRelation}
  P_l = \sqrt{Q_l} \, ,
\end{equation}
i.e.~a primary ideal is also prime only if it is radical. As \(P_l\) is
the prime associated to \(Q_l\), we say that \(Q_l\) is
\(P_l\text{-primary}\). We call the set of primes associated to all
the primary components of \(J\) the set of \textbf{associated primes}. To this
end, we write
\begin{equation}
  \mathrm{assoc}(J) = \big\{ P_1, \ldots,  P_{n_Q(J)} \big\} \, .
\end{equation}
It can be shown that the associated primes 
in a minimal primary decomposition are unique~\cite[Theorem 4.5]{books/daglib/0091700}.
We call an associated prime \(P_i\) of \(J\) a \textbf{minimal prime} if \(P_i \not\supseteq
P_j\) for all \(i \ne j\).
We define the set of \textbf{minimal associated primes} of \(J\)
as 
\begin{equation}\label{eq:MinimalAssociatedPrimes}
  \mathrm{minAssoc}(J) = \Big\{P \in \mathrm{assoc}(J) \text{ where } P \text{ is a minimal prime}\Big\} \, .
\end{equation}
It can be shown that the set of primary components \(Q_k\) of \(J\) for which \(\sqrt{Q_k}\) is a
minimal prime of \(J\) is unique \cite[Theorem 4.10]{books/daglib/0091700}.

Let us now address the relation between the minimal primary decomposition 
of an ideal \(J\) and the minimal decomposition of its associated variety \(U = V(J)\).
First of all, note that the dimension of each primary component may not be the same. 
In fact, it can be shown that
\begin{equation}
   \dim(J) = \max\left(\{\dim(Q_l) \, \, : \,\, 1 \le l \le {n_Q(J)}\}\right) \, .
\end{equation}
If we now interpret Eq.~\eqref{eq:PrimaryDecomposition}
geometrically by taking the variety of both left- and right-hand side and using the
fact that the variety associated to an intersection of a set of ideals
corresponds to the union of the varieties associated to each ideal, we
obtain
\begin{equation} \label{eq:VarietyDecomposition}
   V(J) = \bigcup_{l=1}^{n_Q(J)} V(Q_l) \, . 
\end{equation}
In general, the union of Eq.~\eqref{eq:VarietyDecomposition} may
not be a minimal decomposition of \(V(J)\), because the variety
associated to a primary component may be contained in the variety
associated to another. 
Therefore, we can split set of varieties \(V(Q_l)\) into two distinct subsets:
those who can be removed from the intersection of
Eq.~\eqref{eq:VarietyDecomposition} without changing the result and those
that cannot.
We refer to these as
``embedded'' and ``isolated'', respectively. 
It can be shown that the prime
ideals \(P_k\) in the set \(\mathrm{minAssoc}(J)\) are in one-to-one
correspondence with the irreducible varieties \(U_k\) from
Eq.~\eqref{eq:SingularVarietyReducibility}, i.e.~\(U_k =
V(P_k)\). Therefore, it is clear that \(n_Q(J)\ge n_B(V(J))\). Geometrically, one
can see that the non-uniqueness in a minimal primary decomposition is associated
to the primary ideals \(Q_l\) such that \(V(Q_l)\) is embedded.

Before moving on to some explicit examples in spinor space, let us
remark that there exist general algorithms for the computation of
primary decompositions, see for instance
Ref.~\cite{gianni1988grobner}. A further useful comment is that
for a prime ideal, all maximally independent sets are of the same size
\cite[Proposition 7.26]{becker2012groebner}. This observation can provide
a simple way to show that an ideal is not prime.

\paragraph{Examples of Irreducible Singular Varieties At Three and Four Points.}
\label{sec:orgefbbb80}
To understand Eqs.~\eqref{eq:PrimaryDecomposition} and
\eqref{eq:VarietyDecomposition} in a more physical context, let us turn
to simple examples of the problem at hand: understanding surfaces in
spinor space.  As a first warm-up, we consider there-point phase
space. A well-known fact is that either all angle or all square
brackets must be zero. Formally, this means that the zero ideal in
\(R_3\) (or equivalently the ideal in \(S_3\) generated by momentum
conservation alone) is not primary.  That is, one can compute the
primary decomposition of \(\bigl\langle 0 \bigr\rangle_{R_3}\) to find
\begin{eqnarray}
  \bigl\langle 0 \bigr\rangle_{R_3} & = & \bigl\langle \langle12\rangle, \langle13\rangle, \langle23\rangle \bigr\rangle_{R_3} \cap \bigl\langle [12], [13], [23] \bigr\rangle_{R_3} \, .
\end{eqnarray}
In contrast, four-point phase space, \(V\left(\langle 0 \rangle_{R_4}\right)\), is
irreducible. However, a number of interesting varieties associated to
codimension-one ideals do decompose. This is again a statement that we can
demonstrate with the help of a primary decomposition. For instance, we have
\begin{equation}\label{eq:P1R4}
  \bigl\langle \, [ 1  2 ] \, \bigr\rangle_{R_4} = \bigl\langle [12], \langle 34 \rangle  \bigr\rangle_{R_4} \cap \bigl\langle [1  2], [1  3], [1  4],  [2  3], [2  4], [3  4] \bigr\rangle_{R_4}  \, .
\end{equation}
One way to see why \(\bigl\langle  [1 2] \bigr\rangle_{R_4}\) must decompose is to note that in \(R_4\)
\begin{equation}
\langle 12 \rangle [12] = \langle 34 \rangle [34] \, ,
\end{equation}
i.e.~\(\langle 34 \rangle [34]\) is a member of \(\bigl\langle [12] \bigr\rangle_{R_4}\), but this is not the case for \(\langle 34 \rangle\) nor \([34]\).
Therefore, we must have at least two branches, each one containing one of the two factors
of \(s_{34}\).

A well-known fact that can be interpreted in terms of this splitting
is that four-point massless amplitudes have non-unique common
denominators in terms of spinor brackets. To better understand this
let us consider as a concrete example the Parke-Taylor
expression \cite{PhysRevLett.56.2459} for maximally-helicity-violating (MHV) and
\(\overline{\text{MHV}}\) trees, say
\begin{equation}\label{eq:mhveqmhvbar@4point}
  iA_{g^-g^-g^+g^+} = \frac{\, \langle 12 \rangle^3}{\langle 23 \rangle \langle 34 \rangle \langle 41 \rangle} = \frac{\, [34]^3}{[12][23][41]} \, .
\end{equation}
As shown in Eq.~\eqref{eq:mhveqmhvbar@4point}, at four points MHV
and \(\overline{\text{MHV}}\) coincide. Thus, the denominator is clearly
not unique. To see this algebro-geometrically, let us begin by posing an
apparently legitimate question; that is, whether \(\langle 23 \rangle\)
is a pole of this amplitude. We can now say that this question is
ill-posed because the surface is reducible. The amplitude
\(A_{g^-g^-g^+g^+}\) has a simple pole on \(V\big(\bigl\langle \, \langle 23
 \rangle , [14] \, \bigr\rangle_{R_4}\big)\) but it is regular on
\(V\big( \bigl\langle \langle 1 2 \rangle, \langle 1 3 \rangle, \langle 1 4
 \rangle, \langle 2 3 \rangle, \langle 2 4 \rangle, \langle 3 4 \rangle
 \bigr\rangle_{R_4} \big)\); that is, it has a different behavior on the
different branches of \(V\big(\bigl\langle \, \langle 23 \rangle \,
 \bigr\rangle_{R_4}\big)\). Therefore, \(\langle 23 \rangle\) is both a physical
and a spurious singularity, depending on the branch we look at. 
In conclusion, the physical statement about the singularity is that \(A_{g^-g^-g^+g^+}\)
has a simple pole on \(V\big(\bigl\langle \, \langle 23 \rangle , [14] \,
 \bigr\rangle_{R_4}\big)\) and algebraically we can represent this in two
different ways, either via \(\langle 23 \rangle\) or via \([14]\) in the
denominator. 

As a final warm-up, let us consider the set of ideals at four points generated
by pairs of invariants, together with their primary decompositions. 
We present a set of
such ideals from which all others can be derived by permutations of
the \(n\) momenta and parity. 
\begin{table}[!h]
\centering
\begin{tabular}{r@{\hskip3pt}c@{\hskip3pt}l | r@{\hskip3pt}c@{\hskip3pt}l@{\hskip0.5pt}l}
   $\bigl\langle \langle12\rangle, \langle13\rangle \bigr\rangle_{R_4}$ & $=$ & $P_2 \cap P_3 \cap \overline{P}_3(2341)\, ,$ & $P_1$ & $=$ & $\bigl\langle \langle12\rangle, [34] \bigr\rangle_{R_4}\, ,$ & \\
   $\bigl\langle \langle12\rangle, \langle34\rangle \bigr\rangle_{R_4}$ & $=$ & $P_2 \cap P_4\, ,$ & $P_2$ & $=$ & $\bigl\langle \langle 1 2 \rangle, \langle 1 3 \rangle, \langle 1 4 \rangle,  \langle 2 3 \rangle, \langle 2 4 \rangle, \langle 3 4 \rangle \bigr\rangle_{R_4}\, ,$ & \\
   $\bigl\langle \langle12\rangle, [12] \bigr\rangle_{R_4}$ & $=$ & $P_4 \cap P_5 \cap \overline{P}_5\, ,$ & $P_3$ & $=$ & $\bigl\langle \langle 1  2 \rangle, \langle 1  3 \rangle, \langle 2  3 \rangle, \tilde\lambda_4^{\dot\alpha} \bigr\rangle_{R_4}\, ,$ & \\
   $\bigl\langle \langle12\rangle, [13] \bigr\rangle_{R_4}$ & $=$ & $P_3(1243) \cap \overline{P}_3(1342)\, $ & $P_4$ & $=$ & $\bigl\langle \langle 1  2 \rangle, [12], \langle 3  4 \rangle, [34]\, ,$ & $\refstepcounter{equation}(\theequation)\label{eq:four-point-decompositions}$ \\
   & & $\quad \cap P_5 \cap \overline{P}_5(1324)\, ,$ & & & $\quad \lambda_1^\alpha\tilde\lambda_1^{\dot\alpha} + \lambda_2^\alpha\tilde\lambda_2^{\dot\alpha}, \lambda_3^\alpha\tilde\lambda_3^{\dot\alpha} + \lambda_4^\alpha\tilde\lambda_4^{\dot\alpha} \bigr\rangle_{R_4}\, ,$ & \\ 
   $\bigl\langle \langle12\rangle, [34] \bigr\rangle_{R_4}$ & $=$ & $P_1\, ,$ & $P_5$ & = & $\bigl\langle \langle 1  2 \rangle, [12], [13], [14], [23], [24], [34] \bigr\rangle_{R_4}\, .$ &
\end{tabular}
\end{table}

\vspace{-2mm}
\noindent Here, on the left-hand side, we show ideals generated by
pairs of invariants together with their primary decompositions; on the
right-hand side, we give minimal bases for their primary components.
We remark that \(P_3\), \(P_4\) and \(P_5\) have codimension two, while \(P_1\) and \(P_2\)
have codimension one.

\subsection{Functions Vanishing to Higher Order on Singular Varieties}
\label{sec:org04805e1}
\label{sec:SymbolicPowers}   

Now that we have discussed how to construct the set of irreducible singular
varieties of rational prefactors, our aim is to use these varieties to study the
prefactors and interpret their behavior close to singular varieties as
constraints on the analytic structure of their numerators.
In this section, we review a well-studied class of ideals that we can
use to parameterize these constraints. Specifically, we introduce the so-called
``symbolic power'' of an ideal (see Ref.~\cite{dao2017symbolic} for a recent survey and
Chapter 3.9 of Ref.~\cite{eisenbud1995commutative} for a textbook discussion).
Importantly, we can use the symbolic power to define what we
mean by ``behavior close to a variety''. Our ultimate goal
is to present a numerical algorithm for this study. However, in this
section we content ourselves with the mathematical preliminaries, and
delay the discussion of the numerical procedure to Section \ref{sec:orgc26bb05}.

\paragraph{Vanishing to Higher Order at a Point.}
\label{sec:orgcb18aa8}

Our goal is to discuss a set of elements of \(R_n\) that vanish to \(k^{\mathrm{th}}\)
order on a variety \(U \subset V(J_{\Lambda_n})\). Before we tackle this problem, we begin
by studying the simpler case where the variety is a single point. 
Let us consider a point \((\eta, \tilde{\eta})\) in spinor
space that satisfies momentum conservation, i.e.~\((\eta, \tilde{\eta}) \in
V(J_{\Lambda_n})\). The set of elements of \(R_n\) that vanish on this point is
given by an ideal
\begin{equation}
   \mathfrak{m}_{(\eta, \tilde{\eta})} = \left\langle \lambda_{1 0} - \eta_{1 0}, \lambda_{11} - \eta_{1 1}, \ldots,  \tilde{\lambda}_{1 \dot{0}} - \tilde{\eta}_{1 \dot{0}}, \tilde{\lambda}_{1 \dot{1}} - \tilde{\eta}_{1 \dot{1}}, \ldots \right\rangle_{R_n} .
\label{eq:MaximimalIdealDefinition}
\end{equation}
Here we label such an ideal as \(\mathfrak{m}_{(\eta, \tilde{\eta})}\),
hinting that it is actually a so-called ``maximal ideal'' (see
Appendix \ref{app:AlgGeoGlossary} for the algebraic definition).  
To better understand
Eq.~\eqref{eq:MaximimalIdealDefinition}, let us consider computing the remainder of an element
\(q\in R_n\) modulo a Gröbner basis of
\(\mathfrak{m}_{(\eta, \tilde{\eta})}\). 
From Eq.~\eqref{eq:MaximimalIdealDefinition}, it is easy to see that
the remainder modulo \(\mathcal{G}(\mathfrak{m}_{(\eta, \tilde{\eta})})\) of \(q\) is equivalent to the
evaluation of \(q\) at the point \({(\eta, \tilde{\eta})}\), that is
\begin{equation}
    \Delta_{\mathcal{G}(\mathfrak{m}_{(\eta, \tilde{\eta})})}(q) = q{(\eta, \tilde{\eta})} \, .
    \label{eq:maximalResidueAsEvaluation}
\end{equation}
With this perspective, it is clear that \(q\) vanishes at the
point \({(\eta, \tilde{\eta})}\) if and only if it belongs to the ideal
\(\mathfrak{m}_{(\eta, \tilde{\eta})}\).

With the ideal \(\mathfrak{m}_{(\eta, \tilde{\eta})}\) in hand, we are now in a
position to define a set of elements of \(R_n\) that vanish to \(k^\mathrm{th}\)
order at the point \({(\eta, \tilde{\eta})}\). To motivate the definition, let us start by noting that it is
natural to say
that elements of \(\mathfrak{m}_{(\eta, \tilde{\eta})}\) vanish to (at least)
first order at the point \((\eta, \tilde{\eta})\).
One way
to think of this is to consider a point 
\begin{equation}\label{eq:perturbed_point}
(\eta^{(\epsilon)}, \tilde{\eta}^{(\epsilon)}) = (\eta + \epsilon \delta, \tilde{\eta} + \epsilon \tilde{\delta}) \, ,
\end{equation}
for some small quantity \(\epsilon\) and a point \((\delta, \tilde{\delta})\)
in spinor space which is not required to satisfy momentum conservation
itself, but is chosen such that \((\eta^{(\epsilon)}, \tilde{\eta}^{(\epsilon)})\)
satisfies momentum conservation. At this shifted point
\((\eta^{(\epsilon)}, \tilde{\eta}^{(\epsilon)})\), the generators of
\(\mathfrak{m}_{(\eta, \tilde{\eta})}\) are all proportional to
\(\epsilon\) and we can interpret this as vanishing to first order. It
is easy to see that if we raise \(\mathfrak{m}_{(\eta, \tilde{\eta})}\) to \(k^\mathrm{th}\)
power then all of the generators of this power ideal will be
proportional to \(\epsilon^k\).  This leads us to define that an
element \(q\in R_n\) vanishes to \(k^{\mathrm{th}}\) order at a point \((\eta,
\tilde{\eta})\), if it is an element of \(\mathfrak{m}_{(\eta,
\tilde{\eta})}^k\), i.e.
\begin{equation}
    q \in \mathfrak{m}_{(\eta, \tilde{\eta})}^k \;\; \Rightarrow \;\; q \text{ vanishes to } k^{\mathrm{th}} \text{ order at } {(\eta, \tilde{\eta})} \, .
    \label{eq:PointVanishingDefinition}
\end{equation}
We remind the reader that the ideal power is computed by repeated multiplication
of the generators, see Eqs.~\eqref{eq:IdealProductDefinition} and
\eqref{eq:IdealPowerDefinition}.

\paragraph{Vanishing to Higher Order on a Variety.}
\label{sec:org0f68cc5}

Let us now consider elements of \(R_n\) that vanish not just at a single
point, but on an entire variety \(W\).
We ask an analogous question to the case where the variety was a
single point, that is whether we can construct the set of
elements that vanish to \(k^{\mathrm{th}}\) order at every point on
\(W\). Given our previous discussion, we therefore want to understand elements
which belong to \({\mathfrak m}_{(\eta, \tilde{\eta})}^k\) for every
\((\eta, \tilde{\eta}) \in W\). This is the intersection of each of these ideals, that is
\begin{equation}
\bigcap_{(\eta, \tilde{\eta})  \in W} {\mathfrak m}_{(\eta, \tilde{\eta})}^k \, .
\end{equation}
We will refer to this intersection as the set of elements of \(R_n\) which
vanish to \(k^{\mathrm{th}}\) order on \(W\).
As the set of points in \(W\) is potentially infinite, the computation of this
intersection is a non-trivial exercise. A natural expectation is that the set of
elements which vanish to \(k^{\mathrm{th}}\) order on \(W\) is related to \(I(W)^k\).
However, it turns out that \(I(W)^k\) is insufficient: there can exist elements of
\(R_n\) which vanish to \(k^{\mathrm{th}}\) order but do not belong to \(I(W)^k\). We
must introduce a refined definition of ideal power: the so-called \textbf{symbolic
power} (see Chapter 3.9~\cite{eisenbud1995commutative}).

Let us begin with an irreducible variety \(U\). In the case of a prime ideal such as
\(I(U)\), the symbolic power can be defined as the \(I(U)\text{-primary}\)
component of the ideal power. More precisely, we consider the minimal
primary decomposition of \(I(U)^k\) which we can write as\footnote{It is
perhaps surprising that there could be multiple primary components
\(Q_i\), since \(\sqrt{I(U)^k} = I(U)\), but this is possible because in
general \(V(Q_i) \subseteq U\).  That is, it is possible that 
\(V(Q_i)\) can be embedded. This corresponds to the fact that the ideal \(I(U)^k\) does not
necessarily contain all functions that vanish to \(k^{\mathrm{th}}\)
order on \(U\).}
\begin{equation}
    I(U)^k = \bigcap_{i=1}^m Q_i \, . 
\label{eq:PkPrimaryDecomposition}
\end{equation}
The \(k^{\mathrm{th}}\) symbolic power of an ideal associated to an irreducible variety \(U\) is defined as
\begin{equation}
    I(U)^{\langle k \rangle} = Q_j \, , \;\, \text{where} \;\; \sqrt{Q_j} = I(U) \, ,
    \label{eq:SymbolicDefinitionPrime}
\end{equation}
that is, the \(k^\text{th}\) symbolic power of \(I(U)\) is the unique primary component
\(Q_j\) of \(I(U)^k\) whose associated prime is \(I(U)\).
It is clear from the definition that \(I(U)^{\langle 1 \rangle} = I(U)\). Consider now a
situation where we work with a reducible variety \(W\). Then the 
symbolic power can be defined as the intersection of the symbolic powers of the ideals associated to the irreducible components of \(W\), i.e.
\begin{equation}
   I(W)^{\langle k \rangle} = \bigcap_{P_i \, \in \, \mathrm{assoc}( I(W))} P_i^{\langle k \rangle} \, .
\label{eq:RadicalSymbolicPowerDefinition}
\end{equation}

We are now prepared to describe the set of elements of \(R_n\) which vanish to
\(k^{\mathrm{th}}\) order on a variety. The key theorem we need is the so-called
``Zariski--Nagata theorem'' \cite{Zariski1949, Nagata1962}, in the general form
introduced by Eisenbud and Hochster \cite{EISENBUD1979157}. For our purposes, it
states that for a radical ideal \(J\) in \(R_n\)
\begin{equation}
   \bigcap_{(\eta, \tilde{\eta})  \in V(J)} {\mathfrak m}_{(\eta, \tilde{\eta})}^k \subseteq J^{\langle k \rangle} \, .
  \label{eq:ZariskiNagata}
\end{equation}
where we stress that the powers of the maximal ideals, and the symbolic power of
\(J\), are computed in \(R_n\).
We see that Zariski--Nagata tells us that the set of elements of \(R_n\) that vanish to
\(k^\mathrm{th}\) order on the variety \(V(J)\) is contained within the
\(k^\mathrm{th}\) symbolic power of \(J\).
Therefore, we see that if we
wish to compute the set of polynomials that vanish on a variety \(W\) to
\(k^{\mathrm{th}}\) order, it is sufficient to compute \(I(W)^{ \langle k \rangle}\).

\paragraph{Computing Symbolic Powers.}
\label{sec:orgb260fb7}
A natural question is how one computes the symbolic power in practice. It is clear
from Eq.~\eqref{eq:SymbolicDefinitionPrime} that, for a prime ideal \(P\),
one can calculate the \(k^{\mathrm{th}}\) symbolic power \(P^{\langle k \rangle}\)
by computing the primary decomposition of
\(P^k\). However, obtaining the primary decomposition can be computationally
demanding. In order to circumvent this, we
now introduce a useful lemma.  First, we note an
important technical property of \(R_n\). As it is a quotient
of a polynomial ring by a maximal codimension ideal, \(R_n\) is
``Cohen--Macaulay'' \cite[Proposition 18.13]{eisenbud1995commutative}.
This property has a very useful consequence for certain ideals when computing
symbolic powers. Specifically, if \(A\) is a Cohen--Macaulay ring and \(J\) is a
maximal codimension ideal of \(A\) then the ideal power and symbolic power coincide
\cite[Appendix~6,~Lemma~5]{zariski2013commutative}. That is, for
an ideal \(J\) of \(R_n\) we have
\begin{equation}\label{eq:SimpleSymbolicPowerLemma}
\text{codim}(J) = \mu(J) \;\; \Rightarrow \;\;  J^{\langle n \rangle} = J^n \, .
\end{equation}
One can understand this as follows: if the ideal associated to a
variety \(U\) is of maximal codimension, then the functions which
vanish to \(k^{\mathrm{th}}\) order on \(U\) are simply given by \(I(U)^k\). 
In many cases, strategic application of the lemma in
Eq.~\eqref{eq:SimpleSymbolicPowerLemma} allows one to bypass the computation
of unnecessary primary decompositions when computing symbolic powers.

\paragraph{Examples of Symbolic Powers.}
\label{sec:org31418d4}
To build intuition, let us reconsider the prime ideals \(P_1\) through
\(P_5\) in the four-point quotient ring \(R_4\), as given in
Eq.~\eqref{eq:four-point-decompositions}. It can be shown that
\begin{equation}
   \big\langle \langle 12 \rangle, [34] \big\rangle_{R_4}^{\langle 2 \rangle} = \big\langle \langle 12 \rangle, [34] \big\rangle_{R_4}^{2} = \big\langle \langle 12 \rangle^2, \langle 12 \rangle  [34] , [34]^2 \big\rangle_{R_4} \, .
\end{equation}
That is, in this case, the second symbolic power agrees with the second ideal
power. In fact, for almost all of the \(P_i\) in
Eq.~\eqref{eq:four-point-decompositions} this holds. Specifically,
\begin{equation}
\text{assoc}(P_i^2) = \{P_i\} \quad \Longrightarrow \quad P_i^2 = P_i^{\langle 2 \rangle} \quad \forall \; i \neq 5 \, .
\end{equation}
That is, the second ideal power corresponds to the second symbolic one
in all cases except for \(P_5\). Let us then consider the 
case of \(P_5\), where the symbolic power does not coincide
with the normal power. The associated primes are
\begin{equation}
\text{assoc}(P_5^2) = \{P_5, P_\text{x}\} \quad \text{with} \quad P_\text{x} = \big\langle \langle ij \rangle, [ij] \; \,\, : \,\, 1 \le i < j \le n \big\rangle_{R_4} \, ,
\end{equation}
and the primary decomposition reads 
\begin{equation}
P_5^2 = Q_5 \cap Q_\text{x} \quad \text{with} \quad \sqrt{Q_5} = P_5\, , \; \sqrt{Q_\text{x}} = P_\text{x} \quad \Longrightarrow \quad P_5^{\langle 2 \rangle} = Q_5 \, .
\end{equation}
One finds that the size of the minimal generating sets are given by \(\mu(Q_5) =
16\) and \(\mu(Q_x) = 49\). We, therefore, do not print these ideals in the text, but
they are easily obtainable with computer algebra techniques.
We note that there must be some polynomial which belongs to the symbolic
power \(P_5^{\langle 2 \rangle}\) but not to \(Q_\text{x}\), and hence not
to \(P_5^2\). It is easy to check that
\begin{equation}
[34] \notin Q_\text{x} \quad \text{and} \quad [34] \notin P_5^2 \, , \quad \text{but} \quad [34] \in P_5^{\langle 2 \rangle} \, .
\end{equation}

\section{Numerical Points near Singular Varieties}
\label{sec:orgc26bb05}

In the previous section, we introduced the class of polynomials
which vanish to a \(k^{\mathrm{th}}\) order on a variety. We now wish to understand how to
generate numerical configurations of spinors that are close to irreducible
singular varieties. This will allow us to numerically determine the degree of
vanishing of a numerator polynomial.
One of the important properties of finite fields that makes them useful in computer
algebra applications is that, in contrast to real or complex numbers, 
they can be exactly represented on a computer without approximation.
However, if we wish
to use finite fields to construct configurations of spinors which are close to
some other configuration, this is not possible as it turns out that the
available measure of size is not sufficiently powerful.
To understand this mathematically,
we now review the idea of an \textbf{absolute value} on a field \(\mathbb{F}\) which
will allow us to formalize the notion of size. Absolute values on fields are a
basic idea in the theory of number fields and we refer to
Ref.~\cite{Gouvea1997} for an introduction.
Mathematically, when we wish to discuss the size of elements of a field \(\mathbb{F}\) we
make use of a map \(|\cdot|_{\mathbb{F}}\) from a field \(\mathbb{F}\) to the non-negative real numbers
\(\mathbb{R}_{\ge 0}\), known as an absolute value\footnote{We note that a field \(\mathbb{F}\) may
admit multiple absolute values, but for the cases in this work it will be clear by context the one which we consider.}.
Well-known absolute values
include the standard ones on the real and complex numbers. If we have two elements \(x\) and \(y\) of a field
\(\mathbb{F}\), we will say that \(x\) is smaller than \(y\) if 
\begin{equation}
   |x|_{\mathbb{F}} \, < \, |y|_{\mathbb{F}} \, .
\label{eq:SizeComparison}
\end{equation}
Note that the result of \(|\cdot|_{\mathbb{F}}\) is always a real number, so the
comparison in Eq.~\eqref{eq:SizeComparison} takes place in the real numbers.
An absolute value also induces a metric \(d\) on \({\mathbb{F}}\), given
by
\begin{equation}
   d(x, y) = |x - y|_{\mathbb{F}} \, , 
\end{equation}
where \(x\) and \(y\) are two elements of \({\mathbb{F}}\). We will mostly make use of
Eq.~\eqref{eq:SizeComparison}---the ability to compare sizes of
elements of a field---in order to discuss points close to a variety.

Let us return to the finite-field case. It can be shown that the only absolute value on
\(\mathbb{F}_p\) is the so-called \textbf{trivial absolute value} which takes one of two values~\cite{Gouvea1997}. 
That is, for all \(a \in \mathbb{F}_p\) one can show that\footnote{This is easily proven by using Fermat's little
theorem and by multiplicativity and non-negativity of the absolute
value. Let \(a \in \mathbb{F}_p\), then:
\(a^p=a\;\Rightarrow\;|a^p| - |a| = 0\;\Rightarrow\;|a|^p - |a| =
0\;\Rightarrow\;|a|(|a|^{p-1}-1)=0\;\Rightarrow\;|a| = 0 \;\lor\; |a|
= 1\). Finally, by positive-definiteness we have \(|a| = 0
\;\Rightarrow\; a =0\), and hence Eq.~\eqref{eq:trivialabsval}.}
\begin{equation}\label{eq:trivialabsval}
|a = 0|_{\mathbb{F}_p} = 0 \quad \text{or} \quad |a \neq 0|_{\mathbb{F}_p} = 1 \, .
\end{equation}
Considering the induced metric on \(\mathbb{F}_p\), one can then say that two
elements \(x\) and \(y\) of \(\mathbb{F}_p\) are either \(0\) or \(1\) units apart. This
implies that in \(\mathbb{F}_p\) we can only generate phase-space points which
are either on or away from a given surface. Therefore, the induced metric does
not admit a non-trivial hierarchy of distances.

In order to bypass this issue, in Section \ref{sec:padicIntro} we introduce
another number-theoretical field that admits a more powerful measure of
distance: the \(p\verytiny\text{-adic}\) numbers. Then, in Sections
\ref{sec:FiniteFieldVarietyPoints} and \ref{sec:padicClose}, we show
how to start from a finite-field-valued configuration of spinors that is on a
variety to then construct a \(p\verytiny\text{-adic}\) configuration of spinors
which is close to said variety, by perturbing the finite-field configuration.

\subsection[Beyond Finite Fields: $p\verytiny\text{-adic}$ Numbers]{\boldmath Beyond Finite Fields: $p\verytiny\text{-adic}$ Numbers}\label{sec:padicIntro}

In this section, we review mathematical details of the \(p\verytiny\text{-adic}\) numbers
that are relevant for our applications. These are well-studied objects in
the mathematical literature and we refer the reader to textbooks such as
Ref.~\cite{Gouvea1997} for a pedagogical introduction to the topic.
We begin by introducing the so-called \textbf{\(\boldsymbol{p}\verytiny\text{-adic}\) integers}, which we
denote as \(\mathbb{Z}_p\). These are not to be confused with a finite
field with \(p\) elements, which we denote as \(\mathbb{F}_p\). An element
\(z \in \mathbb{Z}_p\) can be considered as a power series in a
prime number \(p\), i.e.
\begin{equation}\label{eq:padic_integer}
z = \sum_{i=0}^\infty a_i p^i = a_0 + a_1 p + a_2 p^2 + \cdots \, , 
\end{equation}
where the \(a_i\) take integer values in the range \([0, p-1]\).
We call the coefficients \(a_i\) the \textbf{\(\boldsymbol{p}\verytiny\text{-adic}\) digits} of \(z\), in
analogy to a decimal representation of a real number. 
Multiplication and addition of elements of \(\mathbb{Z}_p\) can be defined using
the standard multiplication and addition rules for power series. However, one
must also take into account that the digits of the resulting series must still
live in the range \([0, p-1]\). This can always be achieved by carry rules, analogous to
performing arithmetic with decimal numbers. With this in mind, the first
non-zero \(p\verytiny\text{-adic}\) digit behaves like an element of \(\mathbb{F}_p\). It
can be shown that the set of \(p\verytiny\text{-adic}\) integers forms a ring under multiplication and addition. 
However, \(\mathbb{Z}_p\) is not a field: there exists no multiplicative inverse
for any element of \(\mathbb{Z}_p\) with zero as its first \(p\verytiny\text{-adic}\) digit.

Let us reconsider the power series representation of a \(p\verytiny\text{-adic}\) integer given in
Eq.~\eqref{eq:padic_integer}. Note that extending this representation to
allow for negative powers of \(p\) solves the issue that prevents the
\(p\verytiny\text{-adic}\) integers from being a field. 
This leads us to the \textbf{\(\boldsymbol{p}\verytiny\text{-adic}\) numbers}, which we denote as \(\mathbb{Q}_p\).
Specifically, an element \(x \in \mathbb{Q}_p\) takes the form
\begin{equation}\label{eq:padic_number}
x = \sum_{i=-l}^\infty a_i p^i = a_{-l} p^{-l} + \cdots + a_{-1} p^{-1} + a_0 + a_1 p + a_2 p^2 + \cdots \, ,
\end{equation}
where again \(a_i\) is an integer in the range \([0, p-1]\). Multiplication and
addition are again defined by power series operations with carries.
It is important to note that the \(p\verytiny\text{-adic}\) numbers are not an
algebraically closed field.

As promised, the \(p\verytiny\text{-adic}\) numbers are a field with a more powerful way to measure size.
To discuss this, given a \(p\verytiny\text{-adic}\) number \(x\), we first introduce the \textbf{\(\boldsymbol{p}\verytiny\text{-adic}\) valuation} of \(x\),
which we denote by \(\nu_p(x)\). Considering the
power series representation of \(x\in\mathbb{Q}_p\) of Eq.~\eqref{eq:padic_number}, the valuation of a non-zero \(x\) is the integer \(k\)
such that \(a_k\) is the first non-zero \(p\verytiny\text{-adic}\) digit of \(x\). That is, 
\begin{equation}\label{eq:valuation}
\nu_p(x) =  k \, \text{ such that } \, a_i = 0 \, \text{ for all } \, i < k \, .
\end{equation}
For \(x = 0\) it is conventional to take \(\nu_p(x) = \infty\).
The \textbf{\(\boldsymbol{p}\verytiny\text{-adic}\) absolute value}, which we denote by \(|x|_p\), is defined as
\begin{equation}\label{eq:padic_absolute_value}
|x|_p = p^{-\nu_p(x)} \, ,
\end{equation}
for \(x \ne 0\), and \(|0|_p = 0\). It is this absolute value on \(\mathbb{Q}_p\) that we will use to discuss size.

This measurement of size has a number of interesting implications. Firstly, we see
that \(p\verytiny\text{-adic}\) numbers which are proportional to \(p\) are \(p\verytiny\text{-adically}\) small, and
those proportional to \(\frac{1}{p}\) are \(p\verytiny\text{-adically}\) large. That is,
considering \(p\) as a \(p\verytiny\text{-adic}\) number, we have
\begin{equation}
  |p|_p < 1 < \left| \frac{1}{p} \right|_p \, .
\end{equation}
We emphasize that, when considered \(p\verytiny\text{-adically}\), the quantity \(p\) is to
be regarded as small.
Note that the \(p\verytiny\text{-adic}\) integers form a subset
of the \(p\verytiny\text{-adic}\) numbers whose absolute value is bounded from above. That is,
\begin{equation}
   |x|_p \le 1 \quad \text{for all} \quad x \in \mathbb{Z}_p \, .
\end{equation}
Next, note that the \(p\verytiny\text{-adic}\) absolute value is discrete and
unbounded when considered over the set of \(p\verytiny\text{-adic}\) numbers. 
This is in contrast to the trivial absolute value, which is discrete but bounded to
either 0 or 1; or the standard absolute value over \(\mathbb{R}\) which is unbounded
but continuous.
Finally, we note that, while numbers in \(\mathbb{R}\) satisfy the triangle inequality
\begin{equation}\label{eq:triangle_inequality}
   |x+y|_{\mathbb{R}} \leq |x|_{\mathbb{R}} + |y|_{\mathbb{R}} \quad \text{for} \quad x, y \in \mathbb{R} \, ,
\end{equation}
those in \(\mathbb{Q}_p\) satisfy the strong triangle inequality
\begin{equation}\label{eq:strong_triangle_inequality}
   |x+y|_p \leq \text{max}(|x|_p, |y|_p) \quad \text{for} \quad x, y \in \mathbb{Q}_p \, .
\end{equation}
Eq.~\eqref{eq:strong_triangle_inequality} states that when one sums two \(p\verytiny\text{-adic}\) numbers, the result
cannot be larger than either of the two summands. In practice, this can be
helpful for establishing bounds on the size of intermediate stages of
calculations, which can be important for numerical stability. Furthermore, it
is important to note that, for large \(p\), the bound in
Eq.~\eqref{eq:strong_triangle_inequality} is frequently saturated in practice. This
can be seen by analogy to finite-field computations. Specifically, in
\(\mathbb{F}_p\) it is well-understood that a quantity accidentally
evaluating to zero can be made less probable by raising the value of \(p\). As the
first digit of a \(p\verytiny\text{-adic}\) number behaves like an element of a
finite-field, we see that this implies that, by working with large \(p\), one can
make it improbable that such a quantity becomes accidentally small. As quantities
accidentally becoming small is an important source of precision loss in many
algorithms, this has important practical implications for numerical stability.

\paragraph{\(\boldsymbol{p}\verytiny\text{-adics}\) on a Computer.}
\label{sec:org479b54e}
 Since computers have finite memories, one can consider truncating the power series
expansion of a \(p\verytiny\text{-adic}\) number. Recalling the form of a
\(p\verytiny\text{-adic}\) number \(x\) from Eq.~\eqref{eq:padic_number}, one can truncate the series and write
\begin{equation}\label{eq:truncated_padic_integer}
x =  a_{-l} p^{-l} + \ldots + a_{-1} p^{-1} + a_0 + a_1 p + \cdots + a_{m-1} p^{m-1} + \mathcal{O}(p^m) \, .
\end{equation}
This can be understood as a \(p\verytiny\text{-adic}\)
analogue of real numbers being represented by floating-point numbers of finite
precision.
Comparing the truncated power series in
Eq.~\eqref{eq:truncated_padic_integer} to the full series in
Eq.~\eqref{eq:padic_number}, we see that the error \(\mathcal{O}(p^m)\) made
by truncating the power series can be made small in a \(p\verytiny\text{-adic}\) sense:
increasing \(m\) decreases the error as \(|p^m|_p < |p^{m-1}|_p\).
To make use of this on a computer, we use a floating-point
representation\footnote{A public implementation of \(\mathbb{Q}_p\) can be found in
Sage~\cite{sagemath} or FLINT~\cite{flint}.}. Specifically, we write a
truncated \(p\text{-adic}\) number \(x\) as
\begin{equation}\label{eq:padic_number_implementation}
x = p^{\nu_p(x)} \left( \sum_{i=0}^{k-1} a_i p^i + \mathcal{O}(p^{k}) \right)  \;\, \text{with} \;\, a_i \neq 0 \, ,
\end{equation}
where we call the prefactor \(p^{\nu_p(x)}\) the \textbf{exponent}, the summation part
the \textbf{mantissa} and \(k \in \mathbb{Z}_{>0}\) the \textbf{working precision}.
The mantissa can be stored as a positive integer modulo \(p^k\) and so practical floating-point \(p\verytiny\text{-adic}\) arithmetic is very similar to working modulo \(p^k\).

Let us consider basic arithmetic operations in the floating-point representation.
First, consider multiplying \(x\) by another \(p\verytiny\text{-adic}\) number \(y\), whose digits we
denote as \(b_i\). This is given by
\begin{equation}
 xy = p^{\nu_p(x) + \nu_p(y)} \left[ \left(\sum_{i=0}^{k-1} a_i p^i\right) \left(\sum_{i=0}^{k-1} b_i p^i\right) + \mathcal{O}(p^{k}) \right] \, .
\end{equation}
Here, we can clearly identify the exponent of the product as the sum
of the exponents, and the mantissa of the product as the product of the
mantissae modulo \(p^k\). The multiplicative inverse of \(x\) can computed as
\begin{equation}
   x^{-1} = p^{-\nu_p(x)} \left( \overline{x} + \mathcal{O}(p^k) \right) \, ,
\end{equation}
where \(\overline{x}\) is an integer satisfying
\begin{equation}
   n p^k + \overline{x} \,\, \sum_{i=0}^{k-1} a_i p^i   = 1 \, ,
\end{equation}
for some auxiliary integer \(n\).
Such a pair \((\overline{x}, n)\) can easily be computed through the extended
Euclidean algorithm applied to the mantissa and \(p^k\), in analogy to the finite-field case (see, for example,
Ref.~\cite[Appendix A]{Peraro:2016wsq}).
Note that, as \(p\) is prime, both multiplication and the computation of multiplicative inverse
have the property that the mantissa of the result cannot be proportional to
\(p\), as required in the floating-point representation of
Eq.~\eqref{eq:padic_number_implementation}. 

Let us now consider addition in the floating-point representation. In contrast
to multiplication, questions of stability arise.
Without loss
of generality we can consider \(\nu_p(y) \ge \nu_p(x)\) and write the summation
of \(x\) and \(y\) as
 \begin{equation}
      x + y = p^{\nu_p(x)} \left[ \left(\sum_{i=0}^{k-1} a_i p^i\right) + p^{[\nu_p(y)-\nu_p(x)]} \left(\sum_{i=0}^{k-1} b_i p^i\right) + \mathcal{O}(p^k) \right] \, .
\label{eq:FloatingPadicAddition}
 \end{equation}
In comparison to multiplication, it is more subtle to compute the exponent and
mantissa of the sum from this form. Specifically, for the case where \(\nu_p(x) =
\nu_p(y)\), the part in square brackets in Eq.~\eqref{eq:FloatingPadicAddition} may be proportional to \(p\). 
This
violates the assumption in Eq.~\eqref{eq:padic_number_implementation} that
the leading digit of the mantissa is non-zero. To return to the floating-point
representation, one must then shuffle factors of \(p\) from the mantissa to the
exponent. However, as the mantissa is only known to \(k\) digits, this procedure
introduces an arbitrary choice into the last digits of the new mantissa. In more
traditional terms, one may lose precision when performing addition.
In practice, similarly to working in \(\mathbb{F}_p\), this can be made unlikely
to accidentally happen by increasing the size of the prime \(p\). We note that this
is a generalization of the issue of accidental division by zero in finite
fields.

\subsection{Finite-Field Points on Singular Varieties}
\label{sec:org0fbd25e}
\label{sec:FiniteFieldVarietyPoints}

Let us now discuss how one can generate a point on a variety when working in
\(\mathbb{F}_p\).
To ease the discussion, we will work over the polynomial ring \(\mathbb{F}_p[X_1, \ldots, X_n]\), and denote the tuple of variables as \(\underline{X} = \{X_1, \ldots, X_n\}\).
Given an ideal \(J = \langle q_1, \ldots, q_m \rangle_{\mathbb{F}_p[\underline X]}\), we wish to generate a numerical point
\(\underline X^{(0)} \in \mathbb{F}_p^{n}\) that is a solution to the equations
\begin{equation}
    q_i(\underline{X}) = 0 \quad \text{for} \quad i = 1, \ldots, m \, .
\label{eq:EquationSystem}
\end{equation}
That is, \(\underline X^{(0)} \in V(J)\). Clearly, for a variety that is not zero
dimensional, there are many such points \(\underline{X}^{(0)}\). In the following
we will focus on constructing a single point \(\underline X^{(0)} \in V(J)\).
Geometrically, our strategy is to intersect the variety with a randomly chosen collection of
hyperplanes so that this intersection is a zero-dimensional variety. The
zero-dimensional variety then corresponds to a finite collection of points. We
explicitly construct one such point and take this to be \(\underline{X}^{(0)}\).

In order to build the set of hyperplanes, we begin by constructing a
maximally independent set of \(J\), as discussed in
Section~\ref{sec:IndependentSets}. We denote the maximally independent set
as \(\underline Y\) and the corresponding dependent variables as \(\underline Z = \underline X \,
\backslash \, \underline Y\).
We remind the reader that \(\underline{Y}\) is a tuple of \(\dim(J)\) variables and that
\(\underline{Z}\) is a tuple of \(\mathrm{codim}(J)\) variables.
By definition, the elements of the set \(\underline Y\) can be chosen independently. If we choose
values for \(\underline{Y}\) generically, then they specify a variety such that its intersection with \(V(J)\) is
a zero-dimensional sub-variety of \(V(J)\). To this end, we construct
a point \(\underline{Y}^{(0)} \in \mathbb{F}_p^{\dim(J)}\) by choosing each
component uniformly as integers from the range \([0, p-1]\).
With this point in hand, we now consider the system
\begin{equation}
    q_i(\underline{Z}, \underline{Y}^{(0)}) = 0 \, \quad \text{for} \quad i = 1, \ldots, m \, .
\label{eq:ReducedEquationSystem}
\end{equation}
This system of equations defines our zero-dimensional subvariety
of \(V(J)\). 
Note that the polynomials in Eq.~\eqref{eq:ReducedEquationSystem} \(q_i\)
depend only on the \(\underline Z\) variables, as the components of
\(\underline{Y}^{(0)}\) take values in \(\mathbb{F}_p\). It is useful to introduce
the corresponding ideal in the polynomial ring \(\mathbb{F}_p[\underline Z]\), as
\begin{equation}
J^{(0)} = \langle q_1(\underline{Z}, \underline{Y}^{(0)}), \ldots, q_m(\underline{Z}, \underline{Y}^{(0)}) \rangle_{\mathbb{F}_p[\underline Z]} \, .
\label{eq:ReducedIdeal}
\end{equation}
Clearly, any \(\underline{Z}^{(0)} \in V(J^{(0)})\) can be combined with
\(\underline{Y}^{(0)}\) to find our desired point \(\underline{X}^{(0)}\).

Our problem is now reduced to the simpler task of finding an element of
\(V(J^{(0)})\). However, in general the system of Eqs.~\eqref{eq:ReducedEquationSystem} is non-linear in \(\underline{Z}\), which
makes this a non-trivial exercise.
To this end, we make use of standard tools of elimination
theory, which we now review. We refer the reader to Chapter 3 of Ref.~\cite{cox1994ideals} for a pedagogical introduction.
The key tool we will use is a Gröbner basis with a special monomial ordering.
The ordering that we need is the so-called lexicographic ordering on
the variables \(\underline{Z}\), which we denote as \(\succeq_{\text{lex.}}\).
Specifically, we order the variables as
\begin{equation}
\succeq_{\text{lex.}}: Z_{\mathrm{codim}(J)} \succ \dots \succ Z_1 \, .
\end{equation}
To highlight the use of this monomial order, we will denote the lexicographic
Gröbner basis of \(J^{(0)}\) as \(\mathcal{G}_{\text{lex.}}(J^{(0)})\).
We now consider the subset of \(\mathcal{G}_{\text{lex.}}(J^{(0)})\) which depends
only on the variables \(Z_1\) through \(Z_l\). That is, we define
\begin{equation}
    \mathcal{G}_l = \mathcal{G}_{\text{lex.}}(J^{(0)}) \cap \mathbb{F}_p[Z_1, \ldots, Z_l] \, .
\label{eq:lLexEliminationIdeals}
\end{equation}
The sets of polynomials \(\mathcal{G}_l\) allow one to find a zero of
\(\mathcal{G}_{\text{lex.}}(J^{(0)})\) in an iterative manner, constructing it
variable by variable.
We will call a zero \(\{Z_1^{(0)}, \ldots, Z_l^{(0)}\}\) of the polynomials
\(\mathcal{G}_l\) an \(l^{\mathrm{th}}\) partial solution. 
Note that one can always construct a \(0^{\mathrm{th}}\) partial solution as this
is the empty set.
Given a \((l-1)^{\mathrm{th}}\) partial solution \(\{Z_1^{(0)}, \ldots,
Z_{l-1}^{(0)}\}\), our task is to find a \(Z_l^{(0)} \in \mathbb{F}_p\) such that \(\{Z_1^{(0)},
\ldots, Z_{l}^{(0)}\}\) is an \(l^{\mathrm{th}}\) partial solution. We will refer
to this as extending the \((l-1)^{\mathrm{th}}\) partial solution.
Clearly, repeatedly extending a partial solution will lead to an element of
\(V(J^{(0)})\).

To discuss how to extend a partial solution,  let us consider the ideal
generated by the evaluations of \(\mathcal{G}_l\) on an
\((l-1)^{\mathrm{th}}\) partial solution. That is, we consider the ideal
\begin{equation}
    J^{(0)}_{l, \text{eval}} = 
    \left\langle g(Z_1^{(0)}, \ldots, Z_{l-1}^{(0)}, Z_l ) \,\, : \,\, g \in \mathcal{G}_{l} \right\rangle_{\mathbb{F}_p[Z_l]}.
\end{equation}
It can be shown that \(J^{(0)}_{l, \text{eval}}\) is generated by a single
polynomial \(g_o\) (see Chapter 3.5 of Ref.~\cite{cox1994ideals}). That is,
one can write \(J^{(0)}_{l, \text{eval}}\) as
\begin{equation}
J^{(0)}_{l, \text{eval}} = \langle g_o \rangle_{\mathbb{F}_p[Z_l]} \, .
\label{eq:goDim0Ideal}
\end{equation}
Note that a zero of \(g_o\) is a \(Z_l^{(0)}\) that allows us to extend the
\((l-1)^{\mathrm{th}}\) partial solution.
Importantly, \(g_o\) can be read from \(\mathcal{G}_{l}\).
Specifically, let us write each element \(g\) of \(\mathcal{G}_l\) in the form
\begin{equation}
   g = c_g(Z_1, \ldots, Z_{l-1})Z_l^{N_g} + \text{ terms in which } Z_l \text{ has degree } < N_g \, .
\end{equation}
If we consider the set of polynomials \(g' \in \mathcal{G}_l\) such that \(c_{g'}\) does not evaluate
to zero on the \((l-1)^{\mathrm{th}}\) partial solution, then \(g_o\) can be taken to be a \(g'\) which is a non-constant
polynomial and has lowest degree in \(Z_l\) amongst all such \(g'\). 
To extend a partial solution, we must therefore find a zero of the univariate
polynomial \(g_o\). This can be solved systematically over \(\mathbb{F}_p\) by
general, efficient algorithms such as the Cantor--Zassenhaus algorithm
\cite{cantor1981new}. Note that, in principle, there may be multiple zeros. As
we only want a single point on the variety, it is sufficient to take a single
such zero.

In summary, starting from the trivial \(0^{\mathrm{th}}\) partial solution, we
repeatedly extend the partial solution until we have constructed the
\(\mathrm{codim}(J)^{\mathrm{th}}\) partial solution, which is the desired
\(\underline{Z}^{(0)}\) which satisfies Eq.~\eqref{eq:ReducedEquationSystem}.
This is then combined with \(\underline{Y}^{(0)}\), to give the desired
\(\underline {X}^{(0)}\).
In general, this procedure of extending a partial solution is only guaranteed to
succeed when working in an algebraically closed field, as this means \(g_o\)
must have a zero in the field. This is relevant to our case as we work over
\(\mathbb{F}_p\) which is not algebraically closed.
We find a practical solution to this problem is to repeat the procedure with
different choices of the \(\underline{Y}^{(0)}\).

We make some final remarks. Firstly, we consider applying this procedure in the
case where \(V(J)\) is reducible. This procedure will still generate a point
belonging to \(V(J)\), however it provides no guarantee as to which branch of
\(V(J)\) the point belongs. In practice we solve this issue by only applying the
approach to prime ideals.
Secondly, there exist other algorithms to enumerate the elements of \(V(J^{(0)})\)
that avoid the use of the lexicographic monomial ordering: see, for example,
Section 2.4 of Ref.~\cite{cox2006using}. This can prove more
efficient. However, we do not find this to be necessary in this work.

\subsection[$p\verytiny\text{-adic}$ Points Close to Singular Varieties]{\boldmath $p\verytiny\text{-adic}$ Points Close to Singular Varieties}\label{sec:padicClose}

As already mentioned, we aim to evaluate rational functions on points in spinor space
which are \(p\verytiny\text{-adically}\) close to singular varieties. This
will allow us to numerically probe rational functions to learn how fast they
diverge or vanish.
It this section we discuss how to obtain a \(p\verytiny\text{-adic}\) point close to a
given variety by perturbing an exact finite-field solution. Thereafter, we
discuss how one can interpret this behavior in the language of algebraic geometry.

\paragraph{Lifting \(\fakebold{\mathbb{F}}_{\boldsymbol{p}}\) Solutions to the \(\boldsymbol{p}\verytiny\text{-adic}\) Integers.}
\label{sec:org42e6816}

Consider an ideal \(J\) of \(S_n\) that takes the form 
\begin{equation}
    J = \langle q_1, \ldots, q_m, r_1, \ldots, r_4 \rangle_{S_n} \, ,
\end{equation}
where \(\{r_1,  \ldots, r_4 \}\) generate \(J_{\Lambda_n}\). Naturally, \(V(J)
\subset V(J_{\Lambda_n})\) and we further assume that \(J\) is prime. 
We wish to construct a point \((\eta^{(\epsilon)},
\tilde\eta^{(\epsilon)}) \in \mathbb{Z}_p^{4n}\), such that
\begin{align}
\begin{split}
q_i(\eta^{(\epsilon)}, \tilde\eta^{(\epsilon)}) &= \mathcal{O}(p) \quad \text{for} \quad i = \{1, \ldots, m\} \, , \\
r_j(\eta^{(\epsilon)}, \tilde\eta^{(\epsilon)}) &= \mathcal{O}(p^k) \quad \text{for} \quad j = \{1, \ldots, 4\} \, , 
\end{split}
\label{eq:padicSolution}
\end{align}
where \(k\) is a positive integer. 
We stress that a \(p\verytiny\text{-adic}\) integer point will be suitable for our
purposes.
As the evaluations of the generators \(q_i\) are \(p\verytiny\text{-adically}\) small, we consider such a point
to be close to \(V(J)\).
As discussed in Section \ref{sec:padicIntro}, when working with a computer we work with
truncated \(p\verytiny\text{-adic}\) numbers. Therefore a solution to Eq.~\eqref{eq:padicSolution} is a point that is close to \(V(J)\), but on
\(V(J_{\Lambda_n})\) when working to \(k\) digits of precision.

To construct our desired point, we will work digit by digit in the
\(p\verytiny\text{-adic}\) expansion. 
Specifically, the \(p\verytiny\text{-adic}\) point in spinor space reads
\begin{equation}
(\eta^{(\epsilon)}, \tilde{\eta}^{(\epsilon)}) = \left(\eta^{(\epsilon),0} + p \, \eta^{(\epsilon),1} + \ldots + \mathcal{O}(p^k), \tilde{\eta}^{(\epsilon),0} + p \, \tilde{\eta}^{(\epsilon),1} + \ldots + \mathcal{O}(p^k)\right) ,
\label{eq:ClosePointExpansion}
\end{equation}
where each of the \((\eta^{(\epsilon), i},
\tilde{\eta}^{(\epsilon), i})\) are integers in the range \([0, p-1]\). We will determine the \((\eta^{(\epsilon), i},
\tilde{\eta}^{(\epsilon), i})\) starting from \(i=0\) and moving up to the working precision.

The starting observation is that for \((\eta^{(\epsilon)}, \tilde{\eta}^{(\epsilon)})\)
to be near \(V(J)\), it must be on \(V(J)\) when truncated to first digit.
We, therefore, begin with a finite-field-valued configuration which
lives on the variety analogous to \(V(J)\) over the finite fields, that is an
\((\eta, \tilde\eta)_{\mathbb{F}_p} \in \mathbb{F}_p^{4n}\). Clearly, we can use the algorithm of
Section \ref{sec:FiniteFieldVarietyPoints} to generate such a
configuration.
Importantly, \((\eta, \tilde\eta)_{\mathbb{F}_p}\) is a zero of the generators of
\(J\) when considered modulo \(p\). Therefore, we can reinterpret each finite-field
value as the first digit of a \(p\verytiny\text{-adic}\) integer.
That is, we choose
\begin{equation}
   (\eta^{(\epsilon), 0}, \tilde{\eta}^{(\epsilon), 0}) = (\eta, \tilde\eta)_{\mathbb{F}_p} \, ,
\end{equation}
where we consider the components of \((\eta, \tilde\eta)_{\mathbb{F}_p}\) to be integers in the range \([0, p-1]\).
We now have a \(p\verytiny\text{-adic}\) configuration which is a zero of the \(q_i\) and \(r_i\) up to 
\(\mathcal{O}(p)\) corrections. That is,
\begin{equation}\label{eq:padic_point_near_variety_start}
q_i(\eta^{(\epsilon),0}, \tilde\eta^{(\epsilon),0}) = r_i(\eta^{(\epsilon),0}, \tilde\eta^{(\epsilon),0}) = \mathcal{O}(p) \, .
\end{equation}
Note that Eq.~\eqref{eq:padic_point_near_variety_start} implies that
whatever the value of the digits \((\eta^{(\epsilon),i},
\tilde\eta^{(\epsilon),i})\) for \(i>0\), the \(q_i\) conditions of Eq.~\eqref{eq:padicSolution} will be satisfied.
Therefore, we will not need to consider the polynomials \(q_i\) further.
However, the present \(p\verytiny\text{-adic}\) point in spinor space does not yet
satisfy momentum conservation to working precision. 

Our task, therefore, is to choose the remaining digits of \((\eta^{(\epsilon)}, \tilde{\eta}^{(\epsilon)})\) such that momentum
conservation is satisfied to \(k\) digits. 
To achieve this, we work iteratively order by order in \(p\).
For convenience, let us define
\begin{equation}
(\eta^{(\epsilon), \overline{\nu}}, \tilde{\eta}^{(\epsilon), \overline{\nu}}) = \left(\sum_{i=0}^{\nu} p^i \eta^{(\epsilon), i}, \sum_{i=0}^{\nu} p^i \tilde{\eta}^{(\epsilon), i}\right) \, .
\end{equation}
This represents the first \(\nu+1\) digits of \((\eta^{(\epsilon)}, \tilde{\eta}^{(\epsilon)})\). 
Let us assume that we have determined the \(p\verytiny\text{-adic}\) spinors
\((\eta^{(\epsilon), i}, \tilde\eta^{(\epsilon), i})\) up to i = \(\nu\). 
That is, we assume that we have already fixed \(\nu+1\) digits such that
\begin{equation}
    r_i(\eta^{(\epsilon), \overline{\nu}}, \tilde\eta^{(\epsilon), \overline{\nu}}) = \mathcal{O}(p^{\nu+1}) \, .
\label{eq:padicIteration}
\end{equation}
Our aim is to find a value for the next digit, \((\eta^{(\epsilon), \nu+1},
\tilde\eta^{(\epsilon), \nu+1})\) such that each \(r_i\) will vanish to one order higher in \(p\).
It turns out that \((\eta^{(\epsilon), \nu+1}, \tilde\eta^{(\epsilon), \nu+1})\) satisfies a system of linear equations in a finite field.
Let us expand the four \(r_i\) polynomials around 
\((\eta^{(\epsilon), \overline{\nu}}, \tilde{\eta}^{(\epsilon), \overline{\nu}})\) to \(\mathcal{O}(p^{\nu+2})\). 
One finds
\begin{equation}\label{eq:multivariate_taylor_expansion}
r_i(\eta^{(\epsilon)}, \tilde\eta^{(\epsilon)}) = r_i(\eta^{(\epsilon), \overline{\nu}}, \tilde{\eta}^{(\epsilon), \overline{\nu}})
+ p^{\nu+1} \, \left[(\eta^{(\epsilon), \nu+1}, \tilde{\eta}^{(\epsilon), \nu+1}) \cdot \underline \nabla \, r_i \big|_{(\eta^{(\epsilon), \overline{\nu}}, \tilde{\eta}^{(\epsilon), \overline{\nu}})}\right] + \mathcal{O}(p^{\nu+2}) \, ,
\end{equation}
where \(\underline \nabla\) is the vector of derivatives with respect to
\((\lambda, \tilde\lambda)\).
If we now require that \((\eta^{(\epsilon)}, \tilde\eta^{(\epsilon)})\) is a zero
of \(r_i\) up to \(\mathcal{O}(p^{\nu+2})\) then we have a linear equation for the next digit. That is,
\begin{equation}
(\eta^{(\epsilon), \nu+1}, \tilde{\eta}^{(\epsilon), \nu+1}) \cdot \underline \nabla \, r_i\big|_{(\eta^{(\epsilon), \overline{\nu}}, \tilde{\eta}^{(\epsilon), \overline{\nu}})} = - \frac{1}{p^{\nu+1}}r_i(\eta^{(\epsilon), \overline{\nu}}, \tilde\eta^{(\epsilon), \overline{\nu}}) + \mathcal{O}(p) \, .
\label{eq:ExtensionConstraint}
\end{equation}
This is a linear system of equations for the next digit \((\eta^{(\epsilon), \nu+1}, \tilde{\eta}^{(\epsilon), \nu+1})\).
At a practical level, note that the constraints are modulo
\(p\), so they effectively give a linear system in \(\mathbb{F}_p\).

Importantly, the constraints in Eq.~\eqref{eq:ExtensionConstraint} always have a
solution, given an appropriate choice of \((\eta^{(\epsilon), 0}, \tilde{\eta}^{(\epsilon),0})\). We can see this as follows. Note that the derivatives of the
\(r_i\) are being evaluated close to the variety \(V(J_{\Lambda_n})\) and so up to
\(O(p)\) corrections we can replace them with their evaluations on the variety.
That is,
\begin{equation}
    \underline{\nabla} \, r_i \big|_{(\eta^{(\epsilon), \overline{\nu}}, \tilde{\eta}^{(\epsilon), \overline{\nu}})} = \underline{\nabla} \, r_i \big|_{(\eta, \tilde{\eta})} + O(p) \, ,
\end{equation}
where \((\eta, \tilde{\eta})\) is a point on \(V(J_{\Lambda_n})\).
Therefore, up to \(O(p)\) corrections, the derivative vectors in Eq.~\eqref{eq:ExtensionConstraint} span the cotangent space of \(V(J_{\Lambda_n})\) at \((\eta, \tilde{\eta})\). 
As \(J_{\Lambda_n}\) is a maximal codimension ideal this implies that, if \((\eta,
\tilde{\eta})\) is not a singular point of \(V(J_{\Lambda_n})\), the linear system of
equations in Eq.~\eqref{eq:ExtensionConstraint} is of full rank and a solution
exists. In practice it is easy to avoid such singular points.
Nevertheless, the system does not uniquely define the value of
\((\eta^{(\epsilon), \nu+1}, \tilde{\eta}^{(\epsilon), \nu+1})\) as it can be freely changed by any element
of the tangent space of the \(V(J_{\Lambda_n})\) at the point \((\eta,
\tilde{\eta})\). We make use of this freedom and pick a random solution to Eq.~\eqref{eq:ExtensionConstraint}.

Having determined the value of \((\eta^{(\epsilon), \nu+1},
\tilde{\eta}^{(\epsilon), \nu+1})\), we are now in a position where we have a
solution of Eq.~\eqref{eq:padicIteration} but with \(\nu\) replaced with \(\nu+1\). It is therefore
clear that we can iterate this procedure until we find a solution with \(\nu =
k-1\), which is thus a solution to Eq.~\eqref{eq:padicSolution}.

We close with a few remarks. Firstly, we point out that an analogous procedure could be
followed to generate points close to singular varieties when working
over \(\mathbb{R}\) or \(\mathbb{C}\). Secondly, let us also remark the similarity of this
multivariate procedure to the univariate Hensel's lifting lemma.
Thirdly, it would be interesting to consider extending this procedure to
generate points in ``asymmetric'' approaches to a variety as employed in
Ref.~\cite{DeLaurentis:2019phz}.

\paragraph{Interpretation of \(\boldsymbol{p}\verytiny\text{-adic}\) Evaluations.}
\label{sec:org47e6437}

Let us now consider how to interpret the evaluation of a numerator
\(\mathcal{N} \in R_n\) at a point
close to an irreducible singular variety \(U\). Let \(k\) be the largest
integer such that \(\mathcal{N} \in \mathfrak{m}^k_{(\eta,
\tilde\eta)}\) holds for all points \((\eta, \tilde\eta) \in U\). For
specific points \((\eta, \tilde{\eta})\), it may be the case that \(\mathcal{N}\)
belongs to \(\mathfrak{m}^{k+1}_{(\eta, \tilde\eta)}\), but these must always live
on higher
codimension sub-varieties. We define
\begin{equation}\label{eq:max_k}
\kappa(\mathcal{N}, U) = k \,\,\text{ s.t. }\, \left(\mathcal{N} \in \mathfrak{m}^k_{(\eta, \tilde\eta)} \; \forall \; (\eta, \tilde\eta) \in U \right) \;\, \text{and} \;\, \left(\exists \;  (\eta, \tilde\eta) \in U \, : \, \mathcal{N} \not\in \mathfrak{m}^{k+1}_{(\eta, \tilde\eta)} \right) \, .
\end{equation}
Note that \(\kappa(\mathcal{N}, U) \ge 0\), as
\(\mathcal{N} \in \mathfrak{m}^0_{(\eta, \tilde\eta)}\) holds trivially.
Importantly, it is clear from the definition of \(\kappa(\mathcal{N}, U)\) that we have
\begin{equation}
\mathcal{N} \in \bigcap_{(\eta,\tilde\eta) \in U}
\mathfrak{m}^{\kappa(\mathcal{N}, U)}_{(\eta,\tilde\eta)} \, ,
\label{eq:NUVanishingConstraint}
\end{equation}
and we cannot replace \(\kappa(\mathcal{N}, U)\) with any higher integer.
By the Zariski--Nagata theorem we conclude that
\begin{equation}
\mathcal{N} \in I(U)^{\langle {\kappa(\mathcal{N}, U)} \rangle} \, .
\end{equation}

We will now argue that, for large \(p\), a \(p\verytiny\text{-adic}\) evaluation of \(\mathcal{N}\)
near \(U\) allow us to determine \(\kappa(\mathcal{N},U)\) with high probability.  
Specifically, we will make use of
a \(p\verytiny\text{-adic}\) point in spinor space \((\eta^{(\epsilon)},
\tilde{\eta}^{(\epsilon)})\) as constructed earlier in this section
to satisfy Eq.~\eqref{eq:padicSolution}. 
The corresponding point \((\eta,
\tilde{\eta})\) on the variety \(U\) can be thought of as any of
the infinitely many \(p\verytiny\text{-adic}\) points on \(U\) with the same
first \(p\verytiny\text{-adic}\) digit as \((\eta^{(\epsilon)},
\tilde{\eta}^{(\epsilon)})\). We will argue that the probability of evaluating
\(\mathcal{N}\) at the point \((\eta^{(\epsilon)},
\tilde{\eta}^{(\epsilon)})\) and finding that its \(p\verytiny\text{-adic}\) valuation
exceeds \(\kappa(\mathcal{N},U)\) is small. We begin by noting that, by Eq.~\eqref{eq:NUVanishingConstraint},  \(\mathcal{N}\) 
is an element of \(\mathfrak{m}^{\kappa(\mathcal{N},U)}_{(\eta, \tilde\eta)}\). We can therefore write \(\mathcal{N}\) as 
\begin{equation}\label{eq:explicit_maximal_ideal_membership_in_Sn}
    \mathcal{N} = \sum_{|\beta| + |\tilde\beta| = \kappa(\mathcal{N},U)} n_{\beta,\tilde\beta}(\lambda,\tilde\lambda)  \prod_{i,\alpha} (\lambda_{i\alpha} - \eta_{i\alpha})^{\beta_{i\alpha}} \prod_{j, \dot{\alpha}} (\tilde{\lambda}_{j\dot{\alpha}} - \tilde{\eta}_{j\dot{\alpha}})^{\tilde\beta_{i\dot{\alpha}}}\, ,
\end{equation}
where the summation runs over all sets of powers \(\beta\) and
\(\tilde\beta\) such that the total degree of the product part of
Eq.~\eqref{eq:explicit_maximal_ideal_membership_in_Sn} is \(\kappa(\mathcal{N}, U)\) and \(n_{\beta,\tilde\beta}(\lambda,\tilde\lambda)\) is a polynomial in the spinor variables. The
indices \(i,\,j\) run from \(1\) to \(n\), with \(n\) being the multiplicity
of phase space, and \(\alpha,\,\dot\alpha\) are either \(0\) or
\(1\). 
Evaluating \(\mathcal{N}\) at \((\eta^{(\epsilon)},
\tilde{\eta}^{(\epsilon)})\) we obtain
\begin{equation}
    \mathcal{N}(\eta^{(\epsilon)}, \tilde{\eta}^{(\epsilon)}) = p^{\kappa(\mathcal{N}, U)} \tilde{\mathcal{N}}(\eta^{(\epsilon)}, \tilde{\eta}^{(\epsilon)}) + \mathcal{O}\left(p^{\kappa(\mathcal{N}, U) + 1}\right) \, ,
\end{equation}
where
\begin{equation}\label{eq:eta_eta_tilde_valuations_of_N}
    \tilde{\mathcal{N}}(\eta^{(\epsilon)}, \tilde{\eta}^{(\epsilon)}) = \sum_{|\beta| + |\tilde\beta| = \kappa(\mathcal{N}, U)} n_{\beta,\tilde\beta}(\eta^{(\epsilon),0},\tilde{\eta}^{(\epsilon),0})  \prod_{i,\alpha} \left(\eta_{i\alpha}^{(\epsilon),1}\right)^{\beta_{i\alpha}} \prod_{j, \dot{\alpha}} \left(\tilde{\eta}_{j\dot{\alpha}}^{^{(\epsilon),1}}\right)^{\tilde\beta_{i\dot{\alpha}}}.
\end{equation}
It is then clear that we can extract \(\kappa(\mathcal{N}, U)\) from the numerical
evaluation \(\mathcal{N}(\eta^{(\epsilon)}, \tilde{\eta}^{(\epsilon)})\), if we
can understand the valuation of \(\tilde{\mathcal{N}}(\eta^{(\epsilon)},
\tilde{\eta}^{(\epsilon)})\). We will now argue that if \(p\) is large, then with high probability 
\begin{equation}
\nu_p\left[\tilde{\mathcal{N}}(\eta^{(\epsilon)}, \tilde{\eta}^{(\epsilon)})\right] = 0 \, .
\end{equation}
Firstly, we argue that there exists, with high probability, some \(n_{\beta,
\tilde{\beta}}(\eta^{(\epsilon),0}, \tilde{\eta}^{(\epsilon),0})\) that is not
\(\mathcal{O}(p)\). 
If all \(n_{\beta, \tilde{\beta}}(\eta^{(\epsilon),0},
\tilde{\eta}^{(\epsilon),0})\) vanish modulo \(p\), this would imply that
\(\mathcal{N} \in \mathfrak{m}_{(\eta', \tilde{\eta}')}^{\kappa(\mathcal{N},
U)+1}\), for some point \((\eta', \tilde{\eta}')\) whose first \(p\verytiny\text{-adic}\) digit is given by
\((\eta^{(\epsilon),0}, \tilde{\eta}^{(\epsilon),0})\).
However, recalling the discussion around Eq.~\eqref{eq:max_k}, points such
as \((\eta', \tilde{\eta}')\) belong to higher codimension varieties.
As \((\eta^{(\epsilon),0}, \tilde{\eta}^{(\epsilon),0})\) has been
chosen randomly and \(p\) is large, such points are chosen with low probability.
Secondly, consider \(\tilde{\mathcal{N}}\) as a polynomial in the \((\eta^{(\epsilon),1},
\tilde{\eta}^{(\epsilon),1})\) given fixed \((\eta^{(\epsilon),0}, \tilde{\eta}^{(\epsilon),0})\). 
The point \((\eta^{(\epsilon),1}, \tilde{\eta}^{(\epsilon),1})\) could then be
close to a zero of this polynomial. However, as this point was also chosen randomly
and \(p\) is large, this also occurs with low probability as well.

In summary, for large \(p\), given a point \((\eta^{(\epsilon)},
\tilde{\eta}^{(\epsilon)})\) close to \(U\), and \(\mathcal{N}(\eta^{(\epsilon)},
\tilde{\eta}^{(\epsilon)})\) we conclude that with high probability
\begin{equation}
\label{eq:KappaCalculation}
\kappa(\mathcal{N}, U) = \nu_p\left(\mathcal{N}(\eta^{(\epsilon)}, \tilde{\eta}^{(\epsilon)})\right) \, .
\end{equation}

\section{Ansatz Construction Algorithm}
\label{sec:orgcdcb75d}
\label{sec:AnsatzConstructionAlgorithm}  

In this section, we leverage the
technology described so far to build an algorithm to construct Ansätze for
rational functions in scattering amplitudes. Specifically, for each coefficient
\(\mathcal{C}_i\) we discuss an algorithm to construct a set of rational functions
\(\{\mathfrak{a}_{i, 1}, \ldots, \mathfrak{a}_{i, d_i}\}\) of spinor variables such
that
\begin{equation}
    \mathcal{C}_i(\lambda, \tilde{\lambda}) = \sum_{k=1}^{d_i} c_{i, k} \mathfrak{a}_{i,k}(\lambda, \tilde{\lambda}) \, ,
\label{eq:PrototypeAnsatz}
\end{equation}
where the \(c_{i,k}\) are rational numbers.
Importantly, this Ansatz has fewer terms than those commonly considered in the
literature based on functional reconstruction techniques as it will take into
account the analytical properties of the rational functions. 
We consider the
coefficients in least common denominator form. That is, 
\begin{equation}
    \mathcal{C}_i(\lambda, \tilde{\lambda}) = \frac{\mathcal{N}_i(\lambda, \tilde{\lambda})}{\prod_{j=1}^{n_i} \mathcal{D}_j(\lambda, \tilde{\lambda})^{q_{ij}}} \, ,
    \label{eq:PrototypeRationalFunction}	   
\end{equation}
where \(\mathcal{N}_i\) and \(\mathcal{D}_j\) are elements of \(\mathcal{R}_n\) and
where \(n_i\) is the number of distinct denominator factors \(\mathcal{D}_j\). We
note that we allow the exponents \(q_{ij}\) to be negative, denoting numerator
factors.
For our procedure, we assume that the set of denominator factors \(\{\mathcal{D}_1, \ldots, \mathcal{D}_{n_i} \}\) in
Eq.~\eqref{eq:PrototypeRationalFunction} is known \emph{a priori}.
In physical applications, where the transcendental functions are pure, it is
conjectured that this set can be
constructed from the symbol alphabet \cite{Abreu:2018zmy}.
We further assume that we can numerically
evaluate the \(\mathcal{C}_i\) \(p\verytiny\text{-adically}\), e.g.~either from some analytic formula or from
an appropriate numerical algorithm.

\subsection{Study of Singular Varieties}
\label{sec:orgc66e9c4}

We begin by considering the behavior of the coefficient function \(\mathcal{C}_i\) on singular
varieties. This procedure has two parts: first, we find all relevant irreducible
singular varieties; second, we perform numerical evaluations near these
varieties and interpret the result.

\paragraph{Analytic Study.}
\label{sec:orga9f4521}
Let us consider the set of codimension-\(m\) varieties on which the rational
functions in Eq.~\eqref{eq:PrototypeRationalFunction} may diverge.
These are naturally associated to ideals generated by the denominator factors in
Eq.~\eqref{eq:PrototypeRationalFunction}. Specifically, the set of ideals
that define the singular varieties at codimension \(m\) is
   \begin{equation}
\label{eq:CodimensionMIdeals}
       \mathfrak{D}^{(m)} = \Big\{ J \text{ such that } J = \langle \mathcal{D}_{j_1}, \ldots, \mathcal{D}_{j_m} \rangle_{R_n} \text{ and } \mathrm{codim}(J) = m \Big\} \, ,
   \end{equation}
where the indices \(j_1, \ldots, j_m\) are all distinct and take values in \(1,
   \ldots, n_i\).
We remind the reader that ideals generated by \(m\) elements are not
necessarily of codimension \(m\). 
The varieties associated to the ideals in \(\mathfrak{D}^{(m)}\) may be reducible. To this end, we consider 
the set of irreducible varieties is given by
   \begin{equation}
       \mathcal{V}^{(m)} = \Big\{ U  \text{ such that }  U = V(P)  \text{ where } P \in \mathrm{minAssoc}\big( J \big) \text{ for some } J \in \mathfrak{D}^{(m)} \Big\} \, ,
\label{eq:CodimensionMVarieties}
   \end{equation}
Where we recall from Section \ref{sec:IrreducibleSingularVarieties}, that the set
of irreducible varieties can be extracted from the primary decomposition of the
associated ideal.

The first step of our algorithm is to construct generating sets of the ideals
associated to each variety in \(\mathcal{V}^{(m)}\), for \(m\) ranging from 1 to some
largest codimension of interest. In this work, we study the rational
functions in Eq.~\eqref{eq:PrototypeRationalFunction} only on varieties
of codimension one and two, and leave the impact of higher codimension
studies to further work. Therefore, we begin by performing the requisite
primary decompositions to construct \(\mathcal{V}^{(1)}\) and
\(\mathcal{V}^{(2)}\).

\paragraph{Numerical Warm-up.}
\label{sec:org56398f3}
\label{sec:warmup}

Given the two sets of varieties, \(\mathcal{V}^{(1)}\) and \(\mathcal{V}^{(2)}\), we
now use \(p\verytiny\text{-adic}\) numerical evaluations in order to determine strongly constraining
information about the function \(\mathcal{C}_i\). 
We do this in a two step
procedure, first working at codimension one, and then at codimension two.

\begin{enumerate}
\item \textbf{Codimension One:} The first step is to evaluate \(\mathcal{C}_i\)
near all codimension-one irreducible varieties whose associated ideals are
generated by one element, \(\mathcal{D}_j\). To each of these 
we associate an element of \(\mathcal{V}^{(1)}\), namely \(U_j = V(\langle \mathcal{D}_j \rangle_{R_n})\).
For each \(U_j\), we employ the procedure in Section \ref{sec:padicClose} to
generate a point \((\eta_{U_j}^{(\epsilon)}, \tilde{\eta}_{U_j}^{(\epsilon)})\) that is
\(p\verytiny\text{-adically}\) close to \(U_j\). As the
associated ideal is generated by the single irreducible element
\(\mathcal{D}_j\) and we work with large \(p\), we infer that
\begin{equation}
    q_{ij} = \nu_p\big(\mathcal{C}_i(\eta_{U_j}^{(\epsilon)}, \tilde{\eta}_{U_j}^{(\epsilon)})\big) \, .
\end{equation}
That is, we deduce the exponent of the denominator factor from the
\(p\verytiny\text{-adic}\) valuation of the coefficient when evaluated on a random point nearby
the associated variety.
\item \textbf{Codimension Two:} The second step is to study the behavior of \(\mathcal{C}_i\)
near all codimension-two irreducible varieties \(U \in \mathcal{V}^{(2)}\). 
Specifically, we make use of the numerical techniques of Section
\ref{sec:padicClose} to compute \(\kappa(\mathcal{N}_i, U)\), i.e.~to show
membership of \(\mathcal{N}_i\) to some symbolic power of \(I(U)\), where we
recall that \(I(U)\) is an ideal of \(R_n\).
To do this, for each \(U\) we again generate a point \((\eta_{U}^{(\epsilon)},
   \tilde{\eta}_{U}^{(\epsilon)})\) , which is close to \(U\).
As we know all \(q_{ij}\) from the codimension-one study, we can use
Eq.~\eqref{eq:PrototypeRationalFunction} to evaluate \(\mathcal{N}_i\) on
this point and thereby numerically compute
\(\nu_p\left(\mathcal{N}_i(\eta_{U}^{(\epsilon)},
   \tilde{\eta}_{U}^{(\epsilon)})\right)\). As we perform this procedure for
large \(p\), by Eq.~\eqref{eq:KappaCalculation}, we have calculated
\(\kappa(\mathcal{N}_i, U)\). Gathering all of these constrains, we conclude
that
\begin{equation}\label{eq:numerator_ideal_membership}
   \mathcal{N}_i \in \mathfrak{J}, \quad \text{where} \quad \mathfrak{J} = \bigcap_{U \in \mathcal{V}^{(2)}} I(U)^{\langle \kappa(\mathcal{N}_i, U) \rangle} \, .
\end{equation}
\end{enumerate}

\subsection{The Space of Vanishing Functions}
\label{sec:org7bbca31}
\label{sec:SpaceOfVanishingFunctions}

We have now shown that \(\mathcal{N}_i\) belongs to both the ideal \(\mathfrak{J}\) defined in Eq.~\eqref{eq:numerator_ideal_membership} and
to the space of polynomials of \(\mathcal{M}_{d, \vec{\phi}}\), defined in Eq.~\eqref{eq:PhysicalBracketSpace}. 
We wish to use these two
statements in order to construct an Ansatz of the form given in Eq.~\eqref{eq:PrototypeAnsatz}. To this end, we construct a basis of the space
\begin{equation}
\label{eq:NaiveAnsatzVectorSpaceIntersectSymbolicPowers}
    \mathfrak{M}_{d, \vec{\phi}}(\mathfrak{J}) = \mathcal{M}_{d, \vec{\phi}} \cap \mathfrak{J} \, .
\end{equation}
Once the denominators are restored, a basis of \(\mathfrak{M}_{d,
\vec{\phi}}(\mathfrak{J})\) can be used as the set of rational functions
\(\{\mathfrak{a}_{i, 1}, \ldots, \mathfrak{a}_{i, d_i}\}\) in
Eq.~\eqref{eq:PrototypeAnsatz}. 
There are a number of ways one can construct a basis of
\(\mathfrak{M}_{d,\vec{\phi}}(\mathfrak{J})\), and they can differ strongly in
computational complexity. 
For example, direct computation a generating set of \(\mathfrak{J}\) using Gröbner
basis methods can be intractable. Instead, we find it more
efficient to reduce the problem to one of vector space intersection.
Specifically, given a set of ideals \(\{J_1, \ldots, J_m\}\) in \(R_n\), it is clear that 
\begin{equation}
  \mathfrak{M}_{d, \vec{\phi}}\left(\bigcap_{i=1}^m J_k \right)  = \bigcap_{i=1}^m \mathfrak{M}_{d, \vec{\phi}}\left( J_k \right) \, .
\label{eq:SpaceDecomposition}
\end{equation}
This allows us to avoid computing a generating set for the ideal \(\mathfrak{J}\)
by Gröbner basis methods.
Instead, we construct a basis of each \(\mathfrak{M}_{d, \vec{\phi}}(J_k)\) and
perform the intersection of vector spaces in
Eq.~\eqref{eq:SpaceDecomposition} to find a basis of \(\mathfrak{M}_{d,
\vec{\phi}}(\mathfrak{J})\).

To practically construct a basis of each of the \(\mathfrak{M}_{d, \vec{\phi}}(J_k)\), we first note that
\begin{equation}
\mathfrak{M}_{d, \vec{\phi}}(J_k) = \mathfrak{M}_{d, \vec{\phi}}(J_k \cap \mathcal{R}_n),
\end{equation}
as \(\mathcal{M}_{d, \vec{\phi}}\) is a subspace of \(\mathcal{R}_n\).
This allows us to construct a basis of \(\mathfrak{M}_{d, \vec{\phi}}(J_k)\) by Gröbner basis techniques.
Specifically, we exploit the isomorphism in Eq.~\eqref{eq:invariant_qring}
and work with the polynomial quotient ring \(\mathcal{R}_n^{(q)}\) and the ideals
\(J_k^{(q)}\) that map to \(J_k \cap \mathcal{R}_n\) under the isomorphism (see
Eq.~\eqref{eq:conversion_covariant_invariant_elimination_ideal}).
By computing a basis of the intersection in \(\mathcal{R}_n^{(q)}\), we construct
a set of spinor bracket polynomials that can be
understood as a basis of \(\mathfrak{M}_{d, \vec{\phi}}(J_k)\) by the isomorphism.
To construct this basis we recall the technology of Section \ref{GroebnerBases} for intersecting ideals with vector spaces.
First we compute the remainders modulo \(\mathcal{G}(J_k^{(q)})\) of the elements
of \(M_{d, \vec{\phi}}\). In practice, the calculation of these remainders can
prove computationally intensive. Nevertheless, we find that the remainders
themselves are often simple. By Eq.~\eqref{eq:LinearizedIntersection}, we
then construct a basis of the nullspace of
\(\Delta_{ij}\big({\mathcal{G}[J_k^{(q)}]}, M_{d, \vec{\phi}} \big)\) to obtain a
basis of \(\mathfrak{M}_{d, \vec{\phi}}(J_k)\).

Finally, in order to compute a basis of \(\mathfrak{M}_{d, \vec{\phi}}(\mathfrak{J})\), we
make use of Eq.~\eqref{eq:SpaceDecomposition}, and perform the vector space
intersection with standard linear algebra techniques. 
We remark that, as the remainders modulo \(\mathcal{G}(J_k^{(q)})\) are simple,
the matrices \(\Delta_{ij}\big({\mathcal{G}[J_k^{(q)}]}, M_{d, \vec{\phi}} \big)\)
are sparse. Therefore we find that sparse linear algebra techniques are
efficient when
performing the relevant vector space intersections.

\paragraph{Organizing the Space.}
\label{sec:org0d05743}
We now have a basis for \(\mathfrak{M}_{d,
     \vec{\phi}}(\mathfrak{J})\). However, the techniques of Section
\ref{GroebnerBases} to intersect a vector space with an ideal make extensive
use of (sparse) linear algebra. This introduces
an arbitrary choice into the basis elements given by the pivoting scheme
made in the linear algebra algorithms. This has a practical downside as,
for large \(\dim\big(\mathfrak{M}_{d, \vec{\phi}}(\mathfrak{J})\big)\),
expressing the numerator \(\mathcal{N}_i\) in Eq.~\eqref{eq:PrototypeRationalFunction}
in terms of this basis leads to large rational
numbers. We wish
to address this by constructing a more compact
basis of \(\mathfrak{M}_{d, \vec{\phi}}(\mathfrak{J})\). To this end, we will
organize the basis in a way reminiscent of a partial-fraction
decomposition.

We begin by recalling that an element of \(\mathfrak{M}_{d,
     \vec{\phi}}(\mathfrak{J})\) is to be interpreted as the numerator of a
rational
function. In a partial-fraction decomposition, one attempts to cancel the
numerator against the denominator. In order for a numerator to
cancel against a factor of \(\mathcal{D}_k\) in the denominator, this numerator
must itself come with a factor of \(\mathcal{D}_k\). Naturally, numerators
which factorize \(\mathcal{D}_k\) form a subspace of
\(\mathfrak{M}_{d, \vec{\phi}}(\mathfrak{J})\). Specifically, they are given by
\begin{equation} 
  \mathfrak{M}_{d, \vec{\phi}}(\mathfrak{J}) \cap \langle \mathcal{D}_k \rangle_{\mathcal{R}_n} \, .
\end{equation} 
Note that, due to the
codimension-one study of Section \ref{sec:warmup}, the space of functions
\(\mathfrak{M}_{d, \vec{\phi}}(\mathfrak{J})\) have no common factors given by the
\(\mathcal{D}_k\). Therefore 
\begin{equation} 
\mathfrak{M}_{d, \vec{\phi}}(\mathfrak{J}) \cap \langle \mathcal{D}_k \rangle_{\mathcal{R}_n} \subsetneq \mathfrak{M}_{d, \vec{\phi}}(\mathfrak{J}) \, ,
\end{equation}
that is, it is a proper subspace.
Next, we recall Eq.~\eqref{eq:SpaceDecompositionIsomorphism} and note
that the \(\mathfrak{M}_{d, \vec{\phi}}(\mathfrak{J})\) can be related to the subspace of terms belonging to
the
ideal \(\langle \mathcal{D}_k \rangle_{\mathcal{R}_n}\) by
\begin{equation} 
 \mathfrak{M}_{d, \vec{\phi}}(\mathfrak{J}) 
 \cong 
 \left[\mathfrak{M}_{d, \vec{\phi}}(\mathfrak{J})  \cap \langle \mathcal{D}_k \rangle_{\mathcal{R}_n} \right]
 \oplus 
 \mathcal{Q}_{d, \vec{\phi}}(\mathfrak{J}, \mathcal{D}_k)\, ,
 \label{eq:subspaceDecomposition}
\end{equation} 
where
\begin{equation}
 \mathcal{Q}_{d, \vec{\phi}}(\mathfrak{J}, \mathcal{D}_k)  = \mathfrak{M}_{d, \vec{\phi}}(\mathfrak{J}) / \left[\mathfrak{M}_{d, \vec{\phi}}(\mathfrak{J})  \cap 
 \langle \mathcal{D}_k \rangle_{\mathcal{R}_n}\right]\, .
\end{equation}
We can therefore use a denominator factor \(\mathcal{D}_k\) to break down the
space into two smaller spaces.
In the context of partial fractions, we can interpret
Eq.~\eqref{eq:subspaceDecomposition} as the standard observation that
the choice of numerator of \(\mathcal{D}_k\) is only fixed up to terms
proportional to \(\mathcal{D}_k\). 

To make practical use of Eq.~\eqref{eq:subspaceDecomposition}, we again
employ the isomorphism in Eq.~\eqref{eq:invariant_qring}. This
allows us to construct a basis of the two spaces in the sum using the
Gröbner basis technology for organizing spaces by ideals described in
Section \ref{GroebnerBases}.
When constructing the basis of
\(\mathcal{Q}_{d, \vec{\phi}}(\mathfrak{J}, \mathcal{D}_k)\), we order the
basis elements by the number of terms in their expressions to prioritize
simpler basis elements. We refer to this as the
naive approach to constructing a basis of \(\mathcal{Q}_{d,
     \vec{\phi}}(\mathfrak{J}, \mathcal{D}_k)\). 
To organize \(\mathfrak{M}_{d, \vec{\phi}}(\mathfrak{J})\) we can recursively
applying Eq.~\eqref{eq:subspaceDecomposition} to its left summand with
different choices of \(\mathcal{D}_k\). In practice, we order the choice of
the \(\mathcal{D}_k\) heuristically, such that the \(\mathcal{Q}_{d,
     \vec{\phi}}(\mathfrak{J}, \mathcal{D}_k)\) are kept of low dimension at each
step.

\paragraph{Generating Simple Basis Elements.}
\label{sec:org6957e9c}

While the space of functions \(\mathfrak{M}_{d, \vec{\phi}}(\mathfrak{J}) \cap
   \langle \mathcal{D}_k \rangle_{\mathcal{R}_n}\) is simpler than \(\mathfrak{M}_{d,
   \vec{\phi}}(\mathfrak{J})\), the naive approach for choosing a basis of
\(\mathcal{Q}_{d, \vec{\phi}}(\mathfrak{J}, \mathcal{D}_k)\) can still result
in complicated basis elements. 
To avoid this problem, we introduce a procedure to generate simple 
elements of \(\mathcal{Q}_{d, \vec{\phi}}(\mathfrak{J}, \mathcal{D}_k)\).
Specifically, we choose to construct monomials of the denominator factors
that are linearly independent modulo \(\langle \mathcal{D}_k
   \rangle_{\mathcal{R}_n}\).
It is clear that such a monomial of denominators cannot be proportional to
\(\mathcal{D}_k\), so we construct
\begin{equation}
  \underline{\mathcal{D}}^{\underline{\beta}} = \prod_{j = 1}^{n_i} \mathcal{D}_j^{\beta_j} \quad \text{such that} \quad \underline{\mathcal{D}}^{\underline{\beta}} \in \mathfrak{M}_{d, \vec{\phi}}(\mathfrak{J})  \quad \text{with} \quad \beta_k = 0 \, ,
\label{eq:DenominatorMonomialDefinition}
\end{equation}
where the product over \(j\) runs over the full list of \(n_i\)
denominator factors, the \(\beta_j \in \mathbb{Z}_{\ge 0}\). It is not clear \emph{a priori} if, considered modulo \(\langle \mathcal{D}_k \rangle_{\mathcal{R}_n}\), this set of monomials
of the denominator factors spans \(\mathcal{Q}_{d, \vec{\phi}}(\mathfrak{J},
   \mathcal{D}_k)\), and indeed we find this not to always be the case.
Nevertheless, as we have generated a basis of \(\mathcal{Q}_{d,
   \vec{\phi}}(\mathfrak{J}, \mathcal{D}_k)\) from the naive approach, we can
always supplement the set of independent denominator-factor monomials with elements of
the naive basis to obtain a basis.
In practice, we find that this is rarely necessary in our applications.
In the following, to generate the \(\underline{\beta}\) described in
Eq.~\eqref{eq:DenominatorMonomialDefinition}, we take a two-step
procedure. We first generate denominator-factor monomials and then
find a subset that is linearly independent modulo \(\langle \mathcal{D}_k
   \rangle_{\mathcal{R}_n}\).

To begin, we discuss our approach to generating an overcomplete set of
denominator-factor monomials without imposing independence modulo \(\langle
   \mathcal{D}_k \rangle_{\mathcal{R}_n}\). Let us consider exponent vectors
\(\underline{\beta} \in \mathbb{Z}^{n_i}_{\ge 0}\) that satisfy the equations
\begingroup
\allowdisplaybreaks
\begin{gather} 
  \label{eq:MonomialScalingRequirement}
    \sum_{j=1}^{n_i} \beta_j \kappa(\mathcal{D}_j, U) \ge \kappa(\mathcal{N}, U) \quad \text{for each} \quad U \in \mathcal{V}^{(2)} \, ,\\
  \label{eq:DenominatorMonomialLittleGroupRequirement}
  \sum_{j = 1}^{n_i} \beta_j \{ \mathcal{D}_j \}_i = \phi_i \, , \\
  \label{eq:DenominatorMonomialDimensionRequirement}
  \sum_{j=1}^{n_i} \beta_j [\mathcal{D}_j] = d \, .
\end{gather}
\endgroup
  Here, Eq.~\eqref{eq:MonomialScalingRequirement} guarantees
  that the monomial \(\underline{\mathcal{D}}^{\underline{\beta}}\) is an
  element of the ideal \(\mathfrak{J}\), while
  Eqs.~\eqref{eq:DenominatorMonomialLittleGroupRequirement}
  and~\eqref{eq:DenominatorMonomialDimensionRequirement} require that
  \(\underline{\mathcal{D}}^{\underline{\beta}}\) have little-group weights
  \(\vec{\phi}\) and mass dimension \(d\) respectively.
  Let us denote the set of solutions \(\underline{\beta}\) to
  these equations as \(B_{d, \vec{\phi}}(\mathfrak{J}, \mathcal{D}_k)\).
  In principle, the set \(B_{d, \vec{\phi}}(\mathfrak{J}, \mathcal{D}_k)\) can be enumerated
  by a computer algebra system. This is analogous to the enumeration of
  independent spinor bracket exponents, \(X_{d, \vec{\phi}}\), in Section
  \ref{sec:IndependentFunctions}.

In practice, we find that direct enumeration of the elements of \(B_{d,
  \vec{\phi}}(\mathfrak{J}, \mathcal{D}_k)\) can be computationally prohibitive.
To address this, we instead consider constructing \(B_{d', \vec{\phi}}\) for
\(d' < d\) and multiplying these elements by appropriate functions to arrive at
a set of denominator monomials of mass dimension \(d\). Note that
Eq.~\eqref{eq:MonomialScalingRequirement} and
Eq.~\eqref{eq:DenominatorMonomialLittleGroupRequirement} are already
satisfied by any element of \(B_{d', \vec{\phi}}(\mathfrak{J}, \mathcal{D}_k)\). 
Therefore, we can generate a valid monomial by multiplying by
any little-group-invariant monomial of spinor brackets of mass
dimension \(d-d'\). Specifically, consider the set of monomials
\begin{equation}
    \Gamma_{d,d',\vec{\phi}}(\mathfrak{J}, \mathcal{D}_k) = \left\{  \, \, m \, {\underline{\mathcal{D}}^{\underline{\beta}}}  \; :  \; m \in M_{d-d', \vec{0}}\, ,  \;\, \underline{\beta} \in B_{d', \vec{\phi}}(\mathfrak{J}, \mathcal{D}_k) \,\, \right\},
\end{equation}
where we recall from Eq.~\eqref{eq:PhysicalBracketSpaceSpan} that \(M_{d-d', \vec{0}}\) is a
monomial basis of \(\mathcal{M}_{d-d', \vec{0}}\). For \(d' < d\), \(\Gamma_{d,d',\vec{\phi}}(\mathfrak{J}, \mathcal{D}_k)\) 
forms a subset of all possible monomials in denominator factors in
Eq.~\eqref{eq:DenominatorMonomialDefinition}. We note that there exists a \(d_{\text{min}} \geq 0\) such that for all \(d'<d_{\text{min}}\) the set \(\Gamma_{d,d',\vec{\phi}}(\mathfrak{J}, \mathcal{D}_k)\) is empty.

When considered modulo \(\langle \mathcal{D}_k \rangle_{\mathcal{R}_n}\), the monomials in \(\Gamma_{d,d',\vec{\phi}}(\mathfrak{J},
\mathcal{D}_k)\) are linearly dependent. We must therefore find a linearly
independent subset. To this end, we again recall the Gröbner basis
technology of Section \ref{GroebnerBases}.
Specifically, we apply Eq.~\eqref{eq:RemainderQuotientSpaceDefinition} and
construct the matrix 
\begin{equation}
\Delta_{d,d', \vec{\phi}}\left(\mathfrak{J}, \mathcal{D}_k\right) = \Delta_{ij}\left[\mathcal{G}\left(\langle \mathcal{D}_k \rangle_{\mathcal{R}_n^{(q)}}\right), \Gamma_{d,d',\vec{\phi}}\left(\mathfrak{J}, \mathcal{D}_k\right)\right],
\label{eq:GammaRemainderConstruction}
\end{equation}
where \(i\) and \(j\) are the row and column indices of \(\Delta_{d,d', \vec{\phi}}\left(\mathfrak{J}, \mathcal{D}_k\right)\) respectively. A
linearly independent subset of \(\Gamma_{d,d',\vec{\phi}}\left(\mathfrak{J}, \mathcal{D}_k\right)\) corresponds to the pivot
columns of \(\Delta_{d,d', \vec{\phi}}\left(\mathfrak{J}, \mathcal{D}_k\right)\).
Note that 
\begin{equation}
\mathrm{rank}\left[\Delta_{d,d', \vec{\phi}}\left(\mathfrak{J}, \mathcal{D}_k\right)\right] \le \dim\left[ \mathcal{Q}_{d, \vec{\phi}}(\mathfrak{J}, \mathcal{D}_k) \right]
\end{equation}
and the inequality is saturated if \(\Gamma_{d,d',\vec{\phi}}(\mathfrak{J},
\mathcal{D}_k)\) spans \(\mathcal{Q}_{d, \vec{\phi}}(\mathfrak{J},
\mathcal{D}_k)\).
In practice, we often find that this occurs even for \(d' < d\). 
As \(\Gamma_{d,d',\vec{\phi}}(\mathfrak{J}, \mathcal{D}_k)\)
contains fewer
elements than \(B_{d, \vec{\phi}}(\mathfrak{J}, \mathcal{D}_k)\) it is therefore
often more efficient to search for such a \(d' < d\). In practice, this \(d'\) can be found
by searching from \(d' = 0\) and increasing \(d'\) in unit steps
until either \(\Gamma_{d,d',\vec{\phi}}(\mathfrak{J}, \mathcal{D}_k)\) spans
\(\mathcal{Q}_{d, \vec{\phi}}(\mathfrak{J}, \mathcal{D}_k)\) or stopping at \(d' =
d\).

Finally, we note an important feature when determining a subset of \(\Gamma_{d,d',\vec{\phi}}(\mathfrak{J}, \mathcal{D}_k)\) that is linearly independent
modulo \(\langle \mathcal{D}_k \rangle_{\mathcal{R}_n}\): we can
prioritize elements when choosing a basis by ordering
\(\Gamma_{d,d',\vec{\phi}}(\mathfrak{J}, \mathcal{D}_k)\) when constructing
\(\Delta_{d,d',\vec{\phi}}(\mathfrak{J}, \mathcal{D}_k)\) in
Eq.~\eqref{eq:GammaRemainderConstruction}. In practice, we choose an
ordering criteria that is inspired by partial fractions.
Specifically, we choose to order \(\Gamma_{d,d',\vec{\phi}}(\mathfrak{J},
\mathcal{D}_k)\) by the mass dimension of the numerator, after cancellation
against the denominator, which is known from the co-dimension one study.
That is, given a numerator exponent \(\underline{\beta}\), and the denominator
exponent \(\underline{\alpha}\), the rational function associated to
\(\underline{\beta}\) takes the form \(\underline{\mathcal{D}}^{\underline{\beta} -
\underline{\alpha}}\). The mass dimension of the numerator of
\(\underline{\mathcal{D}}^{\underline{\beta} - \underline{\alpha}}\) can be calculated through
\begin{equation}
   [\mathrm{Num}(\underline{\mathcal{D}}^{\underline{\beta} - \underline{\alpha}})] = \sum_{j \,:\, \beta_j > \alpha_j} [\mathcal{D}_j] . (\beta_j - \alpha_j) \, .
\label{eq:NumeratorMassDimension}
\end{equation}
By ordering the elements of \(\Gamma_{d,d',\vec{\phi}}(\mathfrak{J}, \mathcal{D}_k)\) with respect to the 
criteria of Eq.~\eqref{eq:NumeratorMassDimension} when constructing
\(\Delta_{d,d',\vec{\phi}}(\mathfrak{J}, \mathcal{D}_k)\), the pivot
columns will be such that chosen basis of denominator factor monomials 
will cancel against the denominator as much as possible.

\section{\boldmath Application to Two-Loop \(0 \rightarrow q\bar q\gamma\gamma\gamma\) Finite-Remainder Coefficients}
\label{sec:org22bda87}
\label{sec:application_to_photons}

As a proof-of-concept application of our approach, we reconsider the collection of
pentagon-function remainder coefficients for the leading-color process \(0\rightarrow q\bar
q\gamma\gamma\gamma\) at two loops, obtained in Ref.~\cite{Abreu:2020cwb}. 
In this way, we will demonstrate that our approach requires fewer
evaluations than the original functional reconstruction technique. We follow the
notation of Ref.~\cite{Abreu:2020cwb} and consider the remainders 
\begin{equation}\label{eq:2l_3y_remainders}
R^{(2,\,j)}_{h} = \sum_{i \in B} r_i h_i \, ,
\end{equation}
with \(j \in \{0, N_f\}\), \(N_f\) being the number of quarks treated as massless, \(h\) representing the helicity configuration and \(h_i\)
denoting elements of the basis of pentagon functions \(B\) of Ref.~\cite{Chicherin:2020oor}. 
The reconstruction approach of Ref.~\cite{Abreu:2020cwb} works with parity
even functions. For this reason, the \(r_i\) were decomposed as
\begin{equation}
    r_i = r_i^+ + \frac{\mathrm{tr}_5}{s_{12}^2} \, r_i^- \, , 
\end{equation}
and the functional reconstruction approach was applied to the parity
even functions \(r_i^\pm\). The result was then presented in terms of a basis 
\(\tilde r_i^{\scriptscriptstyle (\pm)}\) of the combined space spanned by the \(r^+_i\) and
\(r^-_i\). Furthermore, the \(r_i\) in Ref.~\cite{Abreu:2020cwb} were
normalized by an amplitude-dependent helicity weight \(\Phi_h\) in order
to make them little-group invariant.  Our method is able to handle
functions that are not little-group invariant and therefore benefit
from simplifications arising from manifesting this behavior. Therefore, we
 apply our approach to the functions
\begin{equation}
    \overline{r}_i = \Phi_h r_i \, .
\end{equation}
As is by now standard practice, we exploit the fact that the \(\overline{r}_i\)
are linearly dependent and 
thus can be written as
\begin{equation}\label{eq:independent_spinor_basis}
\overline{r}_i = \sum_{j}\tilde{r}_j  M_{ji} \, ,
\end{equation}
where the \(\tilde{r}_j\) are a subset of the \(\overline{r}_i\) such that
they form a basis of the space spanned by the \(\overline{r}_i\) and \(M_{ji}\) is a
matrix of rational numbers. We choose the basis elements \(\tilde{r}_i\) such that
their numerator mass dimension in common denominator form is minimized. We then
determine the matrix \(M_{ji}\) from numerical evaluations of the \(\overline{r}_j\)
and standard linear algebra techniques.

As is natural to expect, we observe that the number of linearly
independent coefficients drops by approximately a factor of two when
considering the mixed-parity \(r_i\) as opposed to the parity-even
\(r_i^{\pm}\). Furthermore, we observe that the mass dimension of the
associated numerators is improved. We
summarize these observations in Table \ref{tab:org332a79d}.
\begin{table}[tbp]
\centering
\begin{tabular}{c c c c c}
\hline
Remainder & \(\text{dim}(\text{span}(\tilde r_i^{\scriptscriptstyle (\pm)}))\) & \(\text{max}([\mathrm{Num}(\tilde r_i^{\scriptscriptstyle (\pm)})])\) & \(\text{dim}(\text{span}(\tilde{r}_i))\) & \(\text{max}([\mathrm{Num}(\tilde{r}_i)])\)\\
\hline
\(R^{(2,0)}_{- + +}\) & 171 & 50 & 87 & 35\\
\hline
\(R^{(2,N_f)}_{- + +}\) & 57 & 24 & 29 & 15\\
\hline
\(R^{(2,0)}_{+ + +}\) & 62 & 32 & 31 & 20\\
\hline
\(R^{(2,N_f)}_{+ + +}\) & 12 & 18 & 6 & 8\\
\hline
\end{tabular}
\caption{\label{tab:org332a79d}Improvements in the number of basis functions and in the mass dimension of the numerators with known factors pulled out when moving from a basis \(\tilde r_i^{\scriptscriptstyle (\pm)}\) with definite parity to a basis \(\tilde r_i\) with mixed parity.}

\end{table}
Finally, we note that the rational numbers appearing in \(M_{ji}\) are
smaller than those appearing in the analogous matrix in the work of
Ref.~\cite{Abreu:2020cwb}, with the largest denominators being \(\sim 10^7\)
and \(\sim 10^9\) respectively.

\subsection{Five-Point Phase-Space Geometry}
\label{sec:orgcdd6994}
\label{sec:FivePointGeometry}

In this section, we report on the set of irreducible singular varieties up to
codimension two that are relevant for the \(0 \rightarrow q \overline{q} \gamma \gamma
\gamma\) amplitudes at two loops. That is, we describe \(\mathcal{V}^{(1)}\) and
\(\mathcal{V}^{(2)}\) of Eq.~\eqref{eq:CodimensionMVarieties}.
This set is expected to be sufficient for any five-point
massless scattering process at two loops, if the basis of
transcendental functions from Ref.~\cite{Chicherin:2020oor} is
employed. The required primary decompositions were computed making use of
\texttt{Singular}.  While \texttt{Singular}
implements a number of general algorithms for primary decomposition, we found it
necessary to supplement such algorithms to perform and prove the decompositions
in the harder cases. We describe the details of our techniques in Appendix
\ref{sec:PrimaryDecompositionTechniques}.

At five point the set of denominator factors that we use to construct
\(\mathfrak{D}^{(1)}\) and \(\mathfrak{D}^{(2)}\) are of the form
\begin{equation}
 \langle ij \rangle \, , \; [ij] \, , \; \langle i|j+k|i] \, .
\end{equation}
In total there are 35 such invariants: \(\binom{5}{2}=10\) angle brackets, an
 equal number of square brackets, and \(5\,\binom{4}{2}\,/\,2=15\)
 three-particle spinor chains. This is the same set of invariants
 employed for planar five-parton scattering in
 Ref.~\cite{DeLaurentis:2020qle} with 5 additional three-particle
 spinor chains obtained from non-cyclic permutations.

Let us now discuss the irreducible varieties in this space, up to the symmetries.
The symmetries of five-point massless phase space are given by all
\(5!=120\) possible permutations of the external legs, as well as parity.  
At codimension one, the set \(\mathfrak{D}^{(1)}\) is generated by two ideals not
related by symmetry, namely
 \begin{align}
\begin{split}
   J_1 & = \bigl\langle \, \langle 1  2 \rangle \, \bigr\rangle_{R_5} \, , \\
   J_2 & = \bigl\langle \langle 1 | 2 + 3 | 1 ] \bigr\rangle_{R_5} \, .
\end{split}
 \end{align}
Using \texttt{Singular} it can be shown that these ideals are prime.
Hence, the associated varieties are irreducible and generate
\(\mathcal{V}^{(1)}\).  
At codimension two, \(\mathfrak{D}^{(2)}\) is generated by the action of the
symmetries on 11 ideals. Five of these ideals are generated
by a pair of two-particle spinor contractions. They, alongside their primary
decompositions, are given by
 \begin{align}
\label{eq:5pt_prim_dec_first}
   \begin{split}
   \bigl\langle \langle12\rangle, \langle13\rangle  \bigr\rangle_{R_5} &= P_1 \cap P_2 \cap P_3 \, , \\
   \bigl\langle \langle12\rangle, \langle34\rangle  \bigr\rangle_{R_5} &= P_3 \cap P_4 \, , \\ 
   \bigl\langle \langle12\rangle, [12]              \bigr\rangle_{R_5} &= P_5 \, , \\ 
   \bigl\langle \langle12\rangle, [13]              \bigr\rangle_{R_5} &= P_6 \, , \\ 
   \bigl\langle \langle12\rangle, [34]              \bigr\rangle_{R_5} &= P_1(12543) \cap \overline{P}_1(34512) \, .
  \end{split}
 \end{align}
This defines \(P_5\) and \(P_6\) while the other \(P_i\) are given by
 \begin{align}
\label{eq:5pt_prim_dec_second}
\begin{split}
   P_1 & =  \bigl\langle \langle12\rangle, \langle13\rangle, \langle23\rangle, [45] \bigr\rangle_{R_5} \, , \\
   P_2 & =  \bigl\langle \lambda_1^\alpha  \bigr\rangle_{R_5} \, , \\
   P_3 & =  \bigl\langle \langle 12 \rangle, \langle 23 \rangle, \langle 34 \rangle, \langle 45 \rangle, \langle 15 \rangle, \langle 13 \rangle, \langle 14 \rangle, \langle 24 \rangle, \langle 25 \rangle, \langle 35 \rangle \bigr\rangle_{R_5} \, , \\
   P_4 & =  \bigl\langle \langle 12\rangle, \langle 34\rangle, \lambda_1^\alpha [15] + \lambda_2^\alpha [25] \bigr\rangle_{R_5} \, .
\end{split}
 \end{align}
Three of the generators of \(\mathfrak{D}^{(2)}\) are ideals that are generated by a two-particle
and a three-particle contraction. These ideals, alongside their primary
decompositions, are given by
 \begin{align}
\label{eq:5pt_prim_dec_third}
   \begin{split} 
   \bigl\langle \langle12\rangle, \langle1|2+3|1]   \bigr\rangle_{R_5} &= P_1 \cap P_2 \cap P_3 \cap P_6 \, , \\ 
   \bigl\langle \langle12\rangle, \langle3|1+2|3]   \bigr\rangle_{R_5} &= P_1 \cap \overline{P}_1(34512) \cap P_3 \cap P_4(12453) \, , \\ 
   \bigl\langle \langle12\rangle, \langle3|1+4|3]   \bigr\rangle_{R_5} &= P_3 \cap P_7 \, ,
  \end{split}
 \end{align}
where 
 \begin{align}
\label{eq:5pt_prim_dec_fourth}
   P_7 & =  \bigl\langle \langle12\rangle, \langle3|1+4|3], \lambda_1^\alpha [14] [35] + \lambda_2^\alpha [25] [34] \bigr\rangle_{R_5} \, .
 \end{align}
Finally, the last three generators of \(\mathfrak{D}^{(2)}\) are ideals that are generated
by a pair of three-particle spinor contractions. They, alongside their primary
decompositions, are given by
 \begin{align}
\label{eq:5pt_prim_dec_fith}
   \begin{split}
   \bigl\langle \langle1|2+3|1], \langle1|2+4|1]    \bigr\rangle_{R_5} &= P_2 \cap \overline{P}_2 \cap P_3 \cap \overline{P}_3 \cap P_8 \, , \\ 
   \bigl\langle \langle1|2+3|1], \langle2|1+3|2] \bigr\rangle_{R_5}    &= P_1 \cap \overline{P}_1 \cap P_3 \cap \overline{P}_3 \cap P_9 \, , \\ 
   \bigl\langle \langle1|2+3|1], \langle2|1+4|2] \bigr\rangle_{R_5}   &= P_3 \cap \overline{P}_3 \cap P_{10} \, ,
  \end{split}
 \end{align}
where
 \begin{align}
\label{eq:5pt_prim_dec_last}
\begin{split}
   P_8 & =  \bigl\langle \langle1|2+3|1], \langle1|2+4|1], \langle1|2+3|2+4|1\rangle, \lambda \leftrightarrow \tilde{\lambda}, \\
          & \phantom{= \,\,\bigl\langle} -\langle54\rangle \langle21\rangle \langle31\rangle - \langle53\rangle \langle21\rangle \langle41\rangle + \langle52\rangle \langle31\rangle \langle41\rangle, \lambda \leftrightarrow \tilde{\lambda} \bigr\rangle_{R_5} \, , \\
   P_9 & = \bigl\langle \langle1|2+3|1], \langle2|1+3|2], \tilde{\lambda}^{\dot{\alpha}}_1 \langle 12\rangle \langle 13\rangle - \tilde{\lambda}^{\dot{\alpha}}_2 \langle 12\rangle  \langle 23\rangle - \tilde{\lambda}^{\dot{\alpha}}_3 \langle 23\rangle  \langle 13\rangle , \lambda \leftrightarrow \tilde{\lambda} \bigr\rangle_{R_5} \, , \\
   P_{10} & =  \bigl\langle \langle1|2+3|1], \langle2|1+4|2], \\
          & \phantom{= \,\, \bigl\langle} \lambda_1^\alpha [13][14][25] + \lambda_2^\alpha [12][24][35] + \lambda_3^\alpha [13][24][35] , \lambda \leftrightarrow \tilde{\lambda}\bigr\rangle_{R_5} \, .
\end{split}
 \end{align}

\begin{table}[tbp]
\centering
\begin{tabular}{c c c c c c c c c c}
\hline
\(P_1\) & \(P_2\) & \(P_3\) & \(P_4\) & \(P_5\) & \(P_6\) & \(P_7\) & \(P_8\) & \(P_9\) & \(P_{10}\)\\
\hline
20 & 10 & 2 & 30 & 10 & 60 & 120 & 15 & 30 & 20\\
\hline
\end{tabular}
\caption{\label{tab:org933518c}Counting of the number of distinct irreducible varieties generated by each of the \(P_i\) that, under the action of parity and permutations of the external momenta, generate the set \(\mathcal{V}^{(2)}\).}

\end{table}

\noindent In the above ideal definitions, \(\lambda \leftrightarrow \tilde{\lambda}\)
 means to add another generator obtained by applying the parity
 operation to the previous one. We use the notation \(P_i\) as it can be shown
 that all the \(P_i\) are prime (see Appendix~\ref{IdealPrimeCheck}). This implies that the
 elements of \(\mathfrak{D}^{(2)}\) are all radical.
 We note that while there are 11
 inequivalent ideals up to symmetries generated by two denominator
 factors, there are only 10 inequivalent irreducible
 varieties up to symmetries.
 By permuting the external momenta, and/or applying the parity operation, the
 varieties associated to the \(P_i\) generate \(\mathcal{V}^{(2)}\). In Table
 \ref{tab:org933518c}, we list the counting of the number of varieties
 generated by each of the prime ideals \(P_1\) through to \(P_{10}\).
 In total there are 317 distinct irreducible varieties.

\subsubsection{Symbolic Powers}
\label{sec:org6663224}

In this section, we discuss the practical computation of symbolic powers of the
\(P_i\) in Section~\ref{sec:FivePointGeometry}. The maximal symbolic power of each
ideal \(P_i\) that we compute is controlled by the structure of the \(\tilde{r}_i\),
which
is determined numerically. We list the largest power of each \(P_i\) that we need
to compute in Table \ref{tab:org0e0935f}. We remark that for \(P_5\), \(P_8\) and
\(P_{10}\), no
computation is required as \(J^{\langle 1 \rangle} = J\) for any ideal \(J\).
We also note that the constraints coming from the symbolic power of \(P_3\)
can be seen to be trivial. Specifically, one can rearrange the constraints of little
group and mass dimension on monomial exponents, \eqref{eq:littleGroupConstraint}
and \eqref{eq:massDimensionConstraint}, to give
\begin{equation}
   \sum_{j=1}^n \sum_{i=1}^j \alpha_{ij} = \frac{1}{2}\left(d + \sum_k \phi_k\right) \quad\; \text{and} \quad\;\,
   \sum_{j=1}^n \sum_{i=1}^j \beta_{ij} = \frac{1}{2}\left(d - \sum_k \phi_k\right).
\end{equation}
That is, the total degree of both the angle bracket and square bracket variables
are fixed independently by little group and mass dimension. This implies that
\begin{equation}
    \mathcal{M}_{d, \vec{\phi}} = \mathcal{M}_{d, \vec{\phi}} \cap P_3^{\langle \frac{1}{2}\left(d + \sum_k \phi_k\right) \rangle} = \mathcal{M}_{d, \vec{\phi}} \cap \overline{P}_3^{\langle \frac{1}{2}\left(d - \sum_k \phi_k\right) \rangle} \, .
\label{eq:P3Triviality}
\end{equation}
This allows us to avoid constructing explicit symbolic powers for \(P_3\).
\begin{table}[tbp]
\centering
\begin{tabular}{c c c c c c c c c c c}
\hline
\(P_i\) & 1 & 2 & 3 & 4 & 5 & 6 & 7 & 8 & 9 & 10\\
\hline
\(k_i\) & 10 & 9 & 18 & 3 & 1 & 4 & 3 & 1 & 3 & 1\\
\hline
\end{tabular}
\caption{\label{tab:org0e0935f}List of highest \(k_i\) such that it is required to compute the symbol power \(P_i^{\langle k_i \rangle}\) in order to construct the Ansatz for the rational prefactors in \(R^{(2,0)}_{-++}\). The \(P_i\) are given in Eqs.~\eqref{eq:5pt_prim_dec_first} to \eqref{eq:5pt_prim_dec_last}.}

\end{table}
To construct the remaining symbolic powers, we make use of the Cohen--Macaulay
property of \(R_n\) in the following two ways.
\begin{enumerate}
\item \textbf{Relation to Maximal-Codimension Ideals:} The ideals \(P_2\) and \(P_6\)
are maximal codimension. Therefore, by Eq.~\eqref{eq:SimpleSymbolicPowerLemma}, the symbolic powers are equivalent to the ideal
powers. Specifically, we use
\begin{equation}
   P_i^{\langle k \rangle} = P_i^k \qquad \text{for} \,\, i \in \{2, 6\} \, .
\end{equation}
Next, as ideal intersection is associative, by Eq.~\eqref{eq:P3Triviality} we have that for any \(P_i\)
\begin{equation}
   \mathfrak{M}_{d, \vec{\phi}} \left(P_i^{\langle k \rangle}\right) = \mathfrak{M}_{d, \vec{\phi}} \left(P_i^{\langle k \rangle} \cap P_3^{\langle k \rangle} \right) \quad \text{for} \quad k \le \frac{1}{2}\left(d + \sum_k \phi_k\right) \, ,
    \label{eq:P3Bound}
\end{equation}
alongside the analogous equation for \(\overline{P}_3\). 
In practice, if we wish to compute \(P_i^{\langle k \rangle}\), then \(k\) is such that the
inequality in Eq.~\eqref{eq:P3Bound} holds. Therefore, for the purposes of our algorithm,
it is sufficient to compute \(P_i^{\langle k \rangle} \cap P_3^{\langle k \rangle}\). 
For two ideals, \(P_4\) and \(P_7\), we see from Eqs.~\eqref{eq:5pt_prim_dec_first} and \eqref{eq:5pt_prim_dec_third} that
this intersection is again of maximal codimension. By definition, the symbolic power
commutes with the intersection (see Eq.~\eqref{eq:RadicalSymbolicPowerDefinition}).
This allows us to apply Eq.~\eqref{eq:SimpleSymbolicPowerLemma} and compute the symbolic power through the ideal power.
Specifically, we use
\begin{align}
    P_4^{\langle k \rangle} \cap P_3^{\langle k \rangle} &= {\big\langle \langle 12 \rangle, \langle 34 \rangle \big\rangle_{R_5}}^{k} \, , \\
    P_7^{\langle k \rangle} \cap P_3^{\langle k \rangle} &= {\big\langle \langle 12 \rangle, \langle 3|1+4|3]   \big\rangle_{R_5}}^{k} \, .
\end{align}

\item \textbf{Saturation of Maximal Codimension Ideals}:
In order to compute symbolic powers of the remaining two ideals, \(P_1\) and
\(P_9\), we make use of ideal saturation.
We refer the reader to Appendix \ref{app:PrimDecSaturate} for
theoretical details on saturation. 
In this context, we exploit that ideal saturation removes primary components.
Specifically, we compute the symbolic power of a (radical) ideal whose set of
associated primes contains \(P_1\) or \(P_7\) and remove extraneous components of
the symbolic power by saturation. If we can find a relevant
maximal-codimension radical ideal, then its symbolic power is again the ideal
power. This strategy is natural as the \(P_i\) are constructed from the primary
decompositions of maximal-codimension ideals. In particular, we use that
\begin{align}
    P_1^{\langle k \rangle} \cap P_3^{\langle k \rangle}  &= {\big\langle \langle 12 \rangle, \langle 13 \rangle \big\rangle_{R_5}}^{k} : \big\langle \lambda_{1, 0} \big\rangle_{R_5}^\infty, \\
    P_9^{\langle k \rangle}  &= {\big\langle \langle 1|2+3|1], \langle 2|1+3|2] \big\rangle_{R_5}}^{k} : \big\langle \langle 45 \rangle [45] \big\rangle_{R_5}^\infty
\end{align}
and these saturations can be computed with Gröbner basis methods.
\end{enumerate}

Finally, with generating sets of each of these ideals in hand, we contract each
generator in all possible ways with spinors to reach a set of 
polynomials in spinor brackets and thereby find a generating set of the associated ideal in
\(\mathcal{R}_5^{(q)}\).

\subsection{Implementation and Results}
\label{sec:org412971a}

We now retrace the steps of the Ansatz construction algorithm
presented in Section \ref{sec:AnsatzConstructionAlgorithm}, and provide
details regarding their implementation. To perform the
algebro-geometric operations we make extensive use of the computer
algebra system \texttt{Singular} \cite{DGPS} through its
\texttt{Mathematica} interface \cite{math_singular} and through the
\texttt{Python} interface \texttt{syngular} \cite{syngular}. We note a generally
useful facility in \texttt{Singular}: the
\href{https://www.singular.uni-kl.de/Manual/4-3-0/sing\_166.htm\#SEC206}{\texttt{qring}} declaration. This allows one to work directly in
the quotient rings \(R_n\) and \(\mathcal{R}_n\).

Let us start from the analytic study of codimension-one and
codimension-two varieties. For one primary decomposition, we 
used the algorithm of Gianni, Trager and Zacharias~\cite{gianni1988grobner} as implemented in
\texttt{Singular} under the command \href{https://www.singular.uni-kl.de/Manual/4-3-0/sing\_1511.htm\#SEC1592}{\texttt{primdecGTZ}}. For the other
decompositions, we supplemented this algorithm with the techniques described in
Appendix \ref{sec:PrimaryDecompositionTechniques}.
To remove sub-varieties, we made use of the command
\href{https://www.singular.uni-kl.de/Manual/4-3-0/sing\_1287.htm\#SEC1368}{\texttt{sat}}, as described in Appendix
\ref{sec:orgb4c6838}. To check primality, we made use of the
test presented in Appendix \ref{IdealPrimeCheck}.

For the numerical warm-up, we implemented in \texttt{Python} both the
algorithm for generating finite-field points on irreducible varieties,
described in Section \ref{sec:FiniteFieldVarietyPoints}, and the lifting
procedure to obtain \(p\verytiny\text{-adic}\) solutions close to singular
varieties, described in Section \ref{sec:padicClose}. The required maximally
independent sets were computed with the \texttt{Singular} command
\href{https://www.singular.uni-kl.de/Manual/4-3-0/sing\_269.htm\#SEC309}{\texttt{indepSet}}. In order to solve the univariate polynomial
equations of Eq.~\eqref{eq:goDim0Ideal} over a finite field, we used the
command \href{https://docs.sympy.org/latest/modules/polys/wester.html\#univariate-factoring-over-various-domains}{\texttt{factor}} from the package \texttt{sympy}
\cite{10.7717/peerj-cs.103}. In practice, we observed that
it is always possible to choose a maximally independent set such that the system of
equations for the dependent variables is linear, which guarantees the
existence of a solution in \(\mathbb{F}_p\).

The algorithm for building the Ansatz from the gathered numerical
data, described in Section \ref{sec:SpaceOfVanishingFunctions}, was
implemented in \(\texttt{Mathematica}\). When performing polynomial
reductions, e.g.~when applying Eq.~\eqref{eq:RemaindersInMonomialBasis} to
intersect vector spaces with ideals, we found it important to tune the choice of
variable ordering to minimize the size of the Gröbner bases. This
increases the speed of polynomial reduction. In order to perform
the linear algebra when constructing the Ansatz we employed private codes for
sparse linear algebra over a
finite field\footnote{We thank Mao Zeng for the use of an in-house
implementation of sparse linear algebra techniques over a finite
field.}. The \(p\verytiny\text{-adic}\) and finite-field evaluations of the
rational prefactors that we used in this work were performed using
in-house implementations of the respective fields and the analytic
results of Ref.~\cite{Abreu:2020cwb}.

\begin{table}[tbp]
\centering
\begin{tabular}{c  c c  c  c}
\hline
Remainder & \(R^{(2, 0)}_{-++}\) & \(R^{(2, N_f)}_{-++}\) & \(R^{(2, 0)}_{+++}\) & \(R^{(2, N_f)}_{+++}\)\\
\hline
max \(\dim[\mathcal{M}_{\tilde{d}, \vec{0}}]\) & 41301 & 2821 & 7905 & 1045\\
\hline
max \(\dim[\mathfrak{M}_{d, \vec{\phi}}(\mathfrak{J})]\) & 566 & 20 & 18 & 6\\
\hline
\end{tabular}
\caption{\label{tab:org060139b}Summary of effect of algorithm on Ansatz dimension. The row labeled ``max \(\dim[\mathcal{M}_{\tilde{d}, \vec{0}}]\)'' corresponds to the largest Ansatz dimension for each basis function when using the little-group invariant techniques of Ref. \cite{Abreu:2020cwb}. The \(\tilde{d}\) correspond to third column of Table \ref{tab:org332a79d}. The row labeled ``max \(\dim[\mathfrak{M}_{d, \vec{\phi}}(\mathfrak{J})]\)'' gives the largest Ansatz dimension for each amplitude when the codimension two scaling constraints are taken into consideration.}

\end{table}

The results of our Ansatz construction procedure are summarized in Table
\ref{tab:org060139b}. For all helicity amplitudes, we observe a large decrease
in the size of the Ansatz when applying our procedure.
As a result of using this Ansatz, we therefore find a large reduction in the
size of the expressions for the rational functions in the helicity amplitudes.
We provide our results in the form of ancillary files of two types:
\begin{enumerate}[label=(\alph*)]
    \item Machine-readable expressions for the primary decompositions of
      Eqs.~\eqref{eq:5pt_prim_dec_first} to \eqref{eq:5pt_prim_dec_last} and, in particular, for the generating
      sets for the associated prime ideals. These can be found in the files
      \texttt{generatingIdeals.m} and \texttt{generatingIdeals.py},
      in the languages \texttt{Mathematica} and \texttt{Python 3.8} respectively.
      The \texttt{Python} file also computationally checks the equalities in 
      Eqs.~\eqref{eq:5pt_prim_dec_first}, \eqref{eq:5pt_prim_dec_third} and \eqref{eq:5pt_prim_dec_fith}.
      Furthermore, it employs the method described in Appendix \ref{IdealPrimeCheck}
      to check the primality of all ideals $P_i$ in Section~\ref{sec:FivePointGeometry}.

\item Expressions for the rational functions in the helicity amplitudes.
      The appropriate helicity
configuration and $N_f$ power are encoded in the first part of the file
names, each taking the form
\begin{equation}\notag
\text{\texttt{2l\_}\{\texttt{helicity configuration}\}\{\texttt{Nf power}\}}
\end{equation}
where \{\texttt{helicity configuration}\} refers to the subscript $h$
from Eq.~\eqref{eq:2l_3y_remainders} and can be either
\texttt{pmmpp} or \texttt{pmppp}, i.e.~one of the two
independent helicity configurations; and where \{\texttt{Nf power}\}
refers to the $j$ superscript in Eq.~\eqref{eq:2l_3y_remainders},
it is blank for $j=0$ or \texttt{\_nf} for $j=N_f$. These files are arranged into two classes:
\begin{enumerate}[label=(b.\arabic*)]
\item Files ending in \texttt{\_coo.json} contain the sparse matrices $M_{ji}$ from
Eq.~\eqref{eq:independent_spinor_basis}. The term \texttt{coo}
refers to the notation used to store the matrix, which is the
coordinate list format, i.e.~a map from tuples of indices $(j,
i)$ to the values $M_{ji}$ for all non zero entries of the matrix. 
\item Files
ending in \texttt{\_spinors.m} contain the bases $\tilde{r}_j$ from
Eq.~\eqref{eq:independent_spinor_basis}. The notation used is
\texttt{SP} for spinor angle brackets, and \texttt{SPT} for spinor
square brackets. Using a package such as \texttt{S@M}
\cite{Maitre:2007jq} it is straightforward to match 
at a random phase-space point the product $\tilde{r}_j M_{ji}$
to the functions $\overline{r}_i$ as given in
Ref.~\cite{Abreu:2020cwb}.
\end{enumerate}
\end{enumerate}

\section{Summary and Outlook}
\label{sec:org5f6ab81}
\label{sec:summary}

In this work, we have developed an algorithm to construct compact Ansätze for
rational prefactors of master integrals and transcendental functions in gauge theories. 
To this end, we made use of tools of algebraic geometry to further our understanding of
rational functions of Weyl spinors associated to 
complexified momentum space.
We began by interpreting the spinor-helicity formalism in the language of
algebraic geometry, discussing that physical polynomials of Weyl spinors belong
to a polynomial quotient ring associated to the variety induced by momentum
conservation. We then showed that elements of this
set with appropriate Lorentz transformation properties live in a further polynomial quotient ring, which we used to systematically
account for the relations arising from momentum-conservation and Schouten
identities.
Next, we discussed the singularities of rational prefactors
in terms of irreducible varieties. In particular, we began a
systematic study of the singular structure of these rational
functions via primary decompositions. We understood the singular structure of a
rational function in terms of a set of irreducible varieties and the order of
vanishing/divergence of the function on those varieties. 
Importantly, this
allows us to study the singular structure of a rational function not only on
codimension-one varieties but also on \emph{higher} codimension varieties.
We understood the analytic consequences of this higher codimension data by
connecting it to an important set of polynomials with well-defined vanishing
behavior on these surfaces: the symbolic power of an ideal. 
This allowed us to construct refined Ansätze for 
rational prefactors that match the singular behavior on higher codimension varieties.
In order to practically apply this strategy, we introduced \(p\verytiny\text{-adic}\)
numbers, which make it possible to balance the stability benefits of finite fields with
the a non-trivial measure of size in the number field. With
this numerical tool, we constructed an algorithm to classify the set of symbolic powers
to which the numerators of rational prefactors belong, thereby allowing us to
construct the refined Ansätze. These Ansätze have a strongly reduced number of
free parameters that need to be fixed by numerical evaluations over finite
fields. 
As an example application, we reconsidered the two-loop finite
remainders for the production of three photons at hadron colliders.
We studied them on both codimension-one and codimension-two singular varieties, and built Ansätze
that we organized in a way reminiscent of a partial-fraction decomposition. 
We then constrained these Ansätze using remarkably few evaluations to reconstruct the analytic form of the rational prefactors.

A number of future directions deserve to be explored. 
Firstly, it is clear that the behavior of rational prefactors in gauge theory
amplitudes on higher codimension singular varieties is highly non-trivial. It
would be interesting to understand the physical origin of this behavior.
Secondly, it would be interesting to
investigate other techniques for constructing primary decompositions, such as
those discussed in Ref.~\cite{bates2014comparison}.
Thirdly, an important question that we leave to future work is to understand how
modern numerical algorithms for scattering amplitude calculation, such as
numerical unitarity, behave when then number field is taken to be
the \(p\verytiny\text{-adic}\) numbers. Finally, it would also be interesting to consider evaluating 
rational prefactors near higher codimension varieties, simplifying the linear
system which needs to be solved to fit the Ansatz in an approach similar to that
of Ref.~\cite{DeLaurentis:2019phz}.
Altogether, we foresee applications to as-yet unknown
scattering amplitudes.
\section*{Acknowledgements}
\label{sec:orgfc1d92e}
We thank Samuel Abreu, Gauthier Durieux, Harald Ita, Fernando Febres Cordero, David Kosower,
Daniel Maître and Vasily Sotnikov for useful discussions and/or collaboration on
related topics. We thank Harald Ita, Daniel Maître and Vasily Sotnikov for comments on the draft.
This project has received funding from the European's Union Horizon 2020
research and innovation programme \textit{LoopAnsatz} (grant agreement
number 896690) as well as the French Agence Nationale pour la Recherche, under grant
ANR–17–CE31–0001–01.

\appendix
\section{Glossary of Algebraic-Geometry Terms}
\label{sec:orgcfeeb03}
\label{app:AlgGeoGlossary} 

In this appendix, we recall some of the algebraic definitions of a number of
properties that ideals may have, alongside some useful operations on ideals. A
pedagogical introduction to this material can be found in
Ref.~\cite{cox1994ideals}. To avoid repetition, in the following let \(A\) be
a commutative Noetherian ring, \(J\) and \(K\) be ideals of \(A\), and \(a\) and \(b\) be
elements of \(A\).

\paragraph{Properties}
\label{sec:org81feac1}

We start by reviewing a number of algebraic properties.
First, recall that in Eq.~\eqref{eq:radicalNotation} we introduced the
concept of an ideal being radical by noting that, in an algebraically-closed
field, it is equivalent to the ideal of the variety of the ideal itself. 
The algebraic definition of radicality is 
\begin{equation}\label{eq:RadicalAlgebraicDefinition}
J\text{ is {\bf radical} if } a^k \in J \;
\Rightarrow \;
a \in J \, .
\end{equation}

Second, recall that in Eq.~\eqref{eq:IrreducibleVariety} we defined the
irreducibility property for varieties, and in
Eq.~\eqref{eq:prime_ideal_from_variety_branch} we introduced prime ideals as
those ideals corresponding to irreducible varieties. The algebraic definition is
\begin{equation}\label{eq:PrimeAlgebraicDefinition}
 J\text{ is {\bf prime} if }a b \in J \;
\Rightarrow \;
\text{either }a \in J\text{ or }b \in J \, .
\end{equation}
While radical ideals can be uniquely decomposed as the intersection of
prime ideals, the same is not true for non-radical ideals.

This leads us to the third property we discuss here, as we recall from
Eq.~\eqref{eq:PrimaryDecomposition} that an arbitrary ideal can be
decomposed as an intersection of primary ideals.
 The algebraic definition is
\begin{equation}\label{eq:PrimaryAlgebraicDefinition}
 J\text{ is {\bf primary} if }a b \in J \;
\Rightarrow \;
\text{either }a \in J\text{ or }b^k \in J \, .
\end{equation}
Note that this definition is secretly symmetric because if \(A\) is commutative,
then the roles of \(a\) and \(b\) can be reversed.
We see from considering Eqs.~\eqref{eq:RadicalAlgebraicDefinition} and
\eqref{eq:PrimeAlgebraicDefinition} that a prime ideal is both radical and
primary.
Also, while the radical of
a primary ideal is a prime ideal, not all primaries are necessarily
powers of primes, e.g. \(\left \langle x^2, y \right \rangle\) is
primary with associated prime \(\left \langle x, y \right \rangle\), but
the former is not a power of the latter.

Finally, in Eq.~\eqref{eq:MaximimalIdealDefinition} we introduced a 
special class of prime ideals: maximal ideals. The
algebraic definition of such an ideal is
\begin{equation}\label{eq:maximal_ideal_algebraic_definition}
 J \text{ is {\bf maximal} if there is no proper ideal }K\text{ of }A\text{ such that } J \subset K \, .
\end{equation}
If the field is algebraically closed then the variety associated to a
maximal ideal corresponds to a single point.

\paragraph{Operations}
\label{sec:org4a1a558}

We now review three relevant operations on ideals. 
First, Eq.~\eqref{eq:RadicalAlgebraicDefinition} suggests a
definition of the \textbf{radical} operation \(\sqrt{J}\) namely
\begin{equation}\label{eq:RadicalOperation}
\sqrt{J} = \{ a : a^k \in J \;\, \text{for some} \;\, k \in \mathbb{Z}_{>0} \} \, .
\end{equation}
We can say that \(J\) is radical if \(J=\sqrt{J}\). Starting from this definition,
the equivalence in Eq.~\eqref{eq:radicalNotation} is Hilbert's Strong
Nullstellensatz (see Chapter 4 of \cite{cox1994ideals}).

Second, we discuss the \textbf{ideal quotient} \(J:K\) of two ideals \(J\) and \(K\), defined as
\begin{equation}\label{eq:ideal-quotient}
 J : K = \{ r \in A \; \quad \text{such that} \quad \; rK \subseteq J\} \, .
\end{equation}
If \(J\) is a radical ideal, then there is a geometric
interpretation of the ideal quotient \(J:K\). Specifically, for ideals \(J\) and \(K\)
over an algebraically closed field, where \(J\) is a radical ideal, it can be
shown that~\cite[Section 4.4, Corollary 11]{cox1994ideals}
\begin{equation}
   V(J:K) = \overline{V(J) \backslash V(K)} \, ,
\label{eq:RadicalVarietyQuotient}
\end{equation}
where the overline on the right hand side denotes the Zariski closure. Geometrically, one can
see that the ideal quotient can be useful to simplify primary decompositions
(see Appendix \ref{sec:orgb4c6838}).

Third, as not all ideals \(J\) are radical, it is
useful to introduce also the concept of ideal \textbf{saturation}. It
consists of iterated ideal quotienting until the result is unchanged
\begin{equation}\label{eq:ideal-saturation}
 J : K^\infty = J : K^s \quad \text{such that} \quad J : K^s = J : K^{s+1} \, .
\end{equation}
The smallest such exponent \(s\) is called the saturation index. We find it useful to  define
\(J : a^\infty\) for a single element \(a\) of \(A\) as \(J : \langle a
\rangle^\infty\).
The geometric analogue of saturation is set difference of varieties.
Specifically, for two ideals \(J\) and \(K\) over an algebraically closed field we
have~\cite[Section 4.4, Theorem 10]{cox1994ideals}
\begin{equation}
   V(J : K^\infty) = \overline{V(J) \backslash V(K)} \, ,
\label{eq:SaturationVarietyRemoval}
\end{equation}
where the overline on the right hand side again denotes the Zariski closure. Note
that Eq.~\eqref{eq:SaturationVarietyRemoval} generalizes
Eq.~\eqref{eq:RadicalVarietyQuotient}, in that there is no requirement of radicality of \(J\).

\section{Primary-Decomposition Techniques}
\label{sec:org12cc996}
\label{sec:PrimaryDecompositionTechniques}
In this section, we provide details of the techniques used to perform
the primary decompositions presented in Eqs.~\eqref{eq:5pt_prim_dec_first} to \eqref{eq:5pt_prim_dec_last}. While there exist
general algorithms for primary decomposition (see e.g.~Chapter 8 of Ref.~\cite{becker2012groebner}), due to the high
dimensionality of the ideals we consider, it is often convenient to
exploit more tailored but less algorithmic approaches. 
We
show that to reproduce the results of Section \ref{sec:orgcdd6994} it
is sufficient to apply a primary decomposition algorithm,  to just the ideal
\(\big\langle \langle12\rangle, \langle13\rangle
\big\rangle_{R_5}\). Using the implementation of the algorithm of Gianni, Trager
and Zacharias (GTZ) \cite{gianni1988grobner} under the \texttt{primdecGTZ} command
of \texttt{Singular}, this
operation only takes a few seconds. 
Subsequently, one can use ideal saturation alongside a check of primality to
find the remaining primary components.

We begin in Section \ref{sec:orgb4c6838} with a discussion of
the effect of ideal saturation on primary decompositions. In Section
\ref{ExtensionContraction} we review the concepts of extension and contraction.
This then allows us to construct a test for proving that an ideal is prime in
Section \ref{IdealPrimeCheck}.

\subsection{Primary Decompositions and Saturation}
\label{sec:orgb4c6838}
\label{app:PrimDecSaturate}

Ideal saturation is intimately related to removal of sub-varieties
from a larger variety. It is then interesting to consider how we can
use this geometric intuition to simplify a primary decomposition
calculation when we already have some understanding of the involved
varieties.

Let us first set up the problem in the algebraic context, and then
consider the geometric interpretation. In the following we will work in a
polynomial ring, and suppress the ring label on ideals for simplicity. Consider
an ideal \(J\), its primary
decomposition reads
\begin{equation}
   J = \bigcap_{i=1}^{n_Q(J)} Q_i \, , \;\, \text{with} \;\; \sqrt{Q_i} = P_i \, ,
\end{equation}
where each \(Q_i\) is a primary ideal.
At this stage, we may or may not know explicit generating sets of the
primaries \(Q_i\) and the associated primes \(P_i\). We wish to consider the effect
that saturation by an ideal \(K\) has on \(J\) and on its primary decomposition. To this end, we recall
that ideal saturation
commutes with ideal intersection, i.e.~for any ideal \(K\) we can write
\begin{equation}\label{eq:quotient_and_ideal_intersection}
  J : K^\infty = \bigcap_{i=1}^{n_Q(J)} (Q_i : K^\infty) \, . 
\end{equation}
It is therefore sufficient to understand the effect of ideal saturation on each
primary component.

Consider now the case where \(K\) is generated by a single polynomial \(f\).
It follows from Ref.~\cite[Lemma 4.4]{books/daglib/0091700} that there are
only two possible outcomes of saturation by \(\langle f \rangle\), either
\begin{align}
\begin{split}
 &1) \,f \in \sqrt{Q_i}, \quad \Rightarrow \quad Q_i : \langle f \rangle^\infty = \langle 1 \rangle \, ,\\
 &2) \,f \not \in \sqrt{Q_i}, \quad \Rightarrow \quad Q_i : \langle f \rangle^\infty = Q_i \, .
\end{split}
\end{align}	
Let us consider this geometrically. 
In case 1, the variety \(V(f)\) contains \(V(P_i)\). 
Hence, by Eq.~\eqref{eq:SaturationVarietyRemoval} it 
removes the variety. In case 2, \(V(f)\) does not contain \(V(P_i)\), but it may
still intersect it on some higher codimension variety. The removal of this
higher codimension variety is, however, ``filled back in'' by the Zariski
closure, and we get the same result back.

Let us now consider what happens if we saturate by a more general ideal. First,
recall that saturation by an ideal is the intersection of the saturation by the generators
\cite[Chapter 4.4, Proposition 13]{cox1994ideals}, i.e.
\begin{equation}
    J : \langle f_1, \ldots, f_m \rangle^\infty = \bigcap_{i=1}^m J : \langle f_i \rangle^\infty \, .
\label{eq:SaturationByIntersection}
\end{equation}
Let \(K = \langle f_1, \ldots, f_m \rangle\). The result of \(Q_i : K ^ \infty\) now has one of two possibilities:
\begin{align}
\begin{split}
&1)\text{ Every generator of }K\text{ is in }\sqrt{Q_i} \quad \Rightarrow \quad Q_i : K ^\infty = \langle 1 \rangle \, , \\
&2)\text{ Some generator of }K\text{ is {\em not} in }\sqrt{Q_i}\quad \Rightarrow \quad Q_i : K ^\infty = Q_i \, .
\end{split}
\end{align}
If we consider these results in the context of Eq.~\eqref{eq:quotient_and_ideal_intersection}, 
we see that we can quite generally use
saturation to remove primary components in a controlled manner. 

We now discuss how we derived the results of Eqs.~\eqref{eq:5pt_prim_dec_first} to \eqref{eq:5pt_prim_dec_last}.
We make use of the following heuristic procedure. Consider an ideal \(J\) and a
sequence of primary ideals \(\{ K_1, \ldots, K_n \}\). Define
\begin{equation}
J_i = J_{i-1} : K_i^{\infty} \quad  \text{and} \quad J_0 = J.
\end{equation}
We know from the above discussion that each of the \(J_i\) admits a primary
decomposition with potentially fewer primary components \(J_{i-1}\). 
This occurs because the saturation by \(K_i\) has potentially removed a subset of
the primary components. We consider the final element of the sequence, \(J_n\), as
a candidate for a primary component of \(J\). We (heuristically) check that \(J_n\)
is primary by checking that it is prime using the test in
Appendix~\ref{IdealPrimeCheck}.
If this is the case, we also consider the \(K_i\) for which \(J_{i} \ne J_{i-1}\) as
other good candidates for primary components of \(J\) and construct a tentative
primary decomposition of \(J\) given by
\begin{equation}
    J' = J_n \cap \left(\bigcap_{i: J_i \ne J_{i-1}} K_i\right).
\label{eq:TentativePrimaryDecomposition}
\end{equation}
We then check if Eq.~\eqref{eq:TentativePrimaryDecomposition} is indeed a primary
decomposition of \(J\) by computing \(J'\) and checking if \(J' = J\). 
To derive the results of Section \ref{sec:orgcdd6994}, we apply the procedure to
each co-dimension two ideal of which we wish to obtain a primary decomposition, taking the set of \(K_i\) to be the set of
permutations/parity conjugates of \(\mathrm{assoc}(\big\langle \langle12\rangle,
\langle13\rangle \big\rangle_{R_5})\), which we compute using the command
\texttt{primdecGTZ} in \texttt{Singular}. While this procedure is heuristic, it
was sufficient to derive and prove the results of Eqs.~\eqref{eq:5pt_prim_dec_first} to \eqref{eq:5pt_prim_dec_last}.

\subsection{Extension and Contraction}
\label{sec:org15749c0}
\label{ExtensionContraction}

In order to set up our approach to check if an ideal is prime, we first introduce 
two concepts of fundamental importance, namely extension and contraction. In
this appendix, we mostly follow Ref.~\cite[Chapter 8.7]{becker2012groebner}.

Let us consider the polynomial ring \(\mathbb{F}[\underline X]\)
with \(\underline{X} = \{X_1, \ldots, X_n\}\), an ideal \(J \subseteq
\mathbb{F}[\underline X]\) generated by the polynomials \(\{p_1, \dots, p_n\}\), and let us split \(\underline X\) into two
disjoint sets \(\underline Y\) and \(\underline Z\). The \textbf{extension} of \(J\),
denoted as \(J^e\), is defined as the ideal generated by the infinite
set \(J\) in \(\mathbb{F}(\underline{Y})[\underline{Z}]\), that is in the
polynomial ring in \(\underline Z\) over the field of fractions
\(\mathbb{F}(\underline Y)\). 
A more practical but nevertheless equivalent definition of \(J^e\) can be taken to be
\begin{equation}
 J^e = \big\langle p_1, \dots, p_n \big\rangle_{\mathbb{F}(\underline Y)[\underline Z]} \, .
\end{equation}
The original polynomial ring is contained
in the new one, i.e.~\(\mathbb{F}[\underline X] \subset
\mathbb{F}(\underline Y)[\underline Z]\), since in the latter
polynomials in the variables \(\underline Y\) are also allowed to be
denominators. We will refer to \(F[\underline X]\) as the original ring,
and to \(F(\underline Y)[\underline Z]\) as the extended ring. 

Given an ideal \(J\) of
\(\mathbb{F}(\underline{Y})[\underline Z]\), we define the \textbf{contraction} of \(J\) as the
ideal \(J^{c}\) through
\begin{equation}
J^{c} = J \cap \mathbb{F}[\underline X] \, .
\label{eq:ContractionDefinition}
\end{equation}
As we return to the original ring, contraction could be considered as a type of
inverse operation to extension.
Considering Eq.~\eqref{eq:ContractionDefinition}, we see that
given an ideal \(J\) of the extended ring, its contraction is
the subset of polynomials in \(J\) which do not involve
denominators. We stress that this definition can of course be applied to any ideal of
\(\mathbb{F}(\underline Y)[\underline Z]\), not just those obtained from
extensions of ideals in \(\mathbb{F}[\underline X]\).  
It can be shown that the
contraction operation commutes with intersection, that is for two
ideals \(J_1, \, J_2\) of \(\mathbb{F}(\underline{Y})[\underline Z]\) we
have \cite[Lemma 8.97]{becker2012groebner}
\begin{equation}\label{eq:contraction_and_intersection}
  \left(J_1 \cap J_2\right)^c = J_1^c \cap J_2^c \, .
\end{equation}
Contraction can be computed by means of ideal saturation \cite[Lemma 8.91]{becker2012groebner}.  
Consider an ideal \(K\) of \(\mathbb{F}(\underline Y)[\underline Z]\), and 
a Gröbner basis \(\mathcal{G}(K)\) such
that \(\mathcal{G}(K)\) does not involve any denominator, \(K^c\) can be
computed through
\begin{equation}
K^c = \big\langle \mathcal{G}(K) \big\rangle_{F[\underline X]} : f^\infty \, . 
\end{equation}
We stress that here \(\mathcal{G}(K)\) is now being used to generate an ideal in the original ring, rather than the extended ring.
Here \(f\) is a polynomial
defined as
\begin{equation}\label{eq:f_poly_definition_contraction}
    f = \text{lcm}\{\text{HC}(g) \in \mathbb{F}[\underline Y] \; : \; g \, \in  \, \mathcal{G}(K)\} \, ,
\end{equation}
where \(\text{HC}\) denotes the head coefficient, which is the
coefficient of the lead monomial (LM) defined in
Eq.~\eqref{eq:lead_monomial}, and \(\text{lcm}\) stands for least
common multiple. Starting from a reduced Gröbner basis
\(\mathcal{G}_R(K)\)\footnote{A reduced Gröbner basis has unit head
coefficients.}, we can obtain a Gröbner basis \(\mathcal{G}(K)\) which
is free of denominators by multiplying through each entry of
\(\mathcal{G}_R(K)\) by the \(\text{lcm}\) of its denominators. Therefore,
the polynomial \(f\) can be thought of as the lcm of all
denominators of \(\mathcal{G}_R(K)\).

It is interesting to consider what happens if one takes an ideal \(J\), performs extension, and then
contraction thereafter. Together, the two operations constitute a map \(J
\subseteq \mathbb{F}[\underline X] \; \rightarrow \; J^{ec} \subseteq
\mathbb{F}[\underline X]\), where we denote the extended-contracted ideal as
\(J^{ec}\). It can be shown that \(J \subseteq J^{ec}\), while the reverse inclusion
is in general not true. Therefore, we see that contraction is not the inverse of
extension, as some information may be lost.

Let us now consider how to compute the ideal \(J^{ec}\) directly from \(J\). 
To this end, we introduce a particular ``block'' ordering \(\succeq\), where
\(\underline Z \succ \underline Y\).
Using this ordering, we construct \(\mathcal{G}_{\succeq}(J)\), a Gröbner basis of
\(J\) in the original ring with respect to \(\succeq\). It can be shown that
\(\mathcal{G}_{\succeq}(J)\) is also a Gröbner basis of \(J^e\) in the extended ring
\cite[Lemma 8.93]{becker2012groebner}. 
The fact that we have a single set of polynomials which is a generating set of
\(J\) and \(J^e\) in their respective rings allows us to relate \(J^{ec}\) to \(J\).
To this end, we turn to
Eq.~\eqref{eq:f_poly_definition_contraction}, and take \(K = J^e\). We see that \(\mathcal{G}(K)\) can be taken as \(\mathcal{G}_{\succeq}(J)\) as
\(\mathcal{G}_{\succeq}(J)\) is a Gröbner
basis of an ideal in the extended ring, \(J^e\), that has no denominators. 
Furthermore, this means that
\(\mathcal{G}_{\succeq}(J)\) can be used to compute the polynomial \(f\) of
Eq.~\eqref{eq:f_poly_definition_contraction}. Finally, recall that 
\(\big\langle \mathcal{G}_{\succeq}(J) \big\rangle_{F[\underline X]} = J\), so we can take the ideal \(J\) itself in the right hand side of Eq.~\eqref{eq:f_poly_definition_contraction}.
Therefore, the extended-contracted ideal can be computed as
\begin{equation}
J^{ec} = J : f^\infty \, .
\end{equation}

It is interesting to ask if we can recover \(J\), given \(J^{ec}\). The answer to this
question lies in the following splitting lemma
\cite[Lemma 8.95]{becker2012groebner}
\begin{equation}\label{eq:splitting-lemma}
    J = (J + f^s) \cap (J : f^s) \, ,
\end{equation}
where \(s\) is the saturation index defined in
Eq.~\eqref{eq:ideal-saturation}.  While this splitting lemma holds
for any polynomial \(f\), if we take \(f\) according to
Eq.~\eqref{eq:f_poly_definition_contraction} then the right-hand
term \(J : f^s\) in the intersection is nothing but the
extended-contracted ideal \(J^{ec}\). Therefore, we obtain an expression for \(J\) in terms of \(J^{ec}\).

\subsection{A Primality Test for Equi-dimensional Ideals}
\label{sec:orgea00ff4}
\label{IdealPrimeCheck}

In this section, we make use of the extension/contraction operations
to arrive at a test for checking whether a certain class of ideals is
prime. Specifically, we will consider ideals which we know to be
equi-dimensional.  These are ideals for which every element of the
primary decomposition has the same dimension as the original ideal,
that is, there are no embedded components.  This class of ideals is
sufficient for our use case as the ideals which form the set
\(\mathfrak{D}^{(2)}\) at five-point are all
equi-dimensional (see Eq.~\eqref{eq:CodimensionMIdeals} and Section \ref{sec:orgcdd6994}). Specifically, consider a maximal-codimension ideal
\(J\) in a Cohen--Macaulay ring, such as \(R_n\). Then, it follows that
\(J\) is equi-dimensional \cite[Theorem 17.6]{matsumura_1987}.
Crucially, as all ideals \(P_i\) in Eqs.~\eqref{eq:5pt_prim_dec_first} to \eqref{eq:5pt_prim_dec_last}
can be computed as saturations of an
equi-dimensional ideal, they must also be equi-dimensional.

To determine if an equi-dimensional ideal \(J\) is prime, we recall that
\begin{equation}
    J = (J + f^s) \cap J^{ec} \, ,
\end{equation}
where the polynomial \(f\) is defined according to
Eq.~\eqref{eq:f_poly_definition_contraction} and \(J^e \subset \mathbb{F}(\underline{Y})[\underline{Z}]\).
For the purposes of our test, we stress that \(\underline{Y}\) is a maximally
independent set of \(J\), such that \(J^e\) is a zero-dimensional ideal.
We are going to
test the left-hand term \((J + f^s)\) for redundancy in the
intersection, and the right-hand term \(J^{ec}\) for
reducibility. Given equi-dimensionality, by definition no primary
component of \(J\) can have lower dimensionality. Therefore, if either
term in the intersection is of lower dimensionality then it must be
redundant. This implies that
\begin{equation}\label{eq:primetest}
J \;\, \text{is prime iff} \;\, J^{ec} \;\, \text{is prime and} \;\,  \text{dim}(J + f^s) < \text{dim}(J) \, .
\end{equation}

To check if \(J^{ec}\) is prime, we can use that \(J^{e}\)
being prime implies that \(J^{ec}\) is prime as well \cite[Lemma 8.97]{becker2012groebner}.  
The easiest way to check if \(J^e \subseteq F(\underline
Y)[\underline Z]\) is prime is to check if a (reduced) lexicographical
Gröbner basis of \(J^e\) takes the form
\begin{equation}
    \mathcal{G}(J^e) = \{Z_1 - \zeta_1(\underline{Y}), \ldots, Z_n - \zeta_n(\underline{Y}) \} \, ,
\label{eq:LinearGroebnerBasisForm}
\end{equation}
where \(\zeta_i \in \mathbb{F}(\underline{Y})\).
If this is the case then \(J^e\) is a maximal ideal and maximal ideals 
are prime. 
This check can be done semi-numerically by taking \(\underline Y\) in a
finite field, as in Section \ref{sec:FiniteFieldVarietyPoints}.

We then want to show that \((J + f^s)\) is redundant in the
intersection. Since \(J\) is equi-dimensional, this is the case if \((J +
f^s)\) is of lower dimension. So redundancy can be reduced to dimension testing for \((J+f^s)\).
However, computing the dimensionality of \((J +
f^s)\) can be quite computationally expensive if the polynomial \(f\) is
of high degree. However, we can make use of a further splitting lemma
embedded in Ref.~\cite[Lemma 8.52]{becker2012groebner}.  
Consider \(ab \in I\). Let \(\mu\) be the saturation index of \(b\) in \(I\), then
\begin{equation}
    I = (I + a) \cap (I + b^\mu) \, .
    \label{eq:SplittingLemmaTwo}
\end{equation}
Therefore, it is sufficient to check that the dimensionality of \((J + f_i)\)
drops for each factor \(f_i\) of \(f\). Furthermore, there may be several
\(\underline{Y}\) such that the associated \(J^e\) is manifestly
maximal. Thus, there is some freedom in the choice of
\(\underline{Y}\). It can be helpful to iterate through all choices of
\(\underline Y\) such that \(J^e\) is maximal and choose that with the
simplest \(f\).

Lastly, let us stress that failure to find a linear Gröbner basis for
\(J^e\) does not imply reducibility of \(V(J)\), while it can be shown
that failure of \((J + f^s)\) to drop in dimension does imply
reducibility of \(V(J)\). Using this test, together with ideal
intersection, one can easily and efficiently prove the primary
decompositions given in Section \ref{sec:orgcdd6994}.

\section{The Bracket Polynomial Quotient Ring}
\label{sec:org1870a3f}
\label{LorentzInvarianceProof}  

In Section \ref{sec:IndependentFunctions}, we claimed that the ring of polynomials that only pick up little-group scalings under
Lorentz transformations, \(\mathcal{R}_n\), is isomorphic to
\(\mathcal{R}_n^{(q)}\), a quotient ring of the polynomial ring \(\mathcal{S}_n\).
In this appendix, we develop this statement mathematically.

We begin by connecting \(\mathcal{S}_n\) to \(R_n\). We use a ring homomorphism \(\phi: \,\mathcal{S}_n \, \rightarrow \, R_n\), that acts on the variables in \(\mathcal{S}_n\) as
\begin{align}
    \phi(\langle ij \rangle) = \lambda_{i 1} \lambda_{j 0} - \lambda_{i 0} \lambda_{j 1} \quad\; \text{and} \quad\;
    \phi([ij]) = \tilde{\lambda}_{i \dot 0} \tilde{\lambda}_{j \dot 1} - \tilde{\lambda}_{i \dot 1} \tilde{\lambda}_{j \dot 0} \, .
\label{eq:RingHomomorphism}
\end{align}
We note that as \(\phi\) is a ring homomorphism, for all \(a, b \in
\mathcal{S}_n\) it satisfies \(\phi(ab) = \phi(a)\phi(b)\) and \(\phi(a + b) =
\phi(a) + \phi(b)\). Therefore it is sufficient to define the action of \(\phi\) on
the variables of \(\mathcal{S}_n\). 
Physically, it is clear that the map \(\phi\) is re-expressing any
polynomial in spinor brackets in terms of the spinor variables. That is, \(\phi\)
implements Eq.~\eqref{eq:spinorBrackets}. Let us now consider the image of the map
\(\phi\) in \(R_n\). By construction, it is the requisite subset of \(R_n\), i.e.
\begin{equation}
    \mathcal{R}_n = \{\phi(x) : x \in \mathcal{S}_n\} \, .
\end{equation}
By the so-called ``second isomorphism theorem'' \cite[Corollary 1.56]{becker2012groebner} this means that
\begin{equation}
    \mathcal{R}_n \cong \mathcal{S}_n / \mathrm{ker}(\phi) \, .
\label{eq:SubringIsomorphism}
\end{equation} 
That is, the image of \(\phi\), \(\mathcal{R}_n\), is isomorphic to the quotient of
\(\mathcal{S}_n\) by the kernel of the map \(\phi\). In other words, physically
inequivalent polynomials in spinor brackets can be identified with the elements of
\(\mathcal{S}_n\) which are inequivalent modulo the elements of \(\mathcal{S}_n\)
which \(\phi\) maps to zero.

Given an ideal \(J\) of \(R_n\), let us now consider how to construct the ideal \(J \cap \mathcal{R}_n\) of \(\mathcal{R}_n\). 
To this end, consider a homomorphism \(\phi' : \mathcal{S}_n \rightarrow S_n / [\pi^{-1}_{S_n, R_n}(J)]\), where \(\phi'\) takes the same form as
Eq.~\eqref{eq:RingHomomorphism}, but where the right hand side is considered in \(S_n / [\pi^{-1}_{S_n, R_n}(J)]\). 
The kernel of \(\phi'\) is an ideal of \(\mathcal{S}_n\) consisting of all
polynomials in spinor brackets that map to elements of \(J\). Importantly,
\(\ker(\phi')\) contains \(\ker(\phi)\) and so \(\pi_{\mathcal{S}_n, \mathcal{S}_n /\mathrm{ker}(\phi)}[\ker(\phi')]\)
is the associated ideal in the quotient ring \(\mathcal{S}_n /
\ker(\phi)\). We have therefore constructed an ideal of \(\mathcal{R}_n\) by Eq.~\eqref{eq:SubringIsomorphism}.

To be able to make practical computations with functions of spinor brackets, we
must be able to identify the kernel of ring homomorphisms. It turns out that kernels of
homomorphisms similar to \(\phi\) and \(\phi'\) can be computed with Gröbner
basis techniques. In full generality, consider a ring homomorphism 
\begin{equation}
\psi \,\,: \,\, \mathbb{F}[X_1, \ldots, X_n] \quad \longrightarrow \quad \mathbb{F}[Y_1, \ldots, Y_m]/\langle a_1, \ldots, a_k \rangle_{\mathbb{F}[Y_1, \ldots, Y_m]} \, ,
\end{equation} 
where we know explicit representatives in \(\mathbb{F}[Y_1, \ldots, Y_m]\) of the
\(\psi(X_i)\).
Now, define the ideal 
\begin{equation}
    \mathcal{K} = \langle a_1, \ldots, a_k, X_1 - \psi(X_1), \ldots, X_n - \psi(X_n) \rangle_{\mathbb{F}[X_1, \ldots, X_n, Y_1, \ldots, Y_m]} \, ,
\end{equation}
where by \(\psi(X_i)\) we mean a representative in \(\mathbb{F}[Y_1, \ldots, Y_m]\).
It can then be shown that \cite[Proposition 15.30]{eisenbud1995commutative}
\begin{equation}
    \mathrm{ker}(\psi) = \mathcal{K} \cap \mathbb{F}[X_1, \ldots, X_n] \, .
    \label{eq:KernelRelation}
\end{equation}

\bibliography{padicansatz}
\end{document}